\documentclass[12pt, a4paper]{article}
\pdfoutput=1
\usepackage[utf8]{inputenc}
\usepackage{fancybox}
\usepackage[T1]{fontenc}
\usepackage{lmodern, textcomp}
\usepackage{bm}
\usepackage[british]{babel}
\usepackage{csquotes}

\usepackage{eurosym}
\usepackage[nottoc]{tocbibind}
\usepackage{authblk}
\usepackage{fancyhdr}
\usepackage{xspace}
\usepackage{soul}

\usepackage{graphicx}
\usepackage{subcaption}
\usepackage{float}
\usepackage{makecell}
\usepackage{multirow}
\usepackage{multicol}
\usepackage{marginnote}
\usepackage[multiple]{footmisc}

\usepackage[usenames, dvipsnames, svgnames, table]{xcolor}

\usepackage[backend=bibtex8,
            style=ieee,
            dashed=false,
            citestyle=numeric-comp,
            maxbibnames=99,
			sorting=none,
            sortcase=false,
            sortcites=true,
            giveninits=true]{biblatex}

\usepackage[normalbf,normalem]{ulem}
\usepackage{amsmath, amsfonts, amssymb}
\usepackage{mathtools}
\usepackage{mathrsfs}
\usepackage{cancel}
\usepackage{braket}
\usepackage{tensor}
\usepackage{slashed}
\usepackage{siunitx}
\usepackage{listings}

\usepackage{stmaryrd}
\usepackage{caption}
\usepackage{relsize,exscale}
\usepackage{array}
\usepackage[toc,page]{appendix}

\usepackage{enumerate}

\DeclareSymbolFont{largesymbols}{OMX}{cmex}{m}{n}

\setcellgapes{1pt}
\makegapedcells
\newcolumntype{R}[1]{>{\raggedleft\arraybackslash }b{#1}}
\newcolumntype{L}[1]{>{\raggedright\arraybackslash }b{#1}}
\newcolumntype{C}[1]{>{\centering\arraybackslash }b{#1}}

\newcommand{\Tr}{\mathrm{Tr}}

\newtheorem{theorem}{Theorem}
\newtheorem{definition}{Definition}

\newtheorem{remark}{Remark}
\newtheorem{claim}{Claim}

\newcommand{\beq}{\begin{equation}}
\newcommand{\eeq}{\end{equation}}
\newcommand{\bea}{\begin{eqnarray}}
\newcommand{\eea}{\end{eqnarray}}
\newcommand{\dd}{\mathrm{d}}
\definecolor{mygray}{gray}{0.3}

\newcommand{\bes}{\begin{eqnarray}}
\newcommand{\ees}{\end{eqnarray}}

\newcommand\restr[2]{{
  \left.\kern-\nulldelimiterspace 
  #1 
  \vphantom{\big|} 
  \right|_{#2} 
  }}

\usepackage{multicol}
\usepackage{amssymb}

\usepackage[heightrounded, top=3.5cm, bottom=3.5cm, left=2.5cm, right=2.5cm, headheight=14pt]{geometry}

\usepackage[pdftex, breaklinks=true, linktocpage,
	colorlinks=true, urlcolor=blue, linkcolor=blue, citecolor=red
]{hyperref}

\newcommand{\email}[1]{\href{mailto:#1}{\nolinkurl{#1}}}
\newcommand{\emailfoot}[1]{\thanks{\email{#1}}}

\usepackage{marginnote}
\newcounter{draftcommentcnt}
\NewDocumentCommand{\draftcomment}{s O{red} m}{%
	\def\margnote{\IfBooleanTF{#1}{\marginnote}{\marginpar}}%
	\stepcounter{draftcommentcnt}%
	\textcolor{#2}{#3}%
	\margnote{\textcolor{#2}{$\Leftarrow$ \arabic{draftcommentcnt}}}%
}

\fancypagestyle{preprint}{
	\fancyhf{}

	\fancyhead[R]{\textsf{\preprint}}
}

\numberwithin{equation}{section}
\usepackage{biblatex}
\hypersetup{
	pdftitle={Large time effective kinetics $\beta$-functions for quantum
(2 + p)-spin glass II:},
}

\title{Large time effective kinetics $\beta$-functions for quantum
(2 + p)-spin glass II:
\\ \relax
    {\Large Effective vertex expansion, local potential approximation and symmetries}}

\author[2]{B\^em-Bi\'eri Barth\'el\'emy Natta\emailfoot{nattabarth@gmail.com}}
\author[1]{Vincent Lahoche\emailfoot{vincent.lahoche@cea.fr}}
\author[1,2]{Dine Ousmane Samary\emailfoot{ousmanesamarydine@yahoo.fr}}
\author[1]{Parham Radpay\emailfoot{parhamradpay@gmail.com}}

\affil[1]{%
	Université Paris-Saclay, \textsc{Cea}, Palaiseau, F-91120, France
}

\affil[2]{%
	Faculté des Sciences et Techniques (ICMPA-UNESCO Chair)
	\protect\\
	Université d'Abomey-Calavi, 072 BP 50, Benin
}

\begin{document}

\maketitle
\hrule

\hrule
\begin{abstract}
The accurate description of the glassy transition and long-time dynamics in quantum spin systems remains a major theoretical challenge, often beyond the reach of standard perturbation theory. In this paper, we continue our study initiated in [Annals Phys. 480 (2025), 170102] of the functional renormalization group applied to the quantum $(2+p)$-spin dynamics of an $N$-vector $\bm{x}\in \mathbb{R}^N$. Our unconventional approach relies on coarse-graining over the eigenvalues of the disorder matrix, which act as an effective kinetic term whose distribution follows a deterministic law in the large-$N$ limit. This method proved fruitful in our previous study focusing on the symmetric phase (vanishing magnetization). Here, we extend the formalism to the broken-symmetry phase. As in the symmetric case, we demonstrate that the flow exhibits finite-scale singularities. We show that these singularities can be cured by introducing replica-correlating operators, a mechanism that, at zero temperature, is associated with a first-order phase transition. Finally, the momentum-dependent fixing procedure required by the construction of the kinetic term yields non-trivial Ward identities, which we discuss in the large-$N$ limit.

\medskip

\end{abstract}
\hrule
\hrule
\bigskip

\noindent \textbf{keywords:} Functional renormalization group, Stochastic field theory, random matrix theory, non-local field theory.

\setcounter{footnote}{0}
\newpage

\hrule
\pdfbookmark[1]{\contentsname}{toc}
\tableofcontents
\bigskip
\hrule

\clearpage


\section{Introduction}

Over the last two decades, many approaches to quantum spin glasses have been explored in the literature. The most popular approach is the so-called quantum ferromagnet model, which can be viewed as a quantum analogue of the Heisenberg model, with classical three-dimensional spin vectors replaced by quantum spins represented by Pauli matrices \cite{baldwin2017clustering,cugliandolo2004effects}. Another prominent example is the fermionic Sachdev-Ye-Kitaev (SYK) model, known for its maximal chaos and widely studied as an effective quantum model for black holes, with applications in condensed matter physics and quantum gravity \cite{blake2021systems,chowdhury2022sachdev,rosenhaus2019introduction,gurau2018prescription,dartois2019conformality}. In this paper, we investigate a third, less common framework, sometimes called the quantum spherical $p$-spin-glass model, which describes the behavior of a quantum particle moving in an $N$-dimensional random energy landscape \cite{biroli2001quantum,rokni2004dynamical}.

A central theoretical challenge in the study of these systems is providing a robust description of the long-time dynamics and the transition to the glassy phase. In these regimes, standard perturbative methods typically break down due to the emergence of strong correlations and a complex free-energy landscape. To overcome these limitations, we consider a "2+p" random energy landscape, where the couplings are drawn from both a random Gaussian matrix and a random tensor of rank $p$. While the corresponding classical problem of spherically constrained soft spins with "2+p" interactions can be solved exactly using replica symmetry breaking theory \cite{crisanti2006spherical}, the quantum dynamical counterpart requires non-perturbative tools (see \cite{Tarjus,Tarjus:2004wyx,tarjus2008nonperturbative} and references therein).

At the outset, it should be emphasized that our intention in this work, as in our previous one \cite{lahoche2024largetimeeffectivekinetics}, is primarily to investigate certain mechanisms underlying glassy systems through the lens of the renormalization group. This paper thus represents a further step in an ongoing research program. Our medium-term objective is to develop reliable approximations—built upon the models discussed in our earlier works—to address complex problems that currently lack analytical solutions. In particular, we aim to extend, in the future, the method we recently introduced for the detection of classical signals in quasi-continuous spectra \cite{lahoche3, lahoche4, Finotello1, RG7} to the quantum regime, where disorder exhibits more complex, structured patterns.

In this work, we extend our previous investigations \cite{lahoche2024largetimeeffectivekinetics} using the functional renormalization group (FRG) formalism \cite{Berges,Delamotte_2012}. Our approach employs an unconventional FRG scheme based on coarse-graining over the empirical spectrum of the disorder matrix, which acts as an effective kinetic term and converges to the deterministic Wigner semi-circle law in the large-$N$ limit (this methods was also used in the references \cite{lahoche20241,lahoche20242,lahoche20244}). A distinctive feature of this RG scheme is that the canonical dimensions depend not only on the RG scale but also on two-point observables, such as the mass. In our previous work, we established the mathematical framework and focused on simplified scenarios where these dependencies were minimal. Specifically, we addressed: (i) perturbation theory around the Gaussian fixed point; (ii) the vertex expansion, the leading order of the derivative expansion, in the deep infrared (IR) regime; and (iii) the vertex expansion in the critical regime.

In the first and third regimes, the canonical dimension primarily depends on the RG scale rather than the mass. In the deep IR, power counting reveals that the Gaussian theory behaves similar to an ordinary Euclidean field theory in four dimensions. In the second case, while the power counting resembles that of an ordinary field theory, the system exhibits non-local interactions. Our previous investigations revealed notable features driven by the tensorial disorder, including finite-scale singularities—appearing as "cusps" or sharp analytical divergences depending on the regime—and the existence of attractors in phase space for trajectories emerging from IR critical conditions, accompanied by rapid oscillations.

Some of these phenomena have been observed in other glassy systems \cite{Tarjus,Tarjus2} and beyond \cite{Delamotte2}. Previously, relying on a two-particle irreducible (2PI) formalism, we found evidence suggesting that these singularities are inherently linked to the emergence of correlations between replicas. We showed that as the flow approaches these singularities, the effective potential associated with these correlations develops a non-zero minimum without explicit replica symmetry breaking, a mechanism compatible with a first-order phase transition. In the present paper, we will not revisit the 2PI formalism, as it will be addressed comprehensively in future work. Instead, our primary goal is to develop a complete and robust framework to investigate the phase space beyond the symmetric phase. The formalism presented here is specifically designed to distinguish the singularities related to replica correlations from those associated with standard second-order phase transitions to an ordered phase. This investigation forms the core of the second and main part of the paper.

In addition, the first part of the paper is dedicated to improving the symmetric-phase RG equations beyond the standard vertex expansion used in \cite{lahoche2024largetimeeffectivekinetics}, by applying the effective vertex expansion developed, for instance, in \cite{Lahochebeyond}. This method exploits the structure of the leading-order graphs in the large-$N$ limit to close the hierarchy; it is therefore intended to be more precise than—or at least complementary to—the standard vertex expansion. While this section is rather technical, readers primarily interested in the physical implications beyond the symmetric phase may skip it without loss of continuity. Furthermore, we note that this work is complemented by the companion paper \cite{lahoche2024frequency}, which examines the quantum $p$-spin model via an RG scheme based on frequency coarse-graining. Finally, we utilize exact Ward identities to constrain the construction of next-to-leading-order solutions. These identities arise from the momentum-dependent kinematic-fixing procedure required by our RG framework, yielding highly non-trivial relations in the large-$N$ limit.

To conclude this introduction, let us emphasize an important conceptual point. Although canonical dimension analysis suggests that all local couplings are relevant in the deep UV, this does not imply that the renormalization group is ineffective. In fact, in certain discrete quantum gravity models, such as random matrix models \cite{brezin1992renormalization,dupuis2021nonperturbative,eichhorn2014towards,lahoche2020revisited}, all couplings are technically irrelevant, yet the flow admits a non-trivial fixed point with a relevant direction, where the critical exponent governs the double-scaling limit \cite{di19952d}.
\medskip

The paper is organized as follows: In Section \ref{sec1}, we introduce the model, our conventions, and the basics of the Wetterich-Morris formalism, the $1/N$ expansion, and perturbation theory. Section \ref{EVE} applies the effective vertex expansion to improve upon standard vertex approximations, using the large-N structure of Feynman graph to re-sum the leading sector. Section \ref{secLPA} develops a formalism based on an expansion around the vacuum of the local potential, examining an artificially simplified case to isolate and illustrate the key mechanisms. Section \ref{sec6} extends this expansion to include non-local couplings and replica interactions. 
In the last section, Section \ref{sec2}, we discuss Ward identities arising from the gauge fixing arising because of the propagator.
Finally, Section \ref{sec5} outlines open issues and directions for future research.

\section{Technical preliminaries}\label{sec1}

In this section, we introduce the model, establish our conventions, and present the tools that will be used throughout the remainder of this paper. Specifically, we review the basics of the functional renormalization group and the formal large-$N$ solutions of the underlying field theory, following our previous work \cite{lahoche2024largetimeeffectivekinetics}.

\subsection{Definition of the model}

Let us begin by defining the model. We consider a non-relativistic quantum particle in an $N$-dimensional space, moving through a rough landscape characterized by the combination of two types of disorder: one represented by a Gaussian random matrix $K$, and the other by a Gaussian random tensor $J$ of rank $p > 2$. A version of this quantum mechanical model, without the matrix disorder, was investigated in \cite{biroli2001quantum} and references therein. The wave function of the particle, $\psi(\bm x,t)\in \mathbb{C}$ with $\bm x \equiv (x_1,\cdots, x_N)\in \mathbb{R}^N$, evolves according to the Schr\"odinger equation:
\begin{equation}
\boxed{\hbar\frac{\partial}{\partial t} \psi(\bm x,t)=-\hat{\mathcal{H}}\, \psi(\bm x,t)\,,}
\end{equation}
where the Hamiltonian $\hat{\mathcal{H}}$ reads in the generalized position space basis:
\begin{equation}
\hat{\mathcal{H}}=-\frac{\hbar^2}{2m_0} \frac{\partial^2}{\partial \bm x^2}+U_{J,K}(\bm x)+V(\bm x^2)\,.\label{Hamiltonian Model}
\end{equation}
The \textit{deterministic} potential $V(\bm x^2)$ is a polynomial function of the $O(N)$-invariant squared norm $\bm x^2:=\sum_{i=1}^N x_i^2$:
\begin{equation}
V(\bm x^2):=N \sum_{n=1}^{n_0}\, \frac{h_n}{(2n)!} \left(\frac{\bm x^2}{N}\right)^n\,,
\end{equation}
where the coupling constants $h_n$ can be arbitrary real numbers, except for $h_{n_0} > 0$ due to the stability condition. The powers of $N$ ensure the existence of a $1/N$ expansion (see \cite{ZinnJustinReview} and Section \ref{largeN}). The \textit{random} potential $U_{J,K}({\bm x})$ captures the disorder effects:
\begin{equation}
U_{J,K}({\bm x}):=\frac{1}{2}\sum_{i,j=1}^N K_{ij}{x}_{i}{x}_{j}+ \frac{1}{p!} \sum_{i_1,\cdots i_p}\, J_{i_1\cdots i_p} \, {x}_{i_1} \cdots  {x}_{i_p}\,,\label{defUJK}
\end{equation}
where:
\begin{itemize}
    \item The elements $K_{ij}$ are the entries of an $N\times N$ symmetric random matrix drawn from the Gaussian Orthogonal Ensemble (GOE), with zero mean and variance $\sigma/N$.
    \item The elements $J_{i_1\cdots i_p}$ are the components of a symmetric Gaussian random tensor of rank $p$ and size $N^p$, such that:
    \begin{equation}
\overline{J_{i_1\cdots i_p}}=0\,,\qquad \overline{J_{i_1\cdots i_p} J_{j_1\cdots j_p}}=\frac{\lambda}{N^{p-1}}\, \sum_{\pi\in \mathfrak{P}_p}\, \prod_{m=1}^p \delta_{i_m j_{\pi(m)}}\,.
    \end{equation}
\end{itemize}
    
\noindent
The coupling $\lambda$ characterizes the strength of the disorder, $\mathfrak{P}_p$ is the set of permutations of $p$ elements, and the overline denotes the average with respect to the probability distribution of $J$. The potential $V(\bm{x}^2)$ suppresses large-$\bm{x}$ configurations in the classical equations of motion:
\begin{equation}
m_0 \frac{d^2 x_i}{dt^2}= -\frac{\partial}{\partial x_i} \left(U_{J,K}({\bm{x}})+V({\bm{x}}^2)\right)\,.
\end{equation}
In Appendix \ref{App2}, we analytically investigate this equation for $p=0$ in the quenched regime at late times.
\medskip

The operator formalism of quantum mechanics is not well-suited for the renormalization group (RG) approach; therefore, we switch to the path integral formalism. The time evolution of a Hermitian operator is given by:
\begin{equation}
U(t,t_0):=e^{-\frac{t-t_0}{\hbar}\hat{\mathcal{H}}}\,,
\end{equation}
can be "Trotterized" following the standard procedure \cite{kluber2023trotterization} to construct the path integral. Consequently, the Euclidean propagator takes the form of the partition function of a system in contact with a thermal bath at inverse temperature $\beta=t-t_0$ \cite{zinn2010path}:
\begin{equation}
\mathcal{Z}_{\beta}[K,J,\bm L]=\int [\mathcal{D} x(t)]\, e^{-\frac{1}{\hbar} S_{\text{cl}}[\bm x(t)]+\frac{1}{\hbar} \int_{-\beta/2}^{\beta/2} \dd t \sum_{k=1}^N L_k(t) x_k(t)}\,,\label{pathintegralZ}
\end{equation}
where $\bm L=(L_1,\cdots, L_N)$ and the \textit{classical action} $S_{\text{cl}}[\bm x(t)]$ is:
\begin{equation}
\boxed{S_{\text{cl}}[\bm x(t),J,K]:=\int_{-\beta/2}^{\beta/2} \dd t \left(\frac{1}{2}\dot{\bm{x}}^2+U_{J,K}({\bm{x}})+V({\bm{x}}^2)\right)\,,}\label{classicaction}
\end{equation}
subject to periodic boundary conditions $\bm x(t)=\bm x(t+\beta)$. In this paper, we primarily focus on the limit $\beta\to \infty$, which corresponds to the zero-temperature limit. Note that we use the notation $[\mathcal{D} \bm{x}(t)]$ for the path integral measure to distinguish it from the standard Lebesgue integration measure $\dd \bm x$. Furthermore, since the mass $m_0$ does not renormalize in the large-$N$ limit within the symmetric phase (it corresponds to the field strength renormalization, see below), we set $m_0=1$
\medskip

As is standard in quantum field theory (QFT), the path integral \eqref{pathintegralZ} allows us to compute the vacuum expectation values of field correlations at different times\footnote{The hat denotes a quantum operator.} \cite{ZinnJustinBook2}:
\begin{align}
 \langle 0 \vert T\, \hat{x}_{i_1}(t_1)\hat{x}_{i_2}(t_2)\cdots \hat{x}_{i_n}(t_n) \vert 0 \rangle = \frac{1}{\mathcal{Z}[K,J,0]} \frac{\partial^n \mathcal{Z}[K,J,\textbf{L}]}{\partial L_{i_1}(t_1)\partial L_{i_2}(t_2)\cdots \partial L_{i_n}(t_n)}\,\Bigg\vert_{L=0}\,.
\end{align}
These correlations can be computed using perturbation theory, where the expansion is organized as a power series in $\hbar$. The limit $\hbar \to 0$ corresponds to the standard \textit{classical limit}. Finally, for finite $\beta$, the paths can be decomposed into modes,
\begin{equation}
\bm x(t)=\frac{1}{\sqrt{\beta}}\sum_{k=0}^\infty\, \tilde{\bm x}(\omega_k) e^{i \omega_k t}\,,\label{defmodes}
\end{equation}
where the frequencies are :
\begin{equation}
\omega_k:=\frac{2\pi k}{\beta}\,.
\end{equation}

\subsection{Large $N$ limit field theory}

The physical content of the partition function \eqref{pathintegralZ} can be investigated using perturbation theory. Note that although the model is free of divergences, perturbation theory is inadequate to fully define the partition function, since summing the perturbative series is usually a non-trivial issue requiring advanced constructive tools (see \cite{erbin2021constructive} and references therein). Nevertheless, we can set this issue aside and focus strictly on the perturbative definition of observables, as is standard practice in QFT.
\medskip

Consider the $2n$-point 1PI vertex function $\Gamma^{(2n)}$. In standard perturbation theory, the terms in the expansion correspond to Feynman amplitudes, which are indexed by 1PI Feynman diagrams:
\begin{equation}
\Gamma^{(2n)}=\sum_{\mathcal{G}\in \mathbb{G}_{2n}}\, \frac{1}{s(\mathcal{G})}\mathcal{A}(\mathcal{G})\,.
\end{equation}
where $\mathbb{G}_{2n}$ is the set of 1PI Feynman diagrams with $2n$ external points, and the additional factor $s(\mathcal{G})$, called the \textit{symmetry factor}, depends on the dimension of the automorphism group of the graph \cite{ZinnJustinBook1}. Feynman diagrams consist of vertices representing interactions and edges representing Wick contractions. These Wick contractions can be easily understood by exploiting the global $O(N)$ symmetry of the theory. Indeed, because the probability distributions for $K$ and $J$ are both $O(N)$-invariant\footnote{This means, for instance, that $J\equiv \{J_{i_1\cdots i_q} \}$ and its transformed version $O \triangleright J \equiv \{\sum_{j_1\cdots j_q} O_{i_1j_1}\cdots O_{i_qj_q} J_{j_1\cdots j_q} \}$ follow the exact same probability law, for any orthogonal matrix $O$.}, one can always choose a coordinate system such that $K$ is diagonal. Namely, let $\{u^{(\mu)}\}$ be the set of $N$ orthonormal eigenvectors, such that:
\begin{equation}
\sum_{j=1}^N K_{ij} u_j^{(\mu)}=\xi_\mu \, u_i^{(\mu)}\,,\label{eigenequation}
\end{equation}
where eigenvalues $\{\xi_\mu\}$ are reals. We then introduce the orthogonal transformation:
\begin{equation}
x_\mu(t):=\sum_{i=1}^N\, x_i u_i^{(\mu)}\,.\label{Gaugefix}
\end{equation}
In this basis, the bare propagator is diagonal, and its components in Fourier space read:
\begin{equation}
C(\omega_k,\xi_\mu):=\frac{1}{\omega^2+\xi_\mu+h_1}\,. 
\end{equation}
A typical Feynman diagram $\mathcal{G}$ contains subgraphs $g_n\subset \mathcal{G}$ composed of cycles (called faces, see below) that involve a sum over products of propagators, of the form:
\begin{equation}
I_n(\Omega_1,\cdots,\Omega_n):=\sum_{\mu} \prod_{i=1}^n C(\Omega_i,\xi_\mu)\,,
\end{equation}
where the $\Omega_i$ are the external frequencies of the subgraph. In the large-$N$ limit, this sum can be converted into an integral using the well-known Wigner theorem \cite{Mehta2004,potters2020first}:"

\begin{theorem}
Let $M:=\{M_{ij}\}$ be an $N\times N$ symmetric random matrix whose $N(N+1)/2$ independent entries in the upper triangular part are independent and identically distributed (i.i.d.). More precisely, we assume that the entries have zero mean and satisfy $\mathbb{E}(M_{ii}^2) < \infty$ and $\mathbb{E}(M_{ij}^2)=\sigma^2/N<\infty$ for $i\neq j$. These conditions define the Wigner ensemble, and in the large-$N$ limit, the empirical eigenvalue distribution:
\begin{equation}
\mu_{\text{em}}(x):=\frac{1}{N}\sum_{\mu=1}^N\delta(x-\xi_\mu)\,,
\end{equation}
converges weakly toward the \textit{Wigner semicircle law}:
\begin{equation}
\mu_{\text{em}}(x)\underset{N\to \infty}{\longrightarrow}\mu_{\text{W}}(x)
\end{equation}
where:
\begin{equation*}
\mu_{\text{W}}(x):=\left\{
    \begin{array}{ll}
        \frac{\sqrt{4\sigma^2-x^2}}{2\pi \sigma^2} & \mbox{if:} \,\, -2\sigma \leq x \leq 2\sigma \\
        0 & \mbox{if:}\,\, \vert x \vert > 2\sigma\,,
    \end{array}
\right.
\end{equation*}
is normalized such that $\int \mu_{\text{W}}(x) \dd x=1$. 
\end{theorem}\label{Wigth}

\begin{figure}
\begin{center}
\includegraphics[scale=0.8]{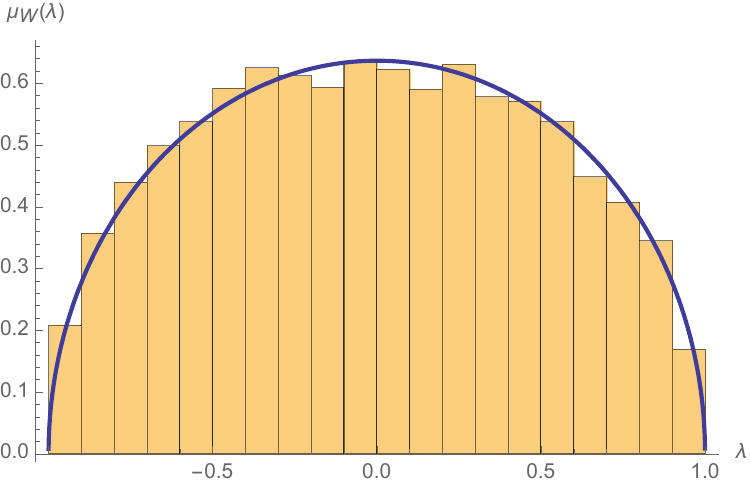}
\end{center}
\caption{Illustration of the convergence toward Wigner distribution. Histogram shows the eigenvalues distribution a $10^4\times 10^4$ i.i.d sample random matrix.}\label{figWig}
\end{figure}

Figure \ref{figWig} illustrates this theorem. The Gaussian Orthogonal Ensemble (GOE) is an example of Wigner matrices for which $\mathbb{E}(M_{ii}^2)=2\sigma^2/N$, and the $n$-point Feynman amplitude $I_{n}$ associated with the subgraph $g_n$ becomes:
\begin{equation}
\frac{1}{N}I_n(\Omega_1,\cdots,\Omega_n) \,\underset{N\to \infty}{\longrightarrow} \, I_n^{(\infty)}(\Omega_1,\cdots,\Omega_n)\equiv \int_{-2\sigma}^{+2\sigma} \mu_{\text{W}}(x) \, \dd x\, \prod_{i=1}^n C(\Omega_i,x)\,.
\end{equation}
For large-$N$ calculations, it is convenient to introduce the \textit{generalized momentum} $p^2$, which behaves as a positive variable in the limit $N\to \infty$:
\begin{equation}
p^2:=x+2\sigma\,.\label{generalizedmomentadef}
\end{equation}
Then, defining the \textit{physical bare mass} $m^2$ as\footnote{Note that the effective mass, much like $p^2$, is defined for arbitrary $N$, but $p^2$ becomes strictly positive only in this limit. A natural way to define $p^2$ for any finite-$N$ sample of $K$ would be to shift it by the smallest eigenvalue—a subtlety that is irrelevant for our present purposes.} :
\begin{equation}
m^2:=h_1-2\sigma\,,\label{generalizedmassdef}
\end{equation}
the Feynman amplitude for the subgraph $g_n$ becomes:
\begin{equation}
I_n^{(\infty)}(\Omega_1,\cdots,\Omega_n)=\int_{0}^{+4\sigma} \rho(p^2) \, \dd p^2\, \prod_{i=1}^n \tilde{C}(\Omega_i,p^2)\,,\label{amplitudeg}
\end{equation}
where the propagator $\tilde{C}$, involving the generalized momentum $p^2$ is:
\begin{equation}
\tilde{C}(\omega,p^2):=\frac{1}{\omega^2+p^2+m^2}\,,
\end{equation}
and where in \eqref{amplitudeg}, the distribution $\rho(p^2)$ is induced from $\mu_{\text{W}}(x)$ from the definition of $p^2$ i.e. $\rho(p^2)$ vanishes outside the interval $p^2\in [0,4\sigma]$, where it takes the value:
\begin{equation}
\rho(p^2)=\frac{\sqrt{p^2(4\sigma-p^2)}}{2\pi \sigma^2}\,.
\end{equation}
Formally, the amplitude \eqref{amplitudeg} resembles the Feynman amplitude associated with a subgraph in an ordinary but non-local field theory, where $p^2$ plays the role of a squared momentum. Note that for such a field theory in $D$ spatial dimensions, the momentum distribution is $\rho_{\mathbb{R}^D}(\vec{p}\,^2)\sim (\vec{p}\,^2)^{\frac{D-2}{2}}$. In particular, for sufficiently small $p^2$, $\rho(p^2)\simeq (p^2)^{\frac{1}{2}}$, and the theory behaves like a $3D$ Euclidean field theory. The analogy need not be pushed too far, however; we can simply note that an ordinary field theory, such as the one we are considering here, can be characterized by a generalized momentum distribution and a definition of locality \cite{lahoche2022generalized,carrozza2016flowing}. This perspective allows us to adopt a more abstract approach to the notion of dimension, which we will discuss in Section \ref{FRG}.

\subsection{Effective replicated non-local field theory}

From the perspective of the previous section, the two disorder components $K$ and $J$ are treated differently. Indeed, due to the properties of Wigner random matrices in the large-$N$ limit, the random spectrum of the matrix-like disorder converges toward a deterministic distribution, acting as an effective kinetic term. No equivalent property exists for its tensorial counterpart, despite efforts to construct one \cite{seddik2024random,goulart2022random,gurau2020generalization,lebeau2024random} (and references therein). We will therefore invoke the self-averaging assumption twice: first, to replace the effect of the disorder $K$ with the asymptotic expression of its eigenvalue density; and second, to average over the disorder $J$ using its explicit Gaussian probability density.

This last point raises the question of which type of average we should consider. An \textit{annealed average} involves directly averaging the partition function \eqref{pathintegralZ}, resulting in a field theory that is non-local with respect to generalized momenta and time. Physically, this kind of averaging corresponds to a situation where the typical timescales for the disorder and the quantum particles are comparable. Conversely, the \textit{quenched average} assumes that the typical timescale for the quantum particles is very small compared to that of $J$, and that the average over $J$ must be performed on the free energy $\mathcal{F}_{\beta}[K,J,\bm L]:=\ln \mathcal{Z}_{\beta}[K,J,\bm L]$ rather than on the partition function. This issue is generally solved by introducing the replica method \cite{castellani2005spin,Parisi,Dominicis,mezard1984nature,MezardInformation} via the elementary formula:
\begin{equation}
\overline{\mathcal{F}_{\beta}[K,J,\bm L]}=\lim_{n\to 0} \frac{\overline{\mathcal{Z}_{\beta}^n[K,J,\bm L]}-1}{n}\,.
\end{equation}
Although formally valid, this formula poses the delicate issue of taking the limit $n\to 0$, since averaging over $n$ copies of the disorder implicitly assumes that $n$ is an integer. This, in particular, raises the possibility of \textit{replica symmetry breaking} (RSB), which is one of the strong indicators of glassy phases. An alternative approach, which is less mathematically debated and more appropriate for functional renormalization group applications, is to perform the average using the \textit{cumulants} of the random quantity $W[K,J,\mathbf{L}]$ \cite{Tarjus,Tarjus2}. This requires explicitly breaking the replica symmetry by providing a different source for each replica, assuming $n\in \mathbb{N}^*$. Hence, the strategy we will adopt takes the form of a \textit{multi-local} expansion, depending on the order of the considered cumulants of $W[K,J,\mathbf{L}]$. These cumulants are defined as usual:
\begin{align*}
\mathcal{W}^{(1)}[\mathbf{L}_\alpha]&= \lim_{N\to \infty}\, \overline{W[K,J,\mathbf{L}_\alpha]}\\
\mathcal{W}^{(2)}[\mathbf{L}_\alpha,\mathbf{L}_\beta]&= \lim_{N\to \infty}\, \overline{W[K,J,\mathbf{L}_\alpha]W[K,J,\mathbf{L}_\beta]}-\mathcal{W}^{(1)}[\mathbf{L}_\alpha]\mathcal{W}^{(1)}[\mathbf{L}_\beta]\\
\mathcal{W}^{(3)}[\mathbf{L}_\alpha,\mathbf{L}_\beta,\mathbf{L}_\gamma]&=\lim_{N\to \infty}\, \overline{W[K,J,\mathbf{L}_\alpha]W[K,J,\mathbf{L}_\beta]W[K,J,\mathbf{L}_\gamma]}-\cdots
\end{align*}
where the Greek indices $\alpha, \beta,\gamma, \cdots$ running from $1$ to $n$ label different sources, namely, different replicas of the system. The $\cdots$ in the last line indicates the subtraction of all disconnected contributions. Furthermore, these cumulants are generated by the functional:
\begin{align}
\nonumber \lim_{N\to \infty}\ln \overline{\mathcal{Z}^n_{\beta}[K,J,\{\bm L_\alpha\}]}&= \sum_\alpha\mathcal{W}^{(1)}[\mathbf{L}_\alpha]+\frac{1}{2!} \sum_{\alpha,\beta}\mathcal{W}^{(2)}[\mathbf{L}_\alpha,\mathbf{L}_\beta]\\
&+\frac{1}{3!}\sum_{\alpha,\beta,\gamma}\mathcal{W}^{(3)}[\mathbf{L}_\alpha,\mathbf{L}_\beta,\mathbf{L}_\gamma]+\cdots\,,
\end{align}
hence, the generating functional $\mathcal{Z}^n_{\beta}[K,J,\{\bm L_\alpha\}]$ is now defined as:
\begin{align}
&\nonumber \mathcal{Z}^n_{\beta}[K,J,\{\bm L_\alpha\}]:= \exp \sum_{\alpha=1}^n W[K,J,\mathbf{L}_\alpha]\\
&\equiv \int \, \prod_{\alpha=1}^n[\mathcal{D} x_\alpha(t)]\, \exp \Bigg(-\frac{1}{\hbar}\sum_{\alpha=1}^nS_{\text{cl}}[\bm{x}_\alpha]+\int_{-\beta/2}^{+\beta/2}\,\dd t\,\sum_{k=1,\alpha=1}^{N,n} L_{k,\alpha}(t)x_{k,\alpha}(t)\Bigg)\,.
\end{align}
Note that, as usual, each replica corresponds to the same realization of the disorder. Averaging over the Gaussian distribution of $J$, we obtain:
\begin{equation}
\overline{\mathcal{Z}^n_{\beta}[K,\{\bm L_\alpha\}]}:=\int \, \prod_{\alpha=1}^n[\mathcal{D} x_\alpha(t)]\, e^{-\frac{1}{\hbar}\overline{S_{\text{cl}}}[\{\bm{x}\}]+\mathcal{J}} \equiv 
 \exp \, \left( \mathcal{W}^{(n)}_{\beta}[K,J,\{\bm L_\alpha\}] \right) \,,\label{partition1}
\end{equation}
where the source term is:
\begin{equation}
\mathcal{J}:=\int_{-\beta/2}^{+\beta/2}\,\dd t\,\sum_{k=1,\alpha=1}^{N,n} L_{k,\alpha}(t)x_{k,\alpha}(t)\,,
\end{equation}
and the \textit{classical averaged action} reads, in the thermodynamic limit:

\begin{equation}
\boxed{\overline{S_{\text{cl}}}[\{\bm{x}_\alpha\}]:=\sum_{\alpha}S_{\text{cl}}[\bm x_\alpha(t),J=0,K]-\frac{\lambda N}{2\hbar}\int_{-\beta/2}^{+\beta/2} \dd t \, \dd t^\prime\sum_{\alpha,\beta}\,  \left(\frac{\bm{x}_\alpha(t)\cdot \bm{x}_\beta(t^\prime)}{N}\right)^p\,,}\label{classicalaveraged}
\end{equation}

\begin{remark}
It is worth noting that the path integral does not converge for an arbitrary $p$ unless the highest-order interaction in the confining potential $V(\bm x^2)$ has a positive coupling and is at least of order $2(p+1)$. In particular, for $p=3$, the path integral converges safely if the confining potential involves a positive octic interaction. Furthermore, note that the coupling corresponding to the non-local contribution must be negative to remain consistent with a tensorial Gaussian disorder.
\end{remark}

\subsection{Functional renormalization group}\label{FRG}

Since the pioneering work of Wilson and Kadanoff \cite{Wilson, kadanoff1966scaling}, various forms of the RG have been developed for different contexts, intersecting with probability and information theories \cite{gordon2021relevance, apenko2012information, beny2015information, pessoa2018exact, cotler2023renormalization}. The RG appears to be a general concept in physics rather than a topic confined to QFT or condensed matter theory. Notably, it enables the exploration of effective infrared (IR) physics by progressively integrating out short-wavelength fluctuations. In field theory, this process is well-defined as long as the theory includes a canonical notion of scale. In QFT, this scale is often derived from the background space or space-time, which provides canonical definitions of \textit{microscopic} and \textit{macroscopic} physics, allowing wavelength modes to be uniquely ordered between these limits \cite{ZinnJustinBook1}. The RG flow can also be defined in a more abstract framework that does not depend on background structures like space-time, as long as quantum fluctuations can be assigned a scale. Here, this scale is determined by the spectrum of the Gaussian kernel, with fluctuations in the corresponding eigenbasis ordered by the magnitude of their eigenvalues \cite{lahoche2022generalized, Bradde, Lahochebeyond, CarrozzaReview}. From an information-theoretic perspective, the Gaussian kernel acts as the Fisher information metric \cite{amari2000methods}, so that intrinsic scaling acquires a geometric interpretation via the eigenvalues of this metric \cite{beny2015information, beny2015renormalization}. For other recent unconventional applications of the RG, see \cite{cheng2023simplex, villegas2023laplacian, lahoche2023functional, lahoche20241, erbin2022non} and references therein.
\medskip

\begin{figure}
\begin{center}
\includegraphics[scale=0.8]{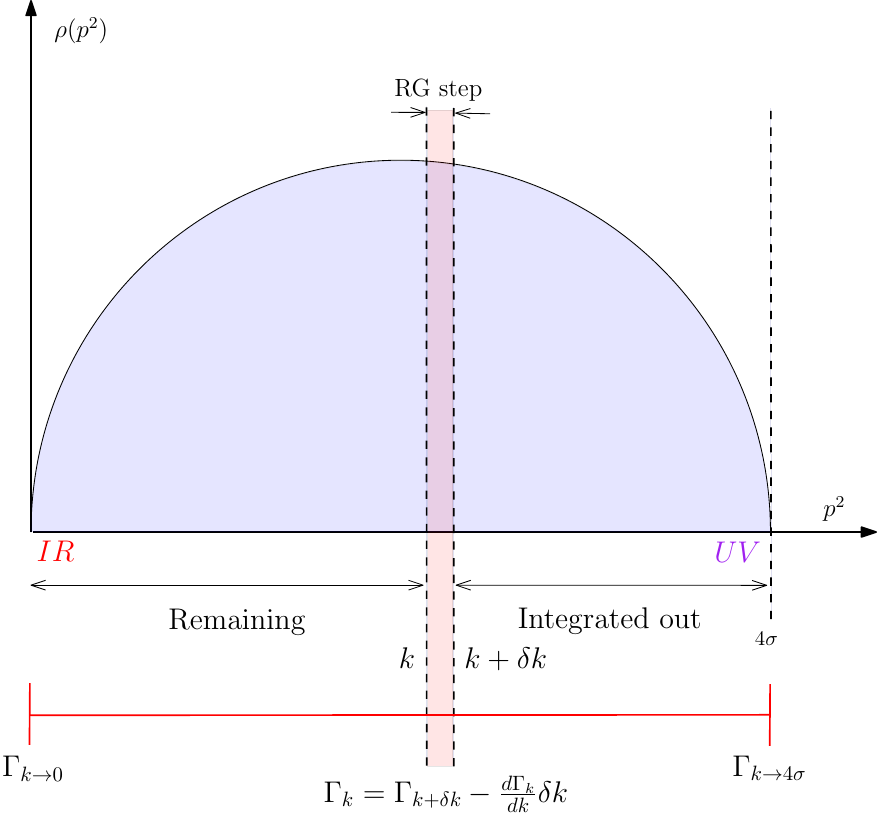}
\end{center}
\caption{Illustration of the coarse-graining definition of the Wigner RG flow. Modes are partially integrated from UV to IR, and the effective average action changes from each step.}\label{figRG1}
\end{figure}

In the field theory considered in this paper, the one-dimensional background manifold—time—provides a canonical notion of scale. Consequently, a renormalization group (RG) scheme can, in principle, be constructed by coarse-graining over frequencies. Such coarse-graining has recently been studied in contexts like classical spin glass kinetics \cite{lahoche2023functional, lahoche2022functional}, out-of-equilibrium field theories such as Model A and the Kardar-Parisi-Zhang (KPZ) equation \cite{duclut2017frequency, squizzato2019kardar}, and quantum mechanical systems \cite{zappala2001improving, synatschke2009flow, heilmann2015convergence}. In this paper, however, we adopt a more abstract RG approach based on coarse-graining over the spectrum of the kinetic disorder $K$, specifically over the generalized momentum (see Figure \ref{figRG1}). Unlike frequencies, this spectrum is not tied to any background geometry, serving as an example of the abstract scaling mentioned earlier. Moreover, it provides a canonical definition of the ultraviolet (UV) and infrared (IR) regimes, corresponding respectively to the regions $p^2 \approx 4\sigma$ and $p^2 \approx 0$. We can thus construct a coarse-grained description of the effective IR physics by partially integrating out the UV fluctuations associated with large generalized momenta.
\medskip

One could also consider a "mixed" coarse-graining approach that incorporates both frequency and momentum. In similar settings, this mixed approximation has been shown to offer significant qualitative advantages \cite{duclut2017frequency}. However, as our primary objective in this paper is to develop the underlying formalism, we opt for a simpler scheme by focusing solely on coarse-graining over the generalized momentum, reserving a more detailed examination of the mixed approach for a forthcoming paper \cite{lahoche2024frequency}. This momentum-only strategy has been explored in our previous work \cite{lahoche2024largetimeeffectivekinetics} and in other studies across various contexts \cite{duclut2017nonuniversality, guilleux2015quantum, canet2010nonperturbative, canet2011nonperturbative, canet2016fully}. Essentially, this amounts to examining the system's behavior \textit{in the long-time limit}, which is the specific regime we focus on in this article.
\medskip 

In the literature, there exist several popular 'exact' equations that describe how the physical representation of a field theory evolves as degrees of freedom within specific momentum windows are integrated out. The Polchinski equation \cite{polchinski1984renormalization, ZinnJustinBook1} is an exact equation describing how the classical (bare) action changes during each infinitesimal RG step. It serves as a formally exact equation and a profound mathematical realization of Wilson's RG. However, there is another, less widely known approach developed by Wetterich and Morris \cite{bagnuls2001exact, Delamotte_2012, Berges, dupuis2021nonperturbative}. This approach, known as the \textit{effective average action} (EAA) method, differs from Wilson's perspective in the choice of the underlying mathematical object. Instead of focusing on the classical action, the Wetterich approach centers on the "effective action of integrated-out modes".

Furthermore, the EAA approach differs from Wilson's in how the fundamental cut-off is implemented. In the Wilson-Polchinski framework, the bare action is varied, and the cut-off scale—around which fluctuations are integrated—is rescaled at each RG step. In contrast, within the EAA formalism, the bare action remains fixed, and a regulator gradually suppresses IR fluctuations until the IR cut-off is reached. The EAA formalism offers several advantages over the Wilson-Polchinski one, particularly for nonperturbative analysis and studies of phases with broken symmetries. Notably, this approach helps regularize the infrared divergences that typically arise in broken-symmetry regions when the mass squared becomes negative.Let us proceed with the explicit construction of the formalism. Following \cite{Berges}, we add to the classical averaged action $\overline{S}$ a scale-dependent mass term, referred to as the \textit{regulator}:
\begin{equation}\Delta S_k[{\bm{x}_\alpha}]:=\frac{1}{2}\int \dd t \sum_{\alpha=1}^n\sum_{\mu=1}^N x_{\mu\alpha}(t) R_k^{(N)}(p_\mu^2)x_{\mu\alpha}(t)\,,
\end{equation}
and denote the modified partition function by $\overline{\mathcal{Z}_{\beta,k}^n}$. Note that the regulator is assumed to be diagonal in both time and replica space. This choice is expected to be sufficiently general; however, other options have been explored in the literature, especially in the context of the random field Ising model \cite{Tarjus, Tarjus2}. We nevertheless believe this regulator is suitable for studying the system at long times \cite{lahoche2024largetimeeffectivekinetics}.

The regulator function $R_k^{(N)}$ depends on the \textit{infrared scale} $k$, and we include an index $N$ since we view $R_k^{(N)}$ as a random quantity, depending on the specific realization of the kinetic matrix $K$. In the limit $N \to \infty$, $R_k^{(N)}$ converges to a deterministic function $R_k$.

The \textit{effective average action} $\Gamma_k$, which represents the effective action for UV modes integrated out up to the scale $k$, is defined by the following relation:

\begin{equation} \Gamma_k[\mathcal{M}]+\Delta S_k [\mathcal{M}]={W}_{k}^{(n)}[\mathcal{L}]-\sum_{\alpha=1}^n \int_{-\beta/2}^{+\beta/2} \dd t \, \textbf{L}_\alpha(t) \cdot \textbf{M}_\alpha(t),\label{defGammak} \end{equation}

where ${W}_{k}^{(n)}[\mathcal{L}]$ is the free energy of the modified partition function:

\begin{equation} \mathcal{W}^{(n)}_{\beta,k}[K,J,\mathcal{L}]:=\ln \overline{\mathcal{Z}_{\beta,k}^n}[K,J,\{\bm L_\alpha\}], \end{equation}

$\mathcal{L}:=\{\mathbf{L}_\alpha \}$ and $\mathcal{M}:=\{\mathbf{M}_\alpha \}$ are the sets of sources and classical fields (each an $N$-dimensional vector per replica). The components of the classical field are given by:

\begin{equation} \frac{\delta }{\delta L_{i\alpha}}\mathcal{W}^{(n)}_{\beta,k}[K,J,\mathcal{L}]=M_{i\alpha}. \end{equation}

Formally, the regulator function ensures that $\Gamma_k[\mathcal{M}]$ provides a smooth transition between the bare action $\overline{S_{\text{cl}}}$ as $k \to 4\sigma$ (where no fluctuations have been integrated out) and the full effective action $\Gamma$ at $k=0$ (where all fluctuations have been integrated out), which corresponds to the full Legendre transform of the free energy \eqref{partition1}. Specifically, the regulator becomes large for $p_\mu \lesssim k$, effectively freezing large-scale degrees of freedom by assigning them a large effective mass. In contrast, microscopic degrees of freedom with $p_\mu \gtrsim k$ remain largely unaffected by the regulator and are thus integrated out. These conditions specifically require:

\begin{enumerate}
    \item $\lim_{k\to 0} R_k(p_\mu^2)=0$, meaning that all the degrees of freedom are integrated out as $k\to 0$ (infrared (IR) limit).
    \item $\lim_{k\to 4\sigma} R_k(p_\mu^2)\to\infty$, meaning that no fluctuations are integrated out in the deep ultraviolet (UV) limit $k\to 4\sigma$.
\end{enumerate}

The equation that describes how the effective function $\Gamma_k$ changes with $k$ is the Wetterich equation \cite{Delamotte_2012}:
\begin{equation} \boxed{\dot{\Gamma}_k=\frac{1}{2\beta} \sum_\omega  \sum_{p_\mu,p_\nu,\alpha} \dot{R}_k(p_\mu^2) G_k(\omega,p_\mu,p_\nu,\alpha),}\label{Wett} 
\end{equation}
where the dot denotes differentiation with respect to $s:=\ln(k/4\sigma)$, and $G_k$ is the \textit{effective $2$-point function} (diagonal with respect to the replica and generalized momentum indices), such that:
\begin{equation} [G_k^{-1}(\omega, p_\mu,p_\nu,\alpha)+R_k(p_\mu^2)\delta_{\mu\nu}]\delta_{\omega,-\omega^\prime}\delta_{\alpha\beta}:=\frac{\delta^2 \Gamma_k}{\delta M_{\mu\alpha}\delta M_{\nu\beta}},.\label{defG} 
\end{equation}
As we focus on the large-$N$ limit to compute the flow equations, only the limiting function $R_k := \lim_{N \to \infty} R_k^{(N)}$ is relevant and needs to be defined in practice. In this paper, we adopt a slightly modified version of the standard Litim regulator \cite{Litim}:
\begin{equation} R_k(p_\mu^2):=\frac{4\sigma}{4\sigma-k^2}(k^2-p_\mu^2)\theta(k^2-p_\mu^2),,\label{LitimReg} 
\end{equation}
where $\theta(x)$ is the Heaviside step function. The prefactor in this expression ensures that the boundary conditions at $k=0$ and $k=2\sqrt{\sigma}$ are satisfied. This regulator has been employed in our previous work \cite{lahoche2024largetimeeffectivekinetics} as well as in recent studies \cite{lahoche20241, lahoche20242, lahoche20244, lahoche2023functional}. Other regulator choices, such as the exponential regulator, are also widely used in the literature \cite{Berges}. Furthermore, the dependence of the results on the choice of regulator could be investigated, as the approximations used to solve the equation may introduce a spurious dependence on the regulator itself \cite{canet2003optimization, duclut2017frequency, lahoche2023stochastic}; we leave these considerations for future work.
\medskip

Equation \eqref{Wett} is exact but impossible to solve analytically; therefore, this paper aims to construct approximate solutions by exploiting the symmetries of the system. To conclude this section, we recall several results from perturbation theory and formal large-$N$ limits in the symmetric phase, which will be relevant for the remainder of this paper. Figure \ref{figRG1} summarizes this construction. At this stage, it is necessary to remark:

\begin{remark}
As explained in \cite{lahoche20241}, the RG construction assumes that the number of degrees of freedom remains fixed by $N$, while the standard deviation $\sigma$ of the matrix-like disorder is rescaled after each RG transformation. More precisely, as we integrate out momenta in the windows $[\Lambda, s\Lambda]$, we rescale $\sigma$ by a factor $s^{-1}$ and define the dimensionless standard deviation $\bar{\sigma} := k\sigma$. In this paper, we rescale the couplings by an appropriate power of $\bar{\sigma}$ so that this parameter is eliminated from our flow equations. This procedure is detailed in \cite{lahoche20241} and is also assumed in \cite{lahoche4} and references therein. This is equivalent to describing the flow in a reference frame that tracks the evolution of $\bar{\sigma}$. Note that the elimination of degrees of freedom is not strictly necessary for the RG construction; as first pointed out by Wegner, a suitable change of variables can achieve an equivalent result—see \cite{latorre2000exact, caticha2016changes, bervillier2014structure} and references therein.One can easily establish the relationship between our perspective and the framework where the number of effective degrees of freedom changes. Considering, for instance, a quartic theory, and denoting $N_{\text{eff}}(k)$ and $g_4(k)$ as the effective degrees of freedom and the corresponding effective quartic coupling at scale $k$, respectively, we find that these must be related to the effective coupling $u_4(k)$ used in this paper as follows (see also \eqref{definitionu4}):
\begin{equation}\frac{1}{N_{\text{eff}}(k)} g_4(k) = \frac{1}{N} \left( \underbrace{\frac{N}{N_{\text{eff}}(k)} g_4(k)}{:=u_4(k)} \right)\,,
\end{equation}
where
\begin{equation}
\frac{N{\text{eff}}(k)}{N} = \int_0^{k^2} \rho(p^2) \dd p^2\,.
\end{equation}
Since neither formulation offers a distinct advantage—in particular, the canonical dimensions are scale-dependent in both cases—we treat them as equivalent. The difference between $g_4$ and $u_4$ remains of order $1$ as long as we stay sufficiently far from the deep IR regime, where the continuous approximation ceases to be valid.
\end{remark}\label{remarkscaling}

\subsection{Perturbation theory and large $N$ closed equations}\label{largeN}

\begin{figure}
\begin{center}
\includegraphics[scale=1]{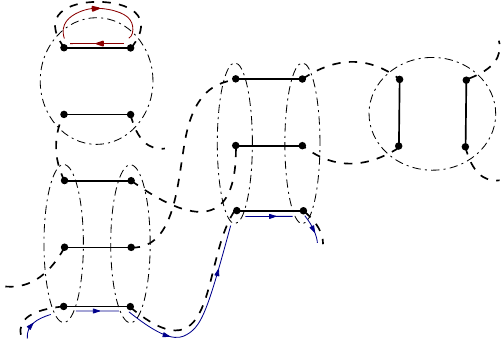}
\end{center}
\caption{A typical Feynman diagram for $p=3$, involving two non-local vertices, two local quartic vertices, and six external (half) edges. The red arrowed path provides an example of a closed face (note that faces are not oriented, and the direction of the arrows is arbitrary. The blue path provides an example of an open face.}\label{FeynmanDiag}
\end{figure}

Perturbation theory for the averaged replicated theory can be constructed using standard Feynman rules. Feynman diagrams consist of vertices representing the interactions defined by the classical averaged action \eqref{classicalaveraged}, with edges representing the corresponding Wick contractions. However, the theory we consider differs from ordinary QFT due to the specific non-local nature of the interactions stemming from the underlying $O(N)$ symmetry, time non-locality, and the replica index structure. Consequently, we must construct a set of graphical rules to represent Feynman diagrams that accommodate these three specificities. We adopt the same conventions as in our previous work \cite{lahoche2024largetimeeffectivekinetics}, which itself draws inspiration from \cite{lahoche2022functional}. Vertices are represented by clusters of black nodes corresponding to the fields $\bm{x}_\alpha$, linked by solid edges representing the Euclidean scalar products of vectors, explicitly:
\begin{equation}
\vcenter{\hbox{\includegraphics[scale=1]{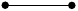}}}\equiv \sum_{i=1}^N \, x_{i\alpha} (t)x_{i\alpha}(t)\,.
\end{equation}
While this accounts for the non-local structure stemming from the underlying $O(N)$ symmetry of the vertices, we must also consider the replica structure and time non-locality. For the latter, it should be noted that the local time components are \textit{always} associated with the same replica index. Consequently, we introduce the graphical convention that any set of dots surrounded by a dashed-dotted circle are both local in time and share the same replica index. For example:
\begin{equation}
\vcenter{\hbox{\includegraphics[scale=1.2]{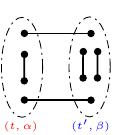}}}\,\equiv \int {\color{red}\dd t}\int {\color{blue}\dd t^\prime} \sum_{{\color{red}\alpha},{\color{blue}\beta}}\, \bm{x}_{{\color{red}\alpha}}^2({\color{red}t})\times (\bm{x}_{{\color{red}\alpha}}({\color{red}t})\cdot \bm{x}_{{\color{blue}\beta}}({\color{blue}t^\prime}))^2\times (\bm{x}_{{\color{blue}\beta}}^2({\color{blue}t^\prime}))^2\,.
\end{equation}
The final component involves the Wick contractions, which we represent as dashed edges connecting the black nodes. A typical Feynman diagram for $p=3$ involving six external edges is shown in Figure \ref{FeynmanDiag}. We refer to the black nodes connected to external (half-)edges as \textit{external nodes}.
\medskip

Feynman graphs for such a non-local field theory involve, in fact, more than just sets of vertices and edges. They also include extended structures known as \textit{faces}, which are defined such that:

\begin{definition}
A face is a cycle made of an alternative sequence of dashed and solid edges. A face can be closed or open. The length of a face is equal to the number of dashed edges involved along its boundary. 
\end{definition}\label{defface}

The concepts of closed and open faces are illustrated in Figure \ref{FeynmanDiag} by the red and blue paths, respectively. Closed faces are particularly relevant for power counting. Indeed, each cycle is essentially a sum over a path involving a product of propagators, which are diagonal in the basis $u^{(\mu)}$. Consequently, a given face $f$ of length $n$ contributes as follows:
\begin{equation}
\sum_{\mu=1}^N\, \prod_{i=1}^n\tilde{C}(\omega_i,p_\mu^2) \to N \left(\int_0^{+4\sigma}\, \rho(p^2) \dd p^2\, \prod_{i=1}^n \tilde{C}(\omega_i,p^2)\right)\,,
\end{equation}
and then contributes a global factor $N$. Considering a given Feynman graph $\mathcal{G}$, with $V_{2s}$ denoting the number of vertices of valence $2s$ and $F$ the number of faces, the corresponding Feynman amplitude scales as:
\begin{equation}
\mathcal{A}({\mathcal{G}}) \sim N^{-\sum_s (s-1)V_{2s}+F}\,.
\end{equation}
In the large-$N$ limit, the leading-order (LO) graphs are those that maximize the number of faces for a fixed number of vertices, while satisfying all constraints imposed by the external dashed edges. For the quartic theory, the structure of these LO graphs is more transparent when using the \textit{loop vertex representation} (LVR) \cite{rivasseau2018loop, Lahochebeyond, lahoche2023functional, rivasseau2010loop, gurau2015multiscale}. In these representations, vertices are effectively mapped to edges, and edges to closed loops; consequently, the LO diagrams become trees.  
\medskip

To conclude, let us recall the main results obtained in our previous work for $p=3$ \cite{lahoche2024largetimeeffectivekinetics}. The self-energy is diagonal in the basis $\{u^{(\mu)}\}$, and in the large-$N$ limit, the diagonal elements $\gamma(\omega)$ depend only on the external frequency $\omega$. Focusing on a deterministic quartic potential, we obtain:
\begin{align}
\nonumber \gamma(\Omega)=&-\int \frac{\dd\omega}{2\pi}\,\rho(p^2) \dd p^2 \bigg(\frac{h_2}{6}\, G_k(\omega,p_\mu,p_\mu,\alpha)\\
&-{6\lambda}\,\int  \rho(q^2)  \dd q^2 G_k(\omega,p ,p,\alpha) G_k(-\omega-\Omega,q,q,\alpha)\bigg)\,,\label{closedeqation}
\end{align}
which can be generalized for higher order local potential \cite{ZinnJustinReview}. Here, $G_k(\omega,p_\mu,p_\mu,\alpha)$ is the effective propagator (including regulator):
\begin{equation}
G_k(\omega,p_\mu,p_\nu,\alpha)=\frac{\delta_{\mu\nu}}{\omega^2+p_\mu^2+m^2-\gamma(\omega)-R_k(p_\mu^2)+i\epsilon}\,.\label{2ptsvertexExp}
\end{equation}
Another relevant quantity is the 1PI $2$-point correlation between replica:
\begin{equation}
q_{\mu,\nu,\alpha\beta}(t,t^\prime):=\, \langle x_{\mu,\alpha}(t)x_{\nu,\beta}(t^\prime) \rangle_{\text{1PI}} \,,
\end{equation}
which is a $\mathcal{O}(N^{-1})$ quantity, satisfying the equation at the leading order of the $1/N$ expansion (in Fourier space):
\begin{equation}
q_{\alpha\beta} (\Omega,\Omega^\prime)=\frac{12 \lambda}{N} \, \int \rho(p^2) \dd p^2\, \left(\int \dd \omega \, G_k(\omega,p,p,\alpha)\right)^2 \delta(\Omega)\delta(\Omega') \,.
\end{equation}

\subsection{Scaling and dimensions}

To conclude these preliminaries, let us discuss the notion of dimension. Usually, dimensionality is determined by external structures such as the background space-time. Here, this structure is one-dimensional, and the scaling dimensions of the couplings are fixed by the requirement that the classical action (in units where $\hbar=1$) remains dimensionless. This leads to the following:
\begin{equation}
[m^2]=2\,,\qquad [\lambda]=4\,,\qquad [h_2]=3\,.
\end{equation}
In standard QFT, the dimensions of the couplings are typically related to the properties of the RG flow near the Gaussian fixed point. This connection between the dimension (fixed by the background space-time) and the flow behavior arises because coarse-graining is performed over the spectrum of the Laplacian, which is an operator defined on the manifold itself. In our case, the situation is distinct: the spectrum used for coarse-graining is that of the matrix $K$, which is independent of the temporal background. Furthermore, the momentum distribution does not exhibit a simple power-law behavior, as is typically the case in standard theories. For such models, where coarse-graining is not tied to a background space, a dimension can still be defined from the flow's behavior near the Gaussian fixed point \cite{Lahochebeyond, lahoche20241, lahoche3}, up to a global rescaling of the couplings designed to cancel the $k$-dependence of the effective loop integrals. Since the momentum distribution does not follow a power law, the resulting dimension is energy-scale dependent—a feature also encountered in other frameworks, such as tensorial field theories \cite{benedetti2015functional}.
\medskip

First, note that the vertex expansion of the Wetterich equation \eqref{Wett} involves a sum of single-loop diagrams of the form (setting external frequencies to zero):
\begin{equation}
L_n(k):=\int \dd \omega \int_0^{4\sigma} \dd p^2 \rho(p^2)\, \dot{R}_k(p^2) \prod_{i=1}^n G(\omega,p,p,\alpha)\,,
\end{equation}
where we assume the limit $\beta \to \infty$. For $k \ll 1$, since the integration window of momenta is centered around $k$ (specifically, between $0$ and $k$ for the Litim regulator), we can approximate $\rho(p^2)$ by $(p^2)^{1/2} / (2\pi\sigma^2)$. In this limit, the dependence of the integral $L_n(k)$ follows a power law, corresponding exactly to what is obtained for a 4-dimensional field theory, as expected. The asymptotic dimensionless couplings are then given by:
\begin{equation}
\bar{\mu}_{2n}=k^{-\omega_{2n}} \mu_{2n}\,,\qquad \bar{\lambda}=k^{-\nu} \lambda\,,
\end{equation}
with $p=3$:
\begin{equation}
\omega_{2n}=4-2n\,,\qquad \nu=-1\,,
\end{equation}
which are respectively the \textit{IR canonical dimensions} for $\mu_{2n}$ and $\lambda$. 
\medskip

\begin{figure}
\begin{center}
\includegraphics[scale=0.8]{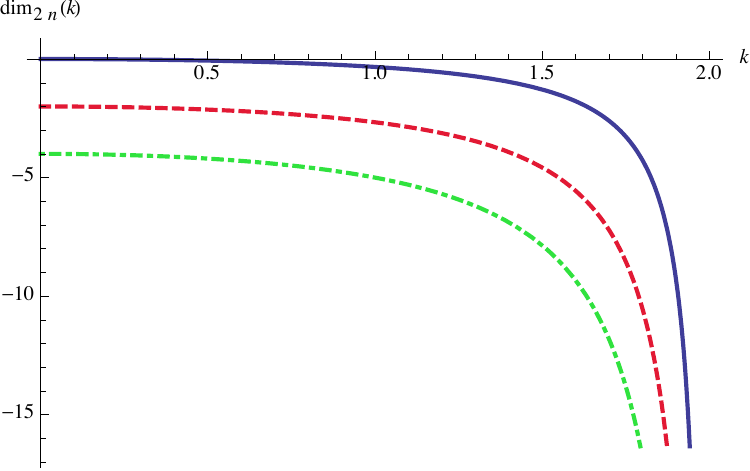}
\end{center}
\caption{Dependency on critical canonical dimensions ($\sigma=1$) for $n=2$ (solid blue curve), $n=3$ (dashed red curve) and $n=4$ (dashed-dotted green curve).}\label{figcanonical}
\end{figure}

The canonical dimension for arbitrary $k$ is non-trivial, and we will return to this issue in Section \ref{secLPA}. However, we can readily define it for the critical theory or in the perturbative regime by assuming that the mass term has a negligible influence on the denominator of the effective propagator $G(\omega, p, p, \alpha)$. In this approximation, each loop integral takes the form:
\begin{equation} 
\Omega^\prime(k):=k^3\int \dd p^2\, \frac{\rho(p^2) \dot{R}_k(p^2)}{(p^2+R_k(p^2))^{3/2}}=\frac{8 k^5}{3 \pi  \sqrt{4-k^2}}\,,
\end{equation}
and to cancel the explicit dependence on $k$ on the loops, it is suitable to define the dimensionless couplings such that:
\begin{equation}
\bar{\mu}_{2n}=\mu_{2n}\, \frac{1}{k^2}\, \left(\frac{\Omega^\prime(k)}{k^3}\right)^{n-1}\,.\label{rescalingGaussian}
\end{equation}
Within this definition holding close enough of the IR regime\footnote{See our previous work arXiv:2408.02602.}, the linear part of flow equations reads:
\begin{equation}
\dot{\bar{u}}_{2n}=-\mathrm{\dim}_{2n}(k) \bar{u}_{2n}+\cdots\,,
\end{equation}
and a straightforward calculation leads to:
\begin{equation}
\boxed{\mathrm{\dim}_{2n}(k):= (n-1)\mathrm{\dim}_4(k)+2(2-n) \,,}
\end{equation}
and:
\begin{equation}
\boxed{\mathrm{\dim}_4(k):=\frac{d}{dt}\, \ln \,\left(k^5(\Omega^\prime)^{-1}(k)\right)=\frac{k^2}{k^2-4}\,.}
\end{equation}
Let us notice that as $k\to 0$, we recover the previous result: 
\begin{equation}
\mathrm{\dim}_{2n}(k)=\omega_{2n}+\frac{k^2}{4} \left(1-n\right)+\mathcal{O}(k^3)\,,
\end{equation}
as expected. This differs significantly from ordinary $4D$ power counting when we trace the flow back towards the UV, as shown in Figure \ref{figcanonical}.

\section{Effective vertex expansion}\label{EVE}

The vertex expansion constructed in our previous work is assumed to be sufficiently accurate for this type of theory, in which local quartic interactions are just-renormalizable in the deep IR. In similar contexts, we have demonstrated good agreement between the leading-order vertex expansion and more advanced methods \cite{lahoche20241, lahoche20242}. We expect this consistency to hold because, sufficiently close to the Gaussian fixed point and far from the deep UV scale, irrelevant couplings become negligible. In the large-$N$ limit and the symmetric phase, however, it is advantageous to employ the non-trivial Schwinger-Dyson relations arising between local observables. This method was first introduced in the context of tensorial field theories \cite{Lahochebeyond}, where the non-localities resemble those encountered here, and has also been applied to spin glasses \cite{lahoche2023functional}. \medskip

\begin{figure}
\begin{center}
\includegraphics[scale=1.2]{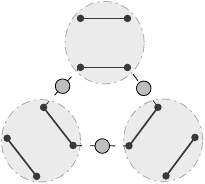}
\end{center}
\caption{Structure of the effective $6$-point function in the large-$N$ limit for the local quartic theory (in the symmetric phase). The dashed edges with grey disks represent the effective $2$-point function $G_k$, and the grey bubble with valence $4$ denotes the $4$-point function $\Gamma_k^{(4)}$.}\label{EVEex1}
\end{figure}

The expected relationship is illustrated in Figure \ref{EVEex1} for a local quartic theory, showing the exact $6$-point function in the large-$N$ limit. In this section, we focus primarily on the sextic theory with a rank-$3$ disorder tensor. The relevant effective diagrams and numerical factors have been calculated in the companion paper \cite{lahoche2024frequency}, from which we draw our results. Following the notation established in that reference, we denote the local and non-local sextic effective vertices as $\Gamma_{k,\text{L}}^{(6)}$ and $\Gamma_{k,\text{NL}}^{(6)}$, respectively. Graphically, this yields:"

\begin{align}
\nonumber \Gamma_{k,\text{L}}^{(6)}&=\vcenter{\hbox{\includegraphics[scale=0.6]{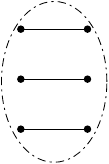}}}+\vcenter{\hbox{\includegraphics[scale=0.7]{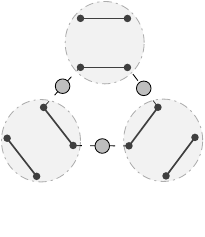}}}+\vcenter{\hbox{\includegraphics[scale=0.7]{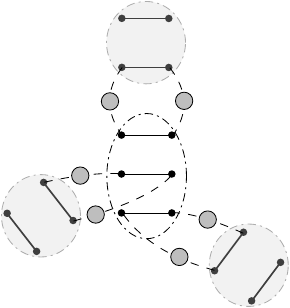}}}+\vcenter{\hbox{\includegraphics[scale=0.7]{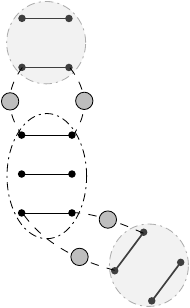}}}+\vcenter{\hbox{\includegraphics[scale=0.7]{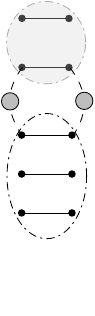}}}\\
&+\vcenter{\hbox{\includegraphics[scale=0.7]{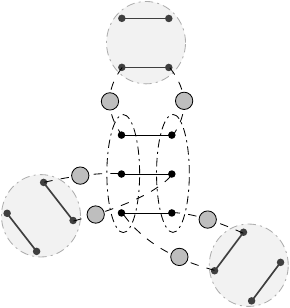}}}+\vcenter{\hbox{\includegraphics[scale=0.7]{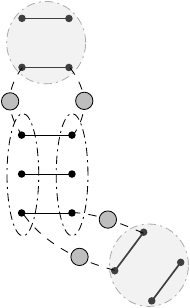}}}+\vcenter{\hbox{\includegraphics[scale=0.7]{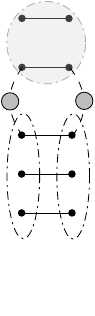}}}\,,
\end{align}
and
\begin{equation}
\Gamma_{k,\text{NL}}^{(6)}=\vcenter{\hbox{\includegraphics[scale=0.7]{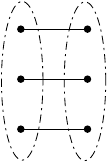}}}\,.
\end{equation}

Note that this relation implies that the non-local vertex does not undergo renormalization. Diagrams can be computed using the leading-order derivative expansion and the local approximation for vertices. We thus obtain a relation between the UV local and non-local couplings $u_6(0)$ and $\tilde{u}_6(0)$ and the effective local coupling $u_6(k)$ in the deep IR\footnote{We use the same notation as in \cite{lahoche2024frequency}; $u_6(0)$ and $\tilde{u}_6(0)$ denote the bare couplings (i.e., those in the deep UV), not those in the deep IR. The context should prevent any ambiguity.}:
\begin{equation}
\boxed{u_6(0)=k^{-2}\frac{\bar{u}_6(k)-24 L_3(\bar{u}_2) \bar{u}_4^3+\bar{\tilde{u}}_6(0) \left(12 \bar{u}_4^3 \tilde{L}_4(\bar{u}_2)-8 \bar{u}_4^2  \tilde{L}_3(\bar{u}_2)+3 \bar{u}_4  \tilde{L}_2(\bar{u}_2)\right)}{1-12 \bar{u}_4^3 L_2^3(\bar{u}_2)+8 \bar{u}_4^2  L_2^2(\bar{u}_2)-3 \bar{u}_4  L_2(\bar{u}_2) }\,,}\label{u60}
\end{equation}
where:
\begin{equation}
  \tilde{L}_2:=
  \frac{1}{6\pi^3 (1+\bar{u}_2)^2} 
  \,,
\end{equation}
\begin{equation}
  \tilde{L}_3:=
  \int \prod_{i=1}^2 \rho(p_{\mu_i}^2)\dd p_{\mu_i}^2 \int \frac{\dd\omega}{2\pi} \, G^2_k(p_{\mu_1}^2,\omega^2) G^2_k(p_{\mu_2}^2,\omega^2)\,,
\end{equation}
\begin{equation}
  \tilde{L}_4:= \int \prod_{i=1}^3 \rho(p_{\mu_i}^2)\dd p_{\mu_i}^2 \int \frac{\prod_{i=1}^2 \dd\omega_i}{(2\pi)^2} \, G^2_k(p_{\mu_1}^2,\omega_1^2) G^2_k(p_{\mu_2}^2,\omega_2^2) G^2_k(p_{\mu_3}^2,(\omega_1+\omega_2)^2)\,,
\end{equation}
\begin{equation}
L_2(\bar{u}_2):=\frac{\frac{6}{\sqrt{\bar{u}_2+1}}+\frac{2}{\left(\bar{u}_2+1\right){}^{3/2}}-3 \left(\log \left(\frac{2 \sqrt{\bar{u}_2+1}+\bar{u}_2+2}{64}\right)+4\right)-6 \log (k)}{12 \pi } 
\,,
\end{equation}
\begin{equation}
L_3(\bar{u}_2):=\frac{1-\frac{1}{\left(1+\bar{u}_2\right){}^{5/2}}}{8 \pi  \bar{u}_2} 
\,.
\end{equation}
We furthermore recall the definition of effective couplings in the local sector:
\begin{align}
\Gamma_k^{(2n)}(\{p_i, \omega_i\})&:=\frac{u_{2n}}{(2n)! N^{n-1}} \delta\left(\sum_{i=1}^{2n}\omega_i\right) \sum_{\pi} \delta_{p_{\pi(1)}p_{\pi(2)}}\cdots \delta_{p_{\pi(2n-1)}p_{\pi(2n)}}\,,\label{localsectorGamma2n}
\end{align}
where $\pi$ denotes a permutation of the $2n$ external indices. Note the presence of the logarithm, which dominates the flow in the deep IR. The flow equation for $\bar{u}_6$ can be derived from the observation that $u_6(0)$ is independent of $k$. Taking the derivative of \eqref{u60} with respect to $s := \ln k$, we obtain:
\begin{align}
\boxed{\dot{\bar{u}}_6=(u_6+a) \left(\frac{\dot{\bar{u}}_2\partial_{\bar{u}_2} b +\dot{\bar{u}}_4\partial_{\bar{u}_4} b}{b} +2\right)-(\dot{\bar{u}}_2\partial_{\bar{u}_2} a +\dot{\bar{u}}_4\partial_{\bar{u}_4} a)+5a\,,}
\end{align}
where we defined:
\begin{align}
a(\bar{u}_2,\bar{u}_4)&:=-24 L_3(\bar{u}_2) \bar{u}_4^3+\bar{\tilde{u}}_6(0) \left(12 \bar{u}_4^3 \tilde{L}_4(\bar{u}_2)-8 \bar{u}_4^2  \tilde{L}_3(\bar{u}_2)+3 \bar{u}_4  \tilde{L}_2(\bar{u}_2)\right)\,,\\
b(\bar{u}_2,\bar{u}_4)&:=1-12 \bar{u}_4^3 L_2^3(\bar{u}_2)+8 \bar{u}_4^2  L_2^2(\bar{u}_2)-3 \bar{u}_4  L_2(\bar{u}_2)\,.
\end{align}

\begin{figure}
\begin{center}
\includegraphics[scale=0.55]{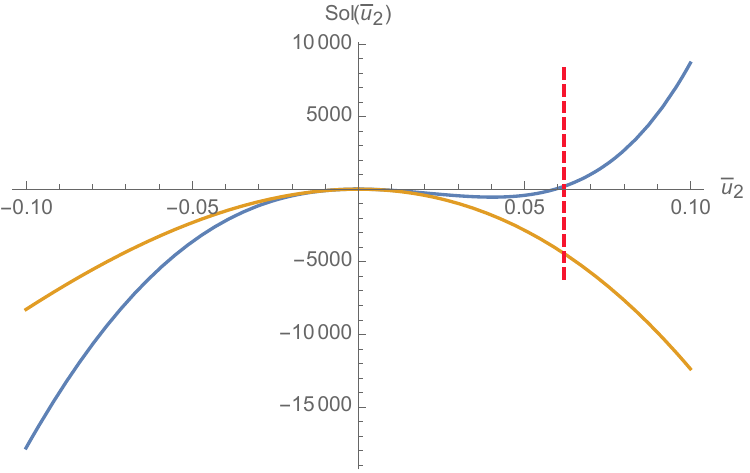}\qquad \includegraphics[scale=0.55]{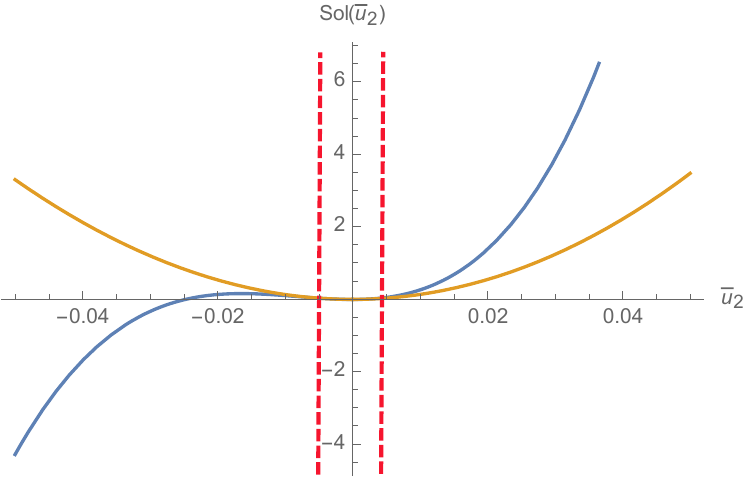}
\end{center}
\caption{The function $\text{Sol}(\bar{u}_2)$ for the sextic theory (on left) and the quartic theory (on right). We set $\bar{\tilde{u}}_6=-10$ for the blue and $\bar{\tilde{u}}_6=0$ for yellow curves, respectively for $k=10^{-2}$ and $k=10^{-4}$.}\label{Solu2}
\end{figure}

The first step is to investigate the existence of asymptotic fixed points. While a global fixed point (incorporating the disorder coupling) might not exist, there may exist 'fixed trajectories' that cause all other $\beta$-functions to vanish. We are primarily interested in fixed trajectories that admit a limit in the deep IR, and we consider the sextic and quartic cases separately, depending on the presence of local sextic interactions in the classical action. Let $\text{Sol}(\bar{u}_2)$ denote the function obtained by substituting the solutions for $\dot{\bar{u}}_2 = \dot{\bar{u}}_4 = 0$ into the $\beta$-function for the sextic coupling. The two $\beta$-functions for $\bar{u}_2$ and $\bar{u}_4$ have been established in our previous work \cite{lahoche2024largetimeeffectivekinetics}:
\begin{align}
\dot{\bar{u}}_2&=-2\bar{u}_2-\frac{\bar{u}_4}{36\pi}\frac{1}{(1+\bar{u}_2)^{\frac{3}{2}}}\,,\\
\dot{\bar{u}}_4&= -\frac{\bar{\tilde{u}}_6 }{30\pi^2}\frac{1}{(1+\bar{u}_2)^2}-\frac{\bar{u}_6 }{60\pi}\frac{1}{(1+\bar{u}_2)^{\frac{3}{2}}}+ \frac{\bar{u}_4^2}{12\pi} \frac{1}{(1+\bar{u}_2)^{\frac{5}{2}}}\,.
\end{align}
Figure \ref{Solu2} illustrates the behavior of the function $\text{Sol}(\bar{u}_2)$ for both the sextic and quartic theories. Interestingly, no asymptotic fixed point is expected for the sextic theory, as the positive-mass fixed point appearing in the IR regime disappears asymptotically in the limit $-\ln(k) \to \infty$. In contrast, two fixed-point solutions are found for the quartic theory at the following values:
\begin{align}
\text{FP1}:=\{\bar{u}_{2,\infty}\approx 0.003\,, \quad \bar{u}_{4,\infty}\approx -0.69\,, \quad \bar{u}_{6,\infty}\approx 2.37\,,\}\\
\text{FP1}:=\{\bar{u}_{2,\infty}\approx -0.003\,, \quad \bar{u}_{4,\infty}\approx 0.68\,, \quad \bar{u}_{6,\infty}\approx 2.35\,,\}
\end{align}
the subscript $\infty$ meaning $-\ln (k)\to \infty$. Their respective critical exponents (in the $(\bar{u}_2,\bar{u}_4)$ plane) are moreover:
\begin{align}
\Theta_{1,\infty}&=\{\theta_1 \approx 2.87,\theta_2 \approx 0.04\}\,,\\
\Theta_{2,\infty}&=\{\theta_1 \approx 1.17,\theta_2 \approx -0.03\}\,.
\end{align}
We have, however, reason to doubt the validity of the fixed point \text{FP2}. The estimated local sextic coupling in the classical action remains quite large (approximately $0.002$, compared to $\approx 10^{-14}$ for the first fixed point). Furthermore, the characteristics of the first fixed point essentially correspond to those obtained via the truncation method in our previous work \cite{lahoche2024largetimeeffectivekinetics}.
\medskip

\begin{figure}
\begin{center}
\includegraphics[scale=0.55]{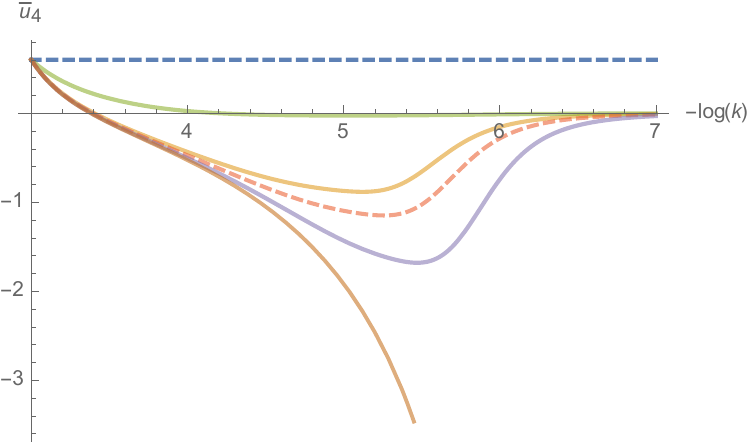}\qquad \includegraphics[scale=0.55]{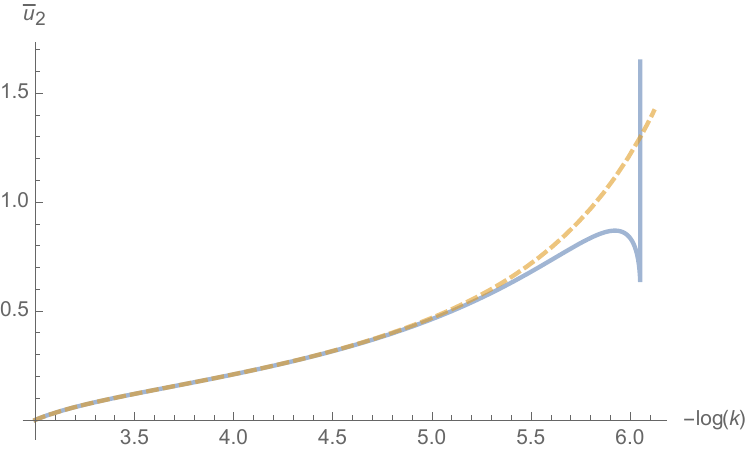}
\end{center}
\caption{On the left: RG trajectories for initial conditions $\bar{u}_2(k_0)=-10^{-5}, \bar{u}_4(k_0)=0.6$ and  $k_0=0.05$. Respectively: $\bar{\tilde{u}}_6(k_0)=0$ (dashed blue curve), $\bar{\tilde{u}}_6=-498$ (yellow curve), $\bar{\tilde{u}}_6(k_0)=-300$ (green curve), $\bar{\tilde{u}}_6(k_0)=-508$ (dashed red curve) and $\bar{\tilde{u}}_6(k_0)=-515$ (purple curve). The remaining curve is for the critical value $\bar{\tilde{u}}_6(k_0)=-519.8$ and is singular for $k_c\approx 0.002$. On the right:  Behavior of $u_2(k)$ for $\bar{\tilde{u}}_6(k_0)=-519.8$ (blue curve) and for  $\bar{\tilde{u}}_6(k_0)=-519.7$ just above the critical value (in dashed yellow).} \label{figsing}
\end{figure}

Finally, it is readily observable that the finite-time singularity phenomenon also emerges within the EVE framework, as illustrated in Figure \ref{figsing}, although this singularity appears less abrupt than in the standard vertex expansion. As shown in the figure, singular points (cusps) form as the disorder increases; for disorder values of the order of $\bar{\tilde{u}}_6 \sim 10^5$ (the precise value depending on the initial conditions), the flow develops a finite-time singularity. Our interpretation of these singularities aligns with our previous work and the discussion in the introduction: they are a consequence of the instability of the potential for observables quantifying correlations between replicas, signaling that certain interactions are missing from the perturbative treatment.

In the remainder of this article, we will explicitly highlight the relationship between these instabilities and the emergence of divergences by considering a flow "improved" by observables that correlate the replicas. To this end, we employ two distinct approximation schemes. The first is an expansion around the vacuum of the local potential for uniform fields in the IR, while the second is a vertex expansion similar to the one developed in our previous work \cite{lahoche2024largetimeeffectivekinetics}. In both cases, we focus on the leading-order derivative expansion and neglect the flow of the anomalous dimension. This choice is justified by the fact that the critical dimension is only reached asymptotically by the flow, and by the large-$N$ limit.

\section{Venturing beyond symmetric phase}\label{secLPA}

In our previous work \cite{lahoche2024largetimeeffectivekinetics} and the companion paper \cite{lahoche2024frequency}, we observed that the presence of disorder leads to the emergence of 'finite-scale' singularities. As previously discussed (see \cite{lahoche2024largetimeeffectivekinetics} for an extended analysis), these singularities likely signal that interactions suppressed by standard perturbation theory—notably correlations between replicas—become significant. However, given our prior focus on the vertex expansion, it remains possible that these singularities are artifacts arising from the choice of an unstable vacuum. In this section, we propose an alternative approximation scheme by expanding the non-local interactions around the vacuum of the local potential in the $N \to \infty$ limit.

\subsection{Expansion around non-zero uniform local vacuum}

The classical action involves a multi-local expansion comprising both local and bi-local interactions. Ignoring derivative interactions, we aim to construct a satisfactory approximation by retaining only the local and bi-local components of the effective action $\Gamma_k$. Note that correlations between replicas are not the only features 'omitted' in this treatment; derivative interactions for local potentials could also be influential and potentially as relevant as the effects currently under investigation. However, since the local quartic sector is marginal in the deep IR, these effects are not necessarily expected to play a major role, even though they may be comparable in magnitude to the effects we aim to highlight. We will therefore neglect derivative interactions entirely in this paper, reserving their study for future work.
\medskip

We will focus on the construction of the flow in the deep IR ($k \ll 1$). Rather than projecting the flow equations around the vacuum $M_{\mu \alpha}(\omega) = 0$, we propose to construct a projection around a non-zero field that is uniform in time, and for which only the macroscopic component $\mu=0$ (corresponding to zero momentum $p_\mu=0$) is non-vanishing:
\begin{equation}
M_{\mu \alpha}^{(0)}(t)= \epsilon_\alpha\sqrt{2 N \rho}_\alpha \, \delta_{\mu 0}\,,\label{vacIR}
\end{equation}
where $\epsilon_\alpha = \pm 1$. In general, the local potential formalism requires a projection onto a homogeneous field \cite{Delamotte_2012}, which corresponds to the macroscopic component $\vec{p} = 0$ of the Fourier modes. Here, the generalized momentum plays this role, as the coarse-graining of the theory is constructed based on its spectrum. Note, moreover, that the number of 'sites' in the underlying network remains invariant, because:
\begin{equation}
2\rho_\alpha=\frac{1}{N}\sum_{\mu} (M_{\mu \alpha}^{(0)}(t))^2=\frac{1}{N}\sum_{i=1}^N M_{i \alpha}^2(t)\,.
\end{equation}

\begin{remark}
One might question the relevance of this approximation, given that the coarse-graining is not performed over the frequencies. However, considering a uniform field is equivalent to identifying the time scales with the abstract scales associated with the eigenvalues of the disorder matrix. This is not an exotic approximation in the literature; see, for instance, \cite{duclut2017frequency, canet2011general}.
\end{remark}

\begin{figure}
\begin{center}
\includegraphics[scale=0.55]{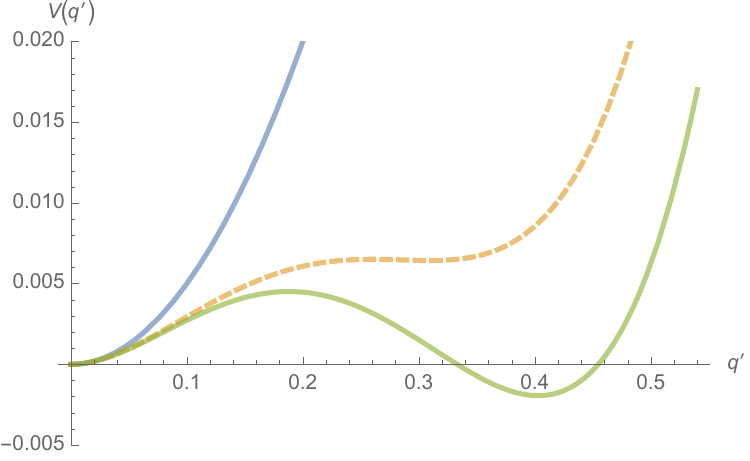}\quad \includegraphics[scale=0.55]{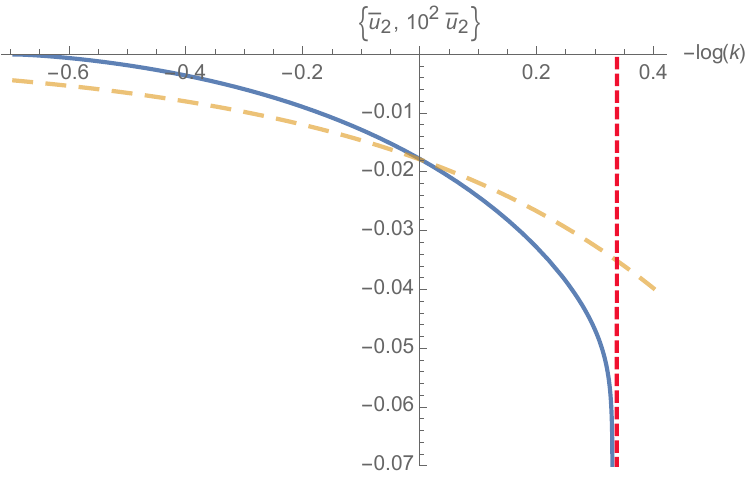}
\end{center}
\caption{Left: Evolution of the critical effective potential ($n=2$) for the quantity $q^\prime$, computed via vertex expansion in the symmetric phase along an RG trajectory exhibiting a finite-scale singularity, for $k=1.50$ (blue), $k=1.45$ (yellow), and $k=1.42$ (green). Right: Behavior of the RG flow near the Gaussian fixed point for vanishing disorder (dashed curve, scaled by a factor of 100) and for non-zero disorder above the critical value (solid blue line).}\label{figplotVq}
\end{figure}

Let us now address the construction of the theory space. For the kinetic action (field power 2), we retain the form provided in \eqref{2ptsvertexExp} for the vertex expansion, while disregarding the flows of the anomalous dimensions. However, we must incorporate non-local and multi-replica contributions:
\begin{align}
\nonumber \Gamma_{k,\text{kin}}&= \frac{1}{2} \int \dd t \sum_{\mu=1}^N \sum_{\alpha=1}^n M_{\mu \alpha}(t)\left(-Y(k)\frac{\dd^2}{\dd t^2} + Z(k)p_\mu^2\right)M_{\mu \alpha}(t) \\\nonumber
&+ \frac{1}{2}\, \int \dd t \int \dd t^\prime \sum_{\mu=1}^N\sum_{\alpha,\beta} q_{\alpha\beta}(k) M_{\mu \alpha}(t) M_{\mu \beta}(t^\prime)\\
&+\frac{1}{2}\, \int \dd t \, \sum_{\mu=1}^N\sum_{\alpha,\beta} q_{\alpha\beta}^\prime(k) M_{\mu \alpha}(t) M_{\mu \beta}(t)\,.
\end{align}
The coupling $q_{\alpha\beta}(k)$ acts between different replicas at different times, whereas $q_{\alpha\beta}^\prime(k)$ represents a coupling between different replicas at the same time. For simplicity, throughout this section, we assume replica symmetry, setting $q_{\alpha\beta}^\prime(k) =: q^\prime(k)$. Mathematically, these couplings govern the RG flow of disorder effects; as highlighted in our previous work \cite{lahoche2024largetimeeffectivekinetics}, their evolution is closely tied to the emergence of finite-scale singularities (see also Figure \ref{figplotVq} and the preceding discussion). Note that the operator $q$, which may be interpreted as a random magnetic field, breaks time-translation invariance in the symmetric phase; however, this symmetry breaking is only asymptotic, as these contributions vanish at frequencies $\omega \neq 0$. If we assume the absence of a random magnetic field, the presence of the $q$ coupling may appear somewhat artificial, and as we shall see, it introduces certain technical difficulties. Physically, one might consider a more realistic operator that also breaks time-translation invariance, analogous to the choice made in \cite{lahoche2023functional}:
\begin{equation}
V^{(2)}_k\propto \frac{1}{2}\, \int \dd \omega \dd \omega^\prime\, \sum_{\mu=1}^N\sum_{\alpha} \Delta(k) M_{\mu \alpha}(\omega) M_{\mu \alpha}(\omega^\prime)\,.
\end{equation}
Although the breaking of time-translation invariance is also a characteristic of glassy phases, we will neglect these effects for the moment to focus on inter-replica correlations.
\medskip 

Now consider the interaction component of the effective average action. Retaining only the bi-local contributions \cite{lahoche2024frequency}, the interaction part of the action is expressed as (see also \cite{tarjus2008nonperturbative}):
\begin{align}
\Gamma_{k,\text{int}}&= \int \dd t \bigg(\sum_{\alpha}\underbrace{U_k[\bm{M}_\alpha^2(t)]}_{\text{Local}}+\frac{1}{2}\sum_{\alpha,\beta} \underbrace{V_k(\Vert \bm{M}_\alpha (t)\Vert,\Vert \bm{M}_\beta(t) \Vert,u_{\alpha \beta}(t,t))}_{\text{Coupling replica}}\\
&+\frac{1}{2} \int \dd t^\prime\sum_{\alpha,\beta} \underbrace{W_k(\Vert \bm{M}_\alpha (t)\Vert,\Vert \bm{M}_\beta(t^\prime) \Vert,u_{\alpha \beta}(t,t^\prime))}_{\text{Non-local}}\bigg)\,,
\end{align}
where $\bm{M}_{\alpha}=(M_{1\alpha},\cdots , M_{N\alpha})$, $\bm{M}_\alpha^2:=\sum_i M_{i\alpha}^2$, and 
\begin{equation}
u_{\alpha \beta}(t,t^\prime):=\frac{\bm{M}_{\alpha}(t)}{\Vert \bm{M}_\alpha(t) \Vert}\cdot \frac{\bm{M}_{\beta}(t^\prime)}{\Vert \bm{M}_\beta (t^\prime)\Vert}. 
\end{equation}
The local potential $U_k$ depends on fields at the same time and replica index; for the remainder of this paper, we assume that $U_k$ is expanded around the global minimum $\kappa(k)$ as:
\begin{equation}
\frac{U_k[\bm{M}_\alpha^2(t)]}{N}=\frac{u_4(k)}{2}\left(\frac{\bm{M}_\alpha^2(t)}{2N}-\kappa(k)\right)^2+\frac{u_6(k)}{3}\left(\frac{\bm{M}_\alpha^2(t)}{2N}-\kappa(k)\right)^3+\cdots\,.
\end{equation}
In contrast, the non-local potential $V_k$ is assumed to be expanded in powers of the field up to order 4, and we restrict ourselves to contributions up to the sextic order. Graphically:
\begin{align}
V_k= \, \underbrace{\vcenter{\hbox{\includegraphics[scale=0.8]{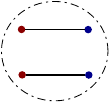}}}}_{v_{4,1}}\,+\, \underbrace{\vcenter{\hbox{\includegraphics[scale=0.8]{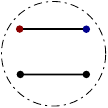}}}}_{v_{4,2}}\,+\, \underbrace{\vcenter{\hbox{\includegraphics[scale=0.8]{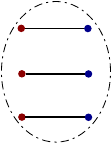}}}}_{v_{6,1}}\,+\, \underbrace{\vcenter{\hbox{\includegraphics[scale=0.8]{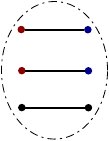}}}}_{v_{6,2}}\,+\, \underbrace{\vcenter{\hbox{\includegraphics[scale=0.8]{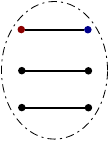}}}}_{v_{6,3}}\,,\label{listinteractions1}
\end{align}
\begin{align}
W_k= \, \underbrace{\vcenter{\hbox{\includegraphics[scale=0.8]{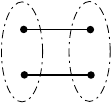}}}}_{w_{4,1}}\,+\, \underbrace{\vcenter{\hbox{\includegraphics[scale=0.8]{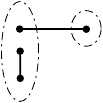}}}}_{w_{4,2}}\,+\, \underbrace{\vcenter{\hbox{\includegraphics[scale=0.8]{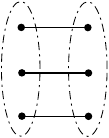}}}}_{w_{6,1}}\,+\, \underbrace{\vcenter{\hbox{\includegraphics[scale=0.8]{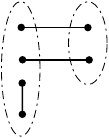}}}}_{w_{6,2}}\,+\, \underbrace{\vcenter{\hbox{\includegraphics[scale=0.8]{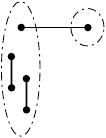}}}}_{w_{6,3}}\,+\, \underbrace{\vcenter{\hbox{\includegraphics[scale=0.8]{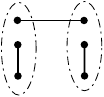}}}}_{w_{6,4}}\,.\label{listinteractions2}
\end{align}
Note that nodes of different colors (red or blue) along a given bubble correspond to distinct replica indices. For instance, the proportionality relation (which defines the coupling constant) is given by:
\begin{equation}
\vcenter{\hbox{\includegraphics[scale=0.8]{Vk41Loc}}}\propto\int \dd t \sum_{\alpha,\beta}\, (\bm{M}_\alpha(t) \cdot \bm{M}_\beta(t))^2\,.
\end{equation}
Prior to the projection onto the vacuum defined in \eqref{vacIR}, all replicated fields are aligned along the same axis, such that $u_{\alpha\beta} \to \epsilon_{\alpha\beta}$ for all $\alpha, \beta$. It follows that:
\begin{align}
\nonumber V_k \big\vert_{M_\alpha =2M^{(0)}_{\alpha}}&=N\Big( v_{4,1} \rho_\alpha \rho_\beta +v_{4,2} \epsilon_{\alpha\beta}\rho_\alpha \sqrt{\rho_\alpha \rho_\beta} + \epsilon_{\alpha\beta}v_{6,1} \sqrt{\rho_\alpha^3 \rho_\beta^3} + v_{6,2} \rho_\alpha^2 \rho_\beta \\
& + \epsilon_{\alpha\beta} v_{6,3}\rho_\alpha^2 \sqrt{\rho_\alpha \rho_\beta} \Big)\,,
\end{align}
and:
\begin{align}
\nonumber W_k \big\vert_{M_\alpha =2 M^{(0)}_{\alpha}}&=N\Big( w_{4,1} \rho_\alpha \rho_\beta + \epsilon_{\alpha\beta} w_{4,2}\rho_\alpha \sqrt{\rho_\alpha \rho_\beta} +  \epsilon_{\alpha\beta}w_{6,1} \sqrt{\rho_\alpha^3 \rho_\beta^3} + w_{6,2} \rho_\alpha^2 \rho_\beta \\
& +  \epsilon_{\alpha\beta} w_{6,3}\rho_\alpha^2 \sqrt{\rho_\alpha \rho_\beta} +  \epsilon_{\alpha\beta} w_{6,4}  \rho_\alpha \rho_\beta \sqrt{\rho_\alpha\rho_\beta}  \Big)\,,
\end{align}
These relations also define the normalization of the various couplings $v_{2n,p}$ and $w_{2n,p}$. Note that the $v$ couplings are introduced for the same reasons as $q^\prime$. This truncation is extensive, as it incorporates bi-local sextic cumulants; however, it involves a large number of couplings, rendering the flow equations computationally demanding to analyze. Regarding the non-local couplings, only $w_{4,1}$ and $w_{6,1}$ admit a physical interpretation in terms of disorder of rank $2$ and $3$, respectively, provided they are negative—analogous to the interpretation of $q < 0$ as disorder of rank $1$. The presence of rank-$2$ disorder in this formalism effectively corrects the variance of the original matrix; the sum of two Wigner matrices with variances $\sigma^2$ and $\tau^2$, respectively, results in a Wigner matrix \cite{potters2020first} with variance $\sigma^2 + \tau^2$. Nevertheless, in the approximation considered here, where the field possesses only a time-independent macroscopic component, these non-local correlations do not provide additional information compared to local interactions that correlate different responses, even though these non-local interactions do exert a back-reaction on the latter.
\medskip

The underlying logic of our approach allows for certain simplifications, provided they yield consistent results—a point we will verify later. In our previous work \cite{lahoche2024largetimeeffectivekinetics} and the companion paper \cite{lahoche2024frequency}, we demonstrated that finite-scale singularities are linked to an instability of the 2PI effective potential for $q^\prime$ (the so-called Ward-Luttinger functional). More precisely, we observed that the potential develops a second stable minimum at $q^\prime_0 \neq 0$, which eventually becomes deeper than the minimum at $q^\prime = 0$. Figure \ref{figplotVq} reproduces the behavior of the Ward-Luttinger functional for $q^\prime$, as computed via vertex expansion in our previous work (within the symmetric phase) for an RG trajectory exhibiting a finite-scale divergence. In this example, with initial conditions for local quartic and quadratic couplings close to the critical region, the first non-zero value for $q^\prime$ is approximately $0.30$. Note that this computation was performed in real time. In imaginary time, the sign of the coupling is negative under our conventions—see also \cite{lahoche2024frequency}.
\medskip

\begin{remark}
To clarify the physical picture for readers familiar with the standard Parisi RSB scheme, it is important to note that the quantum $p$-spin model typically exhibits a Random First-Order Transition (RFOT). In our FRG framework, the finite-scale singularity signals the transition, also accompagned with a first order phase transition, as the figure \ref{figplotVq} shows.
\end{remark}

This order of magnitude appears essentially independent of the initial values of the local couplings. Such behavior is reminiscent of a first-order phase transition, where the final stage—in which the vacuum at $q^\prime = 0$ becomes unstable—is masked by the singularity of the flow\footnote{Recall that to construct the potential, we expanded the Schwinger-Dyson equation and computed the coefficients at $q^\prime = 0$.}. Crucially, we assume that these instabilities drive the growth of the interactions within the potential $V_k$, much as we assume the instabilities of the vacuum $p=0$ are linked to the growth of the non-local interactions generating $W_k$. Since the initial action contains only one non-local sextic interaction and one local propagator, we expect that only a limited set of interactions will dominate the flow as long as $q$ and $q^\prime$ remain sufficiently small, an assumption we maintain throughout this section. Note that this approximation would be highly accurate if the transition were continuous (second-order). However, because the transition is first-order (discontinuous) with a significant discontinuity for $q^\prime$, the perturbative assumption constitutes a strong limitation of our study; a more exhaustive and rigorous calculation, accounting for the currently neglected interactions, would be necessary. Nevertheless, predictions derived from the purely perturbative theory space cannot be regarded as definitive, given that they inevitably lead to divergences. Our hypothesis—the relevance of which can only be assessed through numerical experiments—is that accounting for these non-perturbative (yet sufficiently small) interactions sufficiently early in the flow will prevent or delay the emergence of singularities without significantly biasing the flow.
\medskip

Classical spin glass physics teaches us that there are several distinct characterizations of glass transitions; as noted above, correlations between replicas and the breaking of time-translation invariance are common signatures. In this section, however, we focus primarily on inter-replica correlations, bearing in mind that this approximation limits the complexity of the exact phase space, which we assume also includes regions where time-translation invariance is broken. This choice is further justified by the nature of the approximation (time-uniform field) employed here. Since the flow equations in the Wetterich formalism involve only a single effective loop, it is straightforward to evaluate the relative importance of the vertices $v_{2n,p}$ and $w_{2n,p}$. We find the following scalings: $v_{4,1} = \mathcal{O}(q^\prime)$, $v_{4,2} = \mathcal{O}((q^\prime)^3)$, $v_{6,1} = \mathcal{O}((q^\prime)^4)$, $v_{6,2} = \mathcal{O}((q^\prime)^2)$, and $v_{6,3} = \mathcal{O}((q^\prime)^3)$. Consequently, the approximation we adopt is defined as:

\begin{align}
\nonumber \Gamma_{k}&= \frac{1}{2} \int \dd t \sum_{\mu=1}^N \sum_{\alpha=1}^n M_{\mu \alpha}(t)\left(-\frac{\dd^2}{\dd t^2} +p_\mu^2\right)M_{\mu \alpha}(t) +\frac{1}{2}\, \int \dd t \, \sum_{\mu=1}^N\sum_{\alpha,\beta} q^\prime(k) M_{\mu \alpha}(t) M_{\mu \beta}(t)\\
& +  \int \dd t \,\sum_{\alpha}\underbrace{U_k[\bm{M}_\alpha^2(t)]}_{\text{Local}}+v_{4,1}\,\vcenter{\hbox{\includegraphics[scale=0.8]{Vk41Loc}}}+w_{6,1}\,\vcenter{\hbox{\includegraphics[scale=0.8]{Vk61}}}\,.
\end{align}
\medskip

Now, as a preliminary to the derivation of the flow equations, let us compute the effective 2-point function. By taking the second functional derivative with respect to $M_{\mu\alpha}(t)$ and $M_{\mu\beta}(t^\prime)$ and projecting onto $M_{\alpha}^{(0)}$, we obtain:
\begin{align}
\Gamma_{k}^{(2)}(t,t^\prime)+R_k(p_\mu^2) \delta_{\mu\nu} \delta(t-t^\prime)&=A \delta(t-t^\prime)+B
\end{align}
where:
\begin{align}
\nonumber A&:=\delta_{\alpha \beta}\delta_{\mu \nu} \Big(-\frac{\dd^2}{\dd t^2} + p_\mu^2 + \frac{\partial U_k}{\partial \rho_\alpha}+R_k(p_\mu^2)\Big) +2 \rho_\alpha \frac{\partial^2 U_k}{\partial \rho_\alpha^2}\,\delta_{\alpha \beta}\delta_{\mu 0}\delta_{\nu 0}\\
&+\Big(q^\prime+4v_{4,1}\sqrt{\rho_\alpha \rho_\beta} \Big)\delta_{\mu\nu} +4 v_{4,1}\left(\sum_{\beta^\prime} \rho_{\beta^\prime} \right) \delta_{\mu 0} \delta_{\nu 0}\delta_{\alpha\beta}+12 \beta w_{6,1} \sqrt{\rho_\alpha} \sum_{\beta^\prime} \rho_{\beta^\prime}^{3/2} \delta_{\mu 0} \delta_{\nu 0}\delta_{\alpha\beta}\,,
\end{align}
and:
\begin{equation}
B:=12 w_{6,1} \rho_\alpha \rho_\beta\, \delta_{\mu\nu} \,.
\end{equation}
We define Fourier transform of some function $f(t)$ as:
\begin{equation}
f(t):= \frac{1}{\beta}\sum_{\omega} \tilde{f}(\omega) e^{-i \omega t}\,,\qquad  \tilde{f}(\omega):= \int_{-\beta/2}^{\beta/2} \dd t \, f(t) e^{i \omega t}\,,
\end{equation}
where frequencies are assumed to be quantified $\omega=2 \pi n/\beta$, $n\in \mathbb{N}$. The function $G_k(t,t^\prime)$ is defined by the condition:
\begin{equation}
A G(t,t^\prime)+B \int \dd t^{\prime\prime} G(t^{\prime\prime},t^\prime)= \delta(t-t^\prime)\,,
\end{equation}
which can be rewritten in the Fourier mode as:
\begin{equation}
\tilde{A}(\omega^2) \,G(\omega,\omega^\prime)+B \beta \, G(0,\omega^\prime)\delta_{\omega,0}=\beta \, \delta_{\omega,-\omega^\prime}\,.
\end{equation}
We expect a solution of the form:
\begin{equation}
G_k(\omega,\omega^\prime)=\beta \tilde{A}^{-1}(\omega^2) \delta_{\omega,-\omega^\prime}+ D \delta_{\omega,0} \delta_{\omega^\prime,0}\,,\label{Gminus1}
\end{equation}
and we get the conditions:
\begin{equation}
 \tilde{A}(0)D +\beta^2 B  \tilde{A}^{-1}(0)+ \beta B D=0\,,
\end{equation}
leading to:
\begin{equation}
D=-(\tilde{A}(0)+\beta B)^{-1}B \beta^2 \tilde{A}^{-1}(0)\,,
\end{equation}
and we get:
\begin{equation}
\boxed{G_k(\omega,\omega^\prime)= \beta \left(\tilde{A}^{-1}(\omega^2) \delta_{\omega,-\omega^\prime}-(\tilde{A}(0)+\beta B)^{-1} B \beta \tilde{A}^{-1}(0) \delta_{\omega,0} \delta_{\omega^\prime,0}\right)\,.} 
\end{equation}
Furthermore, the inverse of the matrix $\tilde{A}$ can be readily computed by projecting $\rho_\alpha$ onto the uniform running vacuum $\kappa(k)$:
\begin{align}
\nonumber \left(\tilde{A}^{-1}\right)_{\mu\nu\alpha\beta}(\omega^2)&= \frac{\delta_{\mu\nu}^{(0)}}{\ell_k(p_\mu^2,\omega^2)+R_k(p_\mu^2)} \left[\delta_{\alpha\beta}-\frac{F}{\ell_k(p_\mu^2,\omega^2)+R_k(p_\mu^2)+n F} \right]\\
&+ \frac{\delta_{\mu 0}\delta_{\nu 0}}{\wp_k(\omega^2)+R_k(0)}\left[\delta_{\alpha\beta}-\frac{F}{\wp_k(\omega^2)+R_k(0)+n F} \right]\,.\label{equationAkappa}
\end{align}
where:
\begin{align}
\ell_k(p_\mu^2,\omega^2)&:=\omega^2+p_\mu^2+\partial_{\rho_\alpha} U_k\,,\\
\wp_k(\omega^2)&:=\omega^2+\partial_{\rho_\alpha} U_k+2 \rho_\alpha \partial_{\rho_\alpha}^2 U_k+n \Big[4 v_{4,1} \kappa+\beta \, \Big(12 w_{6,1}  \kappa^{2} \Big)\Big]\\
F&:= q^\prime+4 v_{4,1}\kappa(k)\,,
\end{align}
and:
\begin{equation}
\delta_{\mu\nu}^{(0)}:=\delta_{\mu\nu}-\delta_{\mu 0}\delta_{\nu 0}\,.
\end{equation}
Note that $\partial_{\rho_\alpha} U_k = 0$ for $\rho_\alpha = \kappa$ by construction; nevertheless, we retain the explicit dependence on $\partial_{\rho_\alpha} U_k$ to track this term. Furthermore, in the continuum limit, as we replace the discrete sum with an integral over the Wigner distribution, the contribution of the $\mu = 0$ component vanishes. We thus define:
\begin{equation}
\left(\tilde{A}^{-1}\right)_{\mu\nu\alpha\beta}(\omega^2)=:\delta_{\mu\nu}^{(0)}\left(\mathfrak{A}_{\bar{0}}^{-1}\right)_{\alpha\beta}(p^2,\omega^2)+\delta_{\mu 0}\delta_{\nu 0}\left(\mathfrak{B}_{\bar{0}}^{-1}\right)_{\alpha\beta}(p^2,\omega^2)\,,
\end{equation}
where, projecting along $\kappa$:
\begin{equation}
\left(\mathfrak{A}_{\bar{0}}^{-1}\right)_{\alpha\beta}(p^2,\omega^2)\big\vert_{\rho_\alpha=\kappa}:= \frac{1}{\ell_k(p_\mu^2,\omega^2)+R_k(p_\mu^2)} \left[\delta_{\alpha\beta}-\frac{F}{\ell_k(p_\mu^2,\omega^2)+R_k(p_\mu^2)+n F} \right]\,.\label{defAfrak}
\end{equation}
Now, let us return to the flow equation. The exact flow equation is given by (up to this point, we have kept the dependence on the generalized momenta for $G_k$ explicit):
\begin{equation}
\dot{\Gamma}_k=\frac{N}{2}\,\int \rho(p_\mu^2) \dd p_\mu^2 \int \dd t \dd t^\prime \, G_k(p_\mu^2,t,t^\prime) \dot{R}_k (p_\mu^2,\vert t-t^\prime \vert )\,,\label{floweqWett}
\end{equation}
where for our purpose, 
\begin{equation}
{R}_k (p_\mu^2,\vert t-t^\prime \vert) = R_k(p_\mu^2) \delta(t-t^\prime)\,.
\end{equation}
Then, by projecting both sides onto the uniform field $M_\alpha^{(0)}$, the equation takes the form (an integral over $p_\mu$ is implied on the left-hand side):
\begin{align}
\nonumber f[\{\rho_\alpha\},u_4,u_6,q^\prime,v_{4,1}]&+\beta g[w_{6,1},\{\rho_\alpha\}]=  \frac{1}{2\beta} \sum_{\omega} \int \dd p^2 \rho(p^2) R_k(p^2) \sum_\alpha \left(\mathfrak{A}_{\bar{0}}^{-1}\right)_{\alpha\alpha}(p^2,\omega^2)\\
&- \frac{1}{2}  \int \dd p^2 \rho(p^2) R_k(p^2) \sum_{\alpha,\beta,\gamma} (\mathfrak{A}_{\bar{0}}(p^2,0)+\beta B)^{-1}_{\alpha\beta} B_{\beta \gamma}\left(\mathfrak{A}_{\bar{0}}^{-1}\right)_{\gamma\alpha}(p^2,0)\,,
\end{align}
where the functionals $f$ and $g$ will be defined hereafter. For sufficiently large $\beta$, the discrete sum can be converted into an integral:
\begin{align}
\nonumber f[\{\rho_\alpha\},u_4,u_6,q^\prime,v_{4,1}]&+\beta g[w_{6,1},\{\rho_\alpha\}]=  \frac{1}{4\pi} \int \dd \omega \int \dd p^2 \rho(p^2) R_k(p^2) \sum_\alpha \left(\mathfrak{A}_{\bar{0}}^{-1}\right)_{\alpha\alpha}(p^2,\omega^2)\\
&- \frac{1}{2}  \int \dd p^2 \rho(p^2) R_k(p^2) \sum_{\alpha,\beta,\gamma} (\mathfrak{A}_{\bar{0}}(p^2,0)+\beta B)^{-1}_{\alpha\beta} B_{\beta \gamma} \left(\mathfrak{A}_{\bar{0}}^{-1}\right)_{\gamma\alpha}(p^2,0)\,,
\end{align}
We might consider identifying the left- and right-hand sides by matching powers of $\beta$. However, $\beta$ is treated as a large parameter. If we take the $\beta \to \infty$ limit naively, we conclude that the left-hand side no longer depends on $B$, and consequently, on the non-local coupling. Let us attempt to refine this approach. A moment of reflection reveals that $\tilde{A}(0) = k^2 \mathcal{O}(1)$; furthermore, $B$ has a scaling dimension of $3$ according to power counting, such that $B = k^3 \bar{B}$. Then,
\begin{equation}
\tilde{A}(0)+\beta B= \tilde{A}(0)\left(1+ \frac{\beta k}{\mathcal{O}(1)} \bar{B} \right)\,.
\end{equation}
Then, we have only one interesting regime where the flow of disorder is non-trivial, for
\begin{equation}
\bar{B} \beta k \ll 1\,.
\end{equation}
From this condition, we can distinguish two different sub-regimes. First, if $\bar{B}=\mathcal{O}(1)$, the condition reduces to:
\begin{equation}
\beta k \ll 1\,.
\end{equation}
In other words, the time scale should be much smaller than the generalized momentum scale. Since the eigenvalue spacing is of order $1/N$ for a Wigner matrix of size $N$, we expect the minimal value for $k$ to be $\sim N^{-1/2}$. Consequently, the upper bound for the generalized momentum scale is $\sim \sqrt{N}$, and we obtain the following bound:
\begin{equation}
\beta \ll \sqrt{N}\,.\label{outeq}
\end{equation}
The construction of this continuous limit is highly non-trivial and holds only in the deep IR, for sufficiently large $N$. Note that in classical spin glass kinetics, $\beta \sim \sqrt{N}$ represents the time scale required to reach equilibrium \cite{Dominicis}. It is therefore appropriate to rephrase condition (1). Given that $\bar{B} \sim \bar{w}_{6,1} \bar{\kappa}^2$, we obtain:
\begin{equation}
\bar{\kappa}^2 \ll \mathcal{O}(1) \, \frac{k^{-1}}{\beta \bar{w}_{6,1}}\,,
\end{equation}
or if $\bar{w}_{6,1} \sim k \mathcal{O}(1)$,
\begin{equation}
\boxed{\bar{\kappa}^2 \ll \mathcal{O}(1) \, \frac{k^{-2}}{\beta}\,.}\label{conditioncritical}
\end{equation}
For fixed $k$, taking the limit $\beta \to \infty$ implies $\bar{\kappa} \to 0$. Interestingly, this condition pertains to what may be termed global magnetization. As $\lambda \to 0$ (vanishing tensorial disorder), the system exhibits a second-order phase transition reminiscent of a ferromagnetic transition, in which the $\mu=0$ component of the $M_\mu$ field becomes macroscopic ($M_0 \propto \sqrt{N}$), as demonstrated in our previous work \cite{lahoche2024largetimeeffectivekinetics} in the $N \to \infty$ limit. In other words, we expect that as the magnitude of the vacuum $M^{(0)}$ increases, the flow reaches a regime where the effective matrix-like disorder is much larger than the rank-$3$ disorder $J_{i_1 i_2 i_3}$. Condition \eqref{conditioncritical} implies that the theory must reside within its critical regime:
\begin{equation}
{\kappa}^2  \ll \mathcal{O}(1) \, \frac{k^{2}}{\beta}\,,
\end{equation}
and $\kappa \to 0$ as $k\to 0$. Expanding $G_k$ in power of $\beta$, we have:
\begin{align}
\nonumber f[\{\rho_\alpha\},u_4,u_6,q^\prime,v_{4,1}]&=  \frac{1}{4\pi} \int \dd \omega \int \dd p^2 \rho(p^2) \dot{R}_k(p^2) \sum_\alpha \left(\mathfrak{A}_{\bar{0}}^{-1}\right)_{\alpha\alpha}(p^2,\omega^2)\\
&- \frac{1}{2} \int \dd p^2 \rho(p^2) \dot{R}_k(p^2) \sum_{\alpha,\beta,\gamma} \left(\mathfrak{A}_{\bar{0}}^{-1}\right)_{\alpha \beta}(p^2,0) B_{\beta \gamma} \left(\mathfrak{A}_{\bar{0}}^{-1}\right)_{\gamma\alpha}(p^2,0)\,,\label{flowf}
\end{align}
and:
\begin{align}
\nonumber g[w_{6,1},\{\rho_\alpha\}]&=\frac{1}{2}  \int \dd p^2 \rho(p^2) \dot{R}_k(p^2) \sum_{\alpha,\beta,\gamma,\delta,\eta}  \left(\mathfrak{A}_{\bar{0}}^{-1}\right)_{\alpha \beta}(p^2,0)\\
& \qquad \times B_{\beta \eta}  \left(\mathfrak{A}_{\bar{0}}^{-1}\right)_{\eta\delta}(p^2,0) B_{\delta \gamma}  \left(\mathfrak{A}_{\bar{0}}^{-1}\right)_{\gamma\alpha}(p^2,0)\,.\label{flowg}
\end{align}
We move on to the derivation of the flow equations for the different couplings in the next subsection. 

\begin{remark}
Before concluding this subsection, let us comment on the general case where non-local couplings are included in the truncation. Considering, for instance, the case where $q \neq 0$ and $w_{4,1} \neq 0$, we have:
\begin{align}
\nonumber A&:=\delta_{\alpha \beta}\delta_{\mu \nu} \Big(-\frac{\dd^2}{\dd t^2} + p_\mu^2 + \frac{\partial U_k}{\partial \rho_\alpha}\Big) +2 \rho_\alpha \frac{\partial^2 U_k}{\partial \rho_\alpha^2}\,\delta_{\alpha \beta}\delta_{\mu 0}\delta_{\nu 0}\\\nonumber
&+\Big(q^\prime+4v_{4,1}\sqrt{\rho_\alpha \rho_\beta} \Big)\delta_{\mu\nu} +4 v_{4,1}\sum_{\beta^\prime} \rho_{\beta^\prime} \delta_{\mu 0} \delta_{\nu 0}\delta_{\alpha\beta}\\
&+\beta \,\sum_{\beta^\prime} \Big(4 w_{4,1} \rho_{\beta^\prime}+12 w_{6,1} \sqrt{\rho_\alpha} \rho_{\beta^\prime}^{3/2} \Big)\delta_{\mu 0} \delta_{\nu 0}\delta_{\alpha\beta}\,,
\end{align}
and:
\begin{equation}
B:=\Big( q+4 w_{4,1}\sqrt{\rho_\alpha \rho_\beta}+12 w_{6,1} \rho_\alpha \rho_\beta \Big) \delta_{\mu\nu} \,.
\end{equation}
The condition $\beta k \bar{B} \ll 1$ now takes on a different meaning. In the critical regime, as $\kappa \to 0$, we find that $\bar{q} \ll k^{-2}/\beta$, implying $q \to 0$ as $k \to 0$. This limit is therefore incompatible with the existence of a macroscopic value for $q$ arising from a phase transition.
\end{remark}

\begin{remark}
Let us conclude this section with a remark regarding the anomalous dimension, which is absent from the present truncation. The LPA, fundamentally, relies on the assumption of a weak momentum dependence of the vertices, implying that the anomalous dimension $\eta$ must remain small. It is worth noting that in the symmetric phase, the anomalous dimension is always a next-to-leading-order effect—a direct consequence of the Ward identities (Section \ref{sec2}). However, in the non-symmetric phase and close to the finite-scale singularity, $\eta$ is expected to grow, precisely signaling the breakdown of the derivative expansion at its lowest order. We do not claim that the LPA remains quantitatively exact through the singularity without corrections, but rather that it reliably detects the onset of the instability. Moreover, as explicitly shown below, these singularities can be cured without resorting to the anomalous dimension.
\end{remark}

\subsection{Flow equations}\label{regime1}

In this section, we derive the approximate flow equations for the various couplings. For simplicity, we assume that $\epsilon_{\alpha\beta} = 1$ for all $\alpha, \beta$. We begin by considering the coupling $w_{6,1}$. The functional $g$ can be readily computed as follows:
\begin{equation}
g[w_{6,1}, \{\rho_\alpha \}]  \big \vert_{\rho_\alpha=\kappa}= n^2 \dot{w}_{6,1} \,\kappa^3\,.
\end{equation}
Then, projecting the flow equation \eqref{flowg} along $\kappa$, and using the definition \eqref{defAfrak}, we get, assuming $\kappa \neq 0$:
\begin{align}
\nonumber & n^2 \dot{w}_{6,1} \,=\,288 w_{6,1}^2 \kappa \\
&\qquad \times  \int \dd p^2 \rho(p^2) \dot{R}_k(p^2) \sum_{\alpha,\beta,\gamma,\delta,\eta}  \left(\mathfrak{A}_{\bar{0}}^{-1}\right)_{\alpha \beta}(p^2,0)   \left(\mathfrak{A}_{\bar{0}}^{-1}\right)_{\eta\delta}(p^2,0)   \left(\mathfrak{A}_{\bar{0}}^{-1}\right)_{\gamma\alpha}(p^2,0)\,.
\end{align}
Because we work with the Litim regulator, the integral over $p^2$ factorizes, and we get for $k$ small enough:
\begin{equation}
 \int \dd p^2 \rho(p^2) \dot{R}_k(p^2) \approx \frac{4 k^5}{3 \pi \sigma^{3/2}}\,.
\end{equation}
For $p^2<k^2$, we thus have:
\begin{equation}
\left(\mathfrak{A}_{\bar{0}}^{-1}\right)_{\alpha\beta}(p^2,\omega^2)\big\vert_{\rho_\alpha=\kappa,p^2<k^2}:= \frac{1}{\omega^2+k^2} \left[\delta_{\alpha\beta}-\frac{F}{\omega^2+k^2+n F} \right]\,.\label{defAfrak2}
\end{equation}
Then:
\begin{align}
\boxed{\dot{\bar{w}}_{6,1}\,=\, \bar{w}_{6,1}+ \frac{384}{\pi \sigma^{3/2}} \bar{w}_{6,1}^2 \bar{\kappa}\, \bigg(1-3 n\frac{\bar{F}}{1+n \bar{F}}+3 n^2 \frac{\bar{F}^2}{(1+n \bar{F})^2} - n^3 \frac{\bar{F}^3}{(1+n \bar{F})^3}\bigg)\,,}\label{rel1}
\end{align}
where:
\begin{equation}
\bar{F}:=\bar{q}^\prime+4 \bar{v}_{4,1}\bar{\kappa}(k)\,.
\end{equation}
and $\bar{\kappa}:=k^{-2} \kappa$, $\bar{w}_{6,1}:=k w_{6,1}$. Interestingly, and in contrast with vertex expansion, the disorder renormalizes (we recover the vertex expansion result setting $\kappa=0$). This is particularly opens the possibility of finding a global fixed point in the deep IR. 
\medskip

Now, let us move on to the local counterpart. We have:
\begin{align}
f[\{\rho_\alpha\},u_4,u_6,q^\prime,v_{4,1}]&\,=\,\frac{1}{2} \dot{q}^\prime(k)   \sum_{\alpha \beta} \sqrt{\rho_\alpha \rho_\beta}\\\nonumber
&+\sum_\alpha\Big(\frac{\dot{u}_4(k)}{2}\left(\rho_\alpha-\kappa(k)\right)^2+\frac{\dot{u}_6(k)}{3}\left(\rho_\alpha-\kappa(k)\right)^3+\cdots \Big) \\\nonumber
& - \sum_\alpha \dot{\kappa}(k) \Big(u_4(k)\left(\rho_\alpha-\kappa(k)\right)+{u_6(k)}\left(\rho_\alpha-\kappa(k)\right)^2+\cdots\Big)\\
&+  \sum_{\alpha,\beta}\Big( \dot{v}_{4,1}(k) \rho_\alpha \rho_\beta + \cdots \Big)\,,
\end{align}

First, setting $\rho_\alpha=\kappa$, the local contributions vanish, and we get:
\begin{equation}
f[\{\rho_\alpha\},u_4,u_6,q^\prime,v_{4,1}]\big\vert_{\rho_\alpha=\kappa}= \frac{1}{2} n^2 \dot{q}^\prime   \kappa+ n^2  \dot{v}_{4,1} \kappa^2\,,
\end{equation}
and the flow equation \eqref{flowf} leads to:
\begin{align}
\nonumber  \frac{1}{2} n^2 \dot{q}^\prime   \kappa &+  n^2  \dot{v}_{4,1} \kappa^2=  \frac{n k^5}{3\pi^2 \sigma^{3/2}} \int     \frac{\dd \omega}{\omega^2+k^2} \left[1-\frac{F}{\omega^2+k^2+n F} \right]\\
&- \frac{8 n k w_{6,1} \kappa^2}{\pi \sigma^{3/2}} \bigg(1-2n\frac{\bar{F}}{1+n \bar{F}}+ n^2 \frac{\bar{F}^2}{(1+n \bar{F})^2}\bigg)\,.
\end{align}
Computing the integral, we get finally before subtracting the vacuum energy which does not depend on coupling constants:):
\begin{align}
\nonumber  \frac{1}{2} n^2 (\dot{\bar{q}}^\prime+2{\bar{q}}^\prime)   \bar{\kappa} &+  n^2  \dot{\bar{v}}_{4,1} \bar{\kappa}^2=  \frac{1}{3\pi\sigma^{3/2}}  \left(\frac{1}{\sqrt{1+n\bar{F}}}-1\right)\\
&\qquad - \frac{8 n \bar{w}_{6,1} \bar{\kappa}^2}{\pi \sigma^{3/2}} \bigg(1-2n\frac{\bar{F}}{1+n \bar{F}}+ n^2 \frac{\bar{F}^2}{(1+n \bar{F})^2}\bigg)\,.\label{rel2}
\end{align}
Note that our approximation assumes we remain sufficiently far from the critical regime, with $q^\prime \neq 0$ and $v_{4,1} \neq 0$. At this stage, several strategies exist for deriving the remaining flow equations. One effective method involves the Sherman-Morrison formula, which allows for the inversion of a matrix $\bm{A}$ perturbed by a rank-1 matrix:
\begin{equation}
(\bm A+ \bm u \bm v^T)^{-1}=\bm A^{-1}-\frac{\bm A^{-1} \bm u \bm v^T \bm A^{-1}}{1+\bm v^T \bm A^{-1} \bm u}\,,
\end{equation}
to compute the inverse of a matrix with entries:
\begin{equation}
X_{\alpha \beta}:=a \delta_{\alpha \beta}+b+c \sqrt{\rho_\alpha} \sqrt{\rho_\beta}\,,
\end{equation}
where, for $p^2<k^2$, 
\begin{align}
a&=\omega^2+k^2+U_k^\prime\,,\\
b&=q^\prime\,,\\
c&=4 v_{4,1}\,.
\end{align}
The flow equation for the local sector is then:
\begin{align}
&\nonumber  \frac{1}{2}\dot{q}^\prime   \sum_{\alpha \beta} \sqrt{\rho_\alpha \rho_\beta}+ \sum_{\alpha,\beta}\Big( \dot{v}_{4,1} \rho_\alpha \rho_\beta + \cdots \Big)\\\nonumber
&+\sum_\alpha\Big(\frac{\dot{u}_4(k)}{2}\left(\rho_\alpha-\kappa(k)\right)^2+\frac{\dot{u}_6(k)}{3}\left(\rho_\alpha-\kappa(k)\right)^3+\cdots \Big) \\\nonumber
& - \sum_\alpha \dot{\kappa}(k) \Big(u_4(k)\left(\rho_\alpha-\kappa(k)\right)+{u_6(k)}\left(\rho_\alpha-\kappa(k)\right)^2+\cdots\Big)\\
&= \frac{k^4}{3\pi^2\sigma^{3/2}} \int_{-\infty}^{+\infty} \dd u  \sum_\alpha \bar{X}^{-1}_{\alpha\alpha}(u^2) - \frac{8 \bar{w}_{6,1}k^4}{\pi \sigma^{3/2}} \sum_{\alpha,\beta,\gamma} \bar{X}^{-1}_{\alpha \beta}(0) \bar{\rho}_{\beta} \bar{\rho}_\gamma \bar{X}^{-1}_{\gamma\alpha}(0)\,,\label{fulleq}
\end{align}
where we define the dimensionless quantities $\bar{a} := k^{-2}a$, $\bar{b} := k^{-2}b$, $\bar{c} := c$, and $\bar{X}_{\alpha\beta} := k^{-2}X_{\alpha\beta}$, with $u := \omega/k$ denoting the dimensionless frequency. In principle, all flow equations can be deduced by differentiating both sides of the equation a sufficient number of times with respect to $\sqrt{\rho_\alpha}$. However, the resulting equations are prohibitively complex. It is therefore appropriate to assume that ergodicity is maximally broken (i.e., that replica symmetry holds) and to project the flow equation onto a uniform field $\rho_\alpha = \rho$. In this case (after subtracting the vacuum energy), we obtain:
\begin{align}
&\nonumber  \frac{1}{2} \dot{q}^\prime \rho n^2+  \Big( \dot{v}_{4,1} \rho^2 n^2+ \cdots \Big)\\\nonumber
&+n\Big(\frac{\dot{u}_4(k)}{2}\left(\rho-\kappa(k)\right)^2+\frac{\dot{u}_6(k)}{3}\left(\rho-\kappa(k)\right)^3+\cdots \Big) \\\nonumber
& - n \dot{\kappa}(k) \Big(u_4(k)\left(\rho-\kappa(k)\right)+{u_6(k)}\left(\rho-\kappa(k)\right)^2+\cdots\Big)\\\nonumber
&=  \frac{k^4}{3\pi\sigma^{3/2}}  \left(\frac{1}{\sqrt{1+n\bar{G}+\bar{U}^\prime_k}}+\frac{n-1}{\sqrt{1+\bar{U}_k^\prime}}-n\right)\\
&\qquad - \frac{8 n k^4 \bar{w}_{6,1} \bar{\rho}^2}{(1+\bar{U}^\prime_k)\pi \sigma^{3/2}} \bigg(1-2n\frac{\bar{G}}{1+n \bar{G}+\bar{U}^\prime_k}+ n^2 \frac{\bar{G}^2}{(1+n \bar{G}+\bar{U}^\prime_k)^2}\bigg)\,,
\end{align}
where:
\begin{equation}
\bar{G}:=\bar{q}^\prime+4 \bar{v}_{4,1}\bar{\rho}\,,
\end{equation}
where $\bar{\rho} := k^{-2} \rho$. This approximation is meaningful only because we expand all replicas around the same vacuum $\kappa$. We consider six independent couplings (assuming that all couplings not explicitly listed are zero) and two existing relations given by \eqref{rel1} and \eqref{rel2}. We thus require four additional relations; three of these can be obtained by taking the first, second, and third derivatives of the previous equation with respect to $\rho$. This yields:
\medskip

\begin{align}
\nonumber & 2 n \bar{\kappa } \dot{\bar{v}}_{4,1}+\frac{1}{2} n( \dot{\bar{q}}^\prime+2\bar{q}^\prime)-(\dot{\bar{\kappa}} + 2{\bar{\kappa}} )\bar{u}_4\\\nonumber
&=-\frac{1}{6 n \pi} \bigg[\frac{4 n \left(\bar{v}_{4,1} \left(4 n \bar{\kappa } \bar{v}_{4,1}+n \bar{q}'+1\right){}^{3/2}+24 n \bar{\kappa } \bar{w}_{6,1} \bar{q}'+24 \bar{\kappa } \bar{w}_{6,1}\right)}{\left(4 n \bar{\kappa } \bar{v}_{4,1}+n \bar{q}'+1\right){}^3}\\
&+\bar{u}_4 \left(\frac{1}{\left(4 n \bar{\kappa } \bar{v}_{4,1}+n \bar{q}'+1\right){}^{3/2}}+\frac{48 n \bar{\kappa }^2 \bar{w}_{6,1} \left(4 n \bar{\kappa } \bar{v}_{4,1}+n \bar{q}'-1\right)}{\left(4 n \bar{\kappa } \bar{v}_{4,1}+n \bar{q}'+1\right){}^3}+n-1\right)\bigg]\,,
\end{align}

\begin{align}
\nonumber
& 2 n \dot{\bar{v}}_{4,1}-2 (\dot{\bar{\kappa}}+2\bar{\kappa}) \bar{u}_6+\dot{\bar{u}}_4=-\frac{1}{3 n \pi } \bigg[ (n-1) \bar{u}_6-\frac{3}{4} (n-1) \bar{u}_4^2-\frac{3 \left(4 n \bar{v}_{4,1}+\bar{u}_4\right){}^2}{4 \left(4 n \bar{\kappa } \bar{v}_{4,1}+n \bar{q}'+1\right){}^{5/2}}\\\nonumber
&+\frac{48 n \bar{w}_{6,1}}{\left(4 n \bar{\kappa } \bar{v}_{4,1}+n \bar{q}'+1\right){}^2}-\frac{96 n \bar{\kappa } \bar{u}_4 \bar{w}_{6,1}}{\left(4 n \bar{\kappa } \bar{v}_{4,1}+n \bar{q}'+1\right){}^2}+\frac{48 n \bar{\kappa }^2 \bar{u}_4^2 \bar{w}_{6,1}}{\left(4 n \bar{\kappa } \bar{v}_{4,1}+n \bar{q}'+1\right){}^2}\\\nonumber
&-\frac{48 n \bar{\kappa }^2 \bar{u}_6 \bar{w}_{6,1}}{\left(4 n \bar{\kappa } \bar{v}_{4,1}+n \bar{q}'+1\right){}^2}+\frac{\bar{u}_6}{\left(4 n \bar{\kappa } \bar{v}_{4,1}+n \bar{q}'+1\right){}^{3/2}}\\\nonumber
&+\frac{192 n^2 \bar{\kappa } \bar{w}_{6,1} \left(4 \bar{v}_{4,1} \left(\bar{\kappa } \bar{u}_4-1\right)+\bar{u}_4 \bar{q}'\right)}{\left(4 n \bar{\kappa } \bar{v}_{4,1}+n \bar{q}'+1\right){}^3}-\frac{96 n^2 \bar{\kappa }^2 \bar{u}_4 \bar{w}_{6,1} \left(4 \bar{v}_{4,1} \left(\bar{\kappa } \bar{u}_4-1\right)+\bar{u}_4 \bar{q}'\right)}{\left(4 n \bar{\kappa } \bar{v}_{4,1}+n \bar{q}'+1\right){}^3}\\\nonumber
&+\frac{48 n^2 \bar{\kappa }^2 \bar{w}_{6,1}}{\left(4 n \bar{\kappa } \bar{v}_{4,1}+n \bar{q}'+1\right){}^4}\Big( 16 n \bar{v}_{4,1}^2 \left(\bar{\kappa }^2 \left(\bar{u}_4^2+2 \bar{u}_6\right)-4 \bar{\kappa } \bar{u}_4+3\right)\\\nonumber
&+\bar{q}' \left(\bar{u}_4^2 \left(n \bar{q}'-2\right)+2 \bar{u}_6 \left(n \bar{q}'+1\right)\right)+8\bar{v}_{4,1}\big(\bar{\kappa } \bar{u}_6 \left(2 n \bar{q}'+1\right)\\
&+\bar{u}_4 \left(\bar{\kappa } \bar{u}_4 \left(n \bar{q}'-1\right)-2 n \bar{q}'+1\right)\big) \Big)\bigg]\,,
\end{align}

\begin{align}
&\nonumber  \dot{\bar{u}}_6=2\bar{u}_6+\frac{1}{16 n \pi } \bigg[12 (n-1) \bar{u}_4 \bar{u}_6-5 (n-1) \bar{u}_4^3-\frac{5 \left(4 n \bar{v}_{4,1}+\bar{u}_4\right){}^3}{\left(4 n \bar{\kappa } \bar{v}_{4,1}+n \bar{q}'+1\right){}^{7/2}}+\frac{12 \bar{u}_6 \left(4 n \bar{v}_{4,1}+\bar{u}_4\right)}{\left(4 n \bar{\kappa } \bar{v}_{4,1}+n \bar{q}'+1\right){}^{5/2}}\\\nonumber
&+\frac{384 n \bar{u}_4 \bar{w}_{6,1}}{\left(4 n \bar{\kappa } \bar{v}_{4,1}+n \bar{q}'+1\right){}^2}-\frac{768 n \bar{\kappa } \bar{u}_4^2 \bar{w}_{6,1}}{\left(4 n \bar{\kappa } \bar{v}_{4,1}+n \bar{q}'+1\right){}^2}+\frac{384 n \bar{\kappa }^2 \bar{u}_4^3 \bar{w}_{6,1}}{\left(4 n \bar{\kappa } \bar{v}_{4,1}+n \bar{q}'+1\right){}^2}\\\nonumber
&+\frac{768 n \bar{\kappa } \bar{u}_6 \bar{w}_{6,1}}{\left(4 n \bar{\kappa } \bar{v}_{4,1}+n \bar{q}'+1\right){}^2}-\frac{768 n \bar{\kappa }^2 \bar{u}_4 \bar{u}_6 \bar{w}_{6,1}}{\left(4 n \bar{\kappa } \bar{v}_{4,1}+n \bar{q}'+1\right){}^2}-\frac{768 n^2 \bar{w}_{6,1} \left(4 \bar{v}_{4,1} \left(\bar{\kappa } \bar{u}_4-1\right)+\bar{u}_4 \bar{q}'\right)}{\left(4 n \bar{\kappa } \bar{v}_{4,1}+n \bar{q}'+1\right){}^3}\\\nonumber
&+\frac{1536 n^2 \bar{\kappa } \bar{u}_4 \bar{w}_{6,1} \left(4 \bar{v}_{4,1} \left(\bar{\kappa } \bar{u}_4-1\right)+\bar{u}_4 \bar{q}'\right)}{\left(4 n \bar{\kappa } \bar{v}_{4,1}+n \bar{q}'+1\right){}^3}-\frac{768 n^2 \bar{\kappa }^2 \bar{u}_4^2 \bar{w}_{6,1} \left(4 \bar{v}_{4,1} \left(\bar{\kappa } \bar{u}_4-1\right)+\bar{u}_4 \bar{q}'\right)}{\left(4 n \bar{\kappa } \bar{v}_{4,1}+n \bar{q}'+1\right){}^3}\\\nonumber
&+\frac{768 n^2 \bar{\kappa }^2 \bar{u}_6 \bar{w}_{6,1} \left(4 \bar{v}_{4,1} \left(\bar{\kappa } \bar{u}_4-1\right)+\bar{u}_4 \bar{q}'\right)}{\left(4 n \bar{\kappa } \bar{v}_{4,1}+n \bar{q}'+1\right){}^3}-\frac{1}{\left(4 n \bar{\kappa } \bar{v}_{4,1}+n \bar{q}'+1\right){}^4}\\\nonumber
&\times \Big(768 n^2 \bar{\kappa } \bar{w}_{6,1} \big(8 \bar{v}_{4,1} \left(\bar{\kappa } \bar{u}_6 \left(2 n \bar{q}'+1\right)+\bar{u}_4 \left(\bar{\kappa } \bar{u}_4 \left(n \bar{q}'-1\right)-2 n \bar{q}'+1\right)\right)\\\nonumber
&+16 n \bar{v}_{4,1}^2 \left(\bar{\kappa }^2 \left(\bar{u}_4^2+2 \bar{u}_6\right)-4 \bar{\kappa } \bar{u}_4+3\right)+\bar{q}' \left(\bar{u}_4^2 \left(n \bar{q}'-2\right)+2 \bar{u}_6 \left(n \bar{q}'+1\right)\right)\big) \Big)\\\nonumber
&+\frac{384 n^2 \bar{\kappa }^2 \bar{u}_4 \bar{w}_{6,1}}{\left(4 n \bar{\kappa } \bar{v}_{4,1}+n \bar{q}'+1\right){}^4} \Big( 8 \bar{v}_{4,1} \left(\bar{\kappa } \bar{u}_6 \left(2 n \bar{q}'+1\right)+\bar{u}_4 \left(\bar{\kappa } \bar{u}_4 \left(n \bar{q}'-1\right)-2 n \bar{q}'+1\right)\right)\\\nonumber
&+16 n \bar{v}_{4,1}^2 \left(\bar{\kappa }^2 \left(\bar{u}_4^2+2 \bar{u}_6\right)-4 \bar{\kappa } \bar{u}_4+3\right)+\bar{q}' \left(\bar{u}_4^2 \left(n \bar{q}'-2\right)+2 \bar{u}_6 \left(n \bar{q}'+1\right)\right) \Big)\\\nonumber
&- \frac{768 n^2 \bar{\kappa }^2 \bar{w}_{6,1}}{\left(4 n \bar{\kappa } \bar{v}_{4,1}+n \bar{q}'+1\right){}^5} \Big( 
64 n^2 \bar{v}_{4,1}^3 \left(\bar{\kappa } \bar{u}_4-2\right) \left(\bar{\kappa }^2 \bar{u}_6-\bar{\kappa } \bar{u}_4+1\right)\\\nonumber
&+\bar{u}_4 \bar{q}' \left(\bar{u}_4^2 \left(1-n \bar{q}'\right)+\bar{u}_6 \left(n \bar{q}'-2\right) \left(n \bar{q}'+1\right)\right)\\\nonumber
&-16 n \bar{v}_{4,1}^2\big(\bar{u}_4 \left(2 \bar{\kappa } \bar{u}_4 \left(n \bar{q}'-2\right)-3 n \bar{q}'+\bar{\kappa }^2 \bar{u}_4^2+3\right)+\bar{\kappa } \bar{u}_6 \left(\bar{\kappa } \bar{u}_4 \left(1-3 n \bar{q}'\right)+4 n \bar{q}'+1\right)\big)\\\nonumber
&-4 \bar{v}_{4,1} \big(\bar{u}_6 \left(2 n^2 \left(\bar{q}'\right)^2+\bar{\kappa } \bar{u}_4 \left(n \bar{q}' \left(2-3 n \bar{q}'\right)+2\right)+n \bar{q}'-1\right)\\
&+\bar{u}_4^2 \left(n \bar{q}' \left(n \bar{q}'+2 \bar{\kappa } \bar{u}_4-4\right)-\bar{\kappa } \bar{u}_4+1\right)\big)\Big)\bigg]\,.
\end{align}
Finally, the last relation required to solve the system can be obtained from the full relation \eqref{fulleq} by taking the derivatives with respect to $\sqrt{\rho_1}$ and $\sqrt{\rho_2}$. A straightforward calculation leads to:
\begin{align}
&\nonumber (\dot{\bar{q}}^\prime+2 {\bar{q}}^\prime)+8 \dot{\bar{v}}_{4,1} \bar{\kappa}=\frac{8 \bar{v}_{4,1}}{3 \pi}\frac{\left(n \bar{F} \sqrt{n \bar{F}+1}+\sqrt{n \bar{F}+1}-1\right)}{2 n \left(n \bar{F}+1\right)^{3/2}}\\\nonumber
&+\frac{8 \bar{\kappa} \bar{v}_{4,1}}{3 \pi} (\bar{u}_4+4 v_{4,1}) \Bigg[\bar{u}_4 \frac{\left(n \bar{F} \left(n \bar{F} \left(3 n \bar{F}+2\right) \sqrt{n \bar{F}+1}-5 \sqrt{n \bar{F}+1}+7\right)-4 \sqrt{n \bar{F}+1}+4\right)}{8 n^3 \bar{F} \left(n \bar{F}+1\right)^{5/2}}\\\nonumber
&+ 2 \bar{v}_{4,1} \frac{\left(-2 \sqrt{n \bar{F}+1}+n \bar{F} \left(-2 n \bar{F} \sqrt{n \bar{F}+1}-4 \sqrt{n \bar{F}+1}+5\right)+2\right)}{4 n^2 \bar{F} \left(n \bar{F}+1\right)^{5/2}}\Bigg]\\\nonumber
&+\frac{64 \bar{w}_{6,1} \bar{\kappa }}{\pi \left(1+n \bar{F}\right)^7} \Bigg[ 4 n \left(\bar{F}-3\right) \bar{\kappa }^2 \bar{v}_{4,1}^2 \left(n \bar{F}+1\right)^3\\\nonumber
&+2 \bar{\kappa } \bar{v}_{4,1} \left(2 \bar{\kappa } \bar{u}_4 \left(n \bar{F} \left(n \bar{F}+3\right)-2\right) \left(n \bar{F}+1\right)^3+8 \bar{F} \bar{\kappa }^3 \bar{u}_4^2 \left(n \bar{F} \left(n \bar{F}+2\right)-1\right)+3 \left(n \bar{F}+1\right)^4\right)\\
&- \bar{F} \left(n \bar{F}+1\right)^3 \left(\bar{\kappa } \bar{u}_4 \left(2 n \bar{F}+5\right) \left(n \bar{F}+1\right)-\bar{\kappa }^2 \bar{u}_4^2 \left(n \bar{F}+2\right)-\left(n \bar{F}+1\right)^2 \left(n \bar{F}+2\right)\right)\Bigg]\,.
\end{align}
We study these equations numerically in the following section, including the behavior of trajectories involving these new operators that correlate the replicas. Some analytical results can also be obtained in various limits; we will focus on one of them: the $n \to \infty$ limit. Assuming that the couplings are independent of $n$, the flow equations reduce to\footnote{Note that these equations assume $\kappa \neq 0$.}:

\begin{equation}
\dot{\bar{\kappa}}\,=\,\frac{2 \left(\bar{v}_{4,1}-48 \bar{\kappa } \bar{w}_{6,1}-9 \bar{\kappa } \bar{u}_4\right)}{9 \bar{u}_4}+\frac{1}{6 \pi }\,,
\end{equation}
\begin{equation}
\dot{\bar{q}}\,=\,-2 \bar{q}'\,,
\end{equation}
\begin{equation}
\dot{\bar{u}}_4\,=\,\frac{4 \left(\bar{\kappa } \bar{u}_6-\bar{u}_4\right) \left(\bar{v}_{4,1}-48 \bar{\kappa } \bar{w}_{6,1}\right)}{9 \bar{\kappa } \bar{u}_4}+\frac{\bar{u}_4^2}{4 \pi }\,,
\end{equation}
\begin{equation}
\dot{\bar{v}}_{4,1}\,=\,\frac{8  \bar{\kappa } \left(\bar{v}_{4,1}-48 \bar{\kappa } \bar{w}_{6,1}\right)}{9  \bar{\kappa}^2 n}\,,
\end{equation}
\begin{equation}
\dot{\bar{u}}_6\,=\,+2 \bar{u}_6 -\frac{5 \bar{u}_4^3}{16 \pi }+\frac{3 \bar{u}_6 \bar{u}_4}{4 \pi }\,,
\end{equation}
\begin{equation}
\dot{\bar{w}}_{6,1}\,=\,+ \bar{w}_{6,1}\,.
\end{equation}
Note that the equation for $\bar{v}_4$ must be considered at order $1/n$ to ensure a non-trivial flow (as the next-to-leading order is $O(1/n^2)$). In this limit, the flow of $q^\prime$ decouples from the other equations; while this parameter necessarily played a significant role in the UV, its contribution is masked in this regime, where only the parameter $\bar{v}_{4,1}$ remains.

\subsection{Numerical investigations}\label{regime2}

\begin{figure}
\includegraphics[scale=0.55]{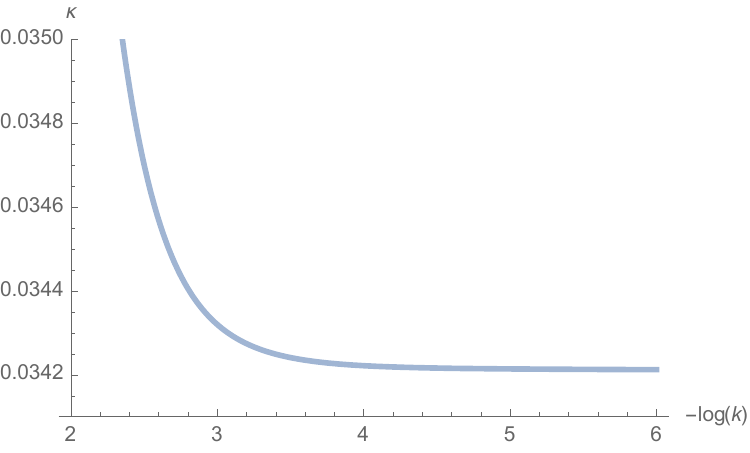}\qquad \includegraphics[scale=0.55]{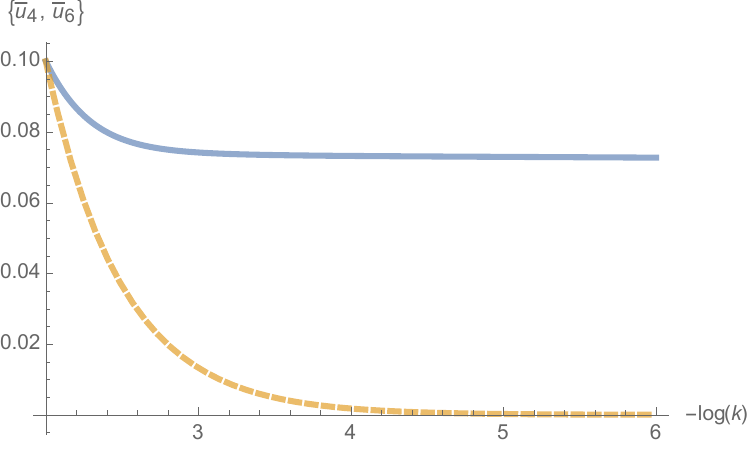}
\caption{Behavior of the RG flow for $n=1$ and initial condition $S_0$ and $w_{6,1}=v_{4,1}=q^\prime=0$. On the right for $\kappa(k)$ and on the left for $\bar{u}_4$ (the solid blue line) and $\bar{u}_6$ (the dashed yellow line).}\label{flowzero}
\end{figure}

\begin{figure}
\includegraphics[scale=0.55]{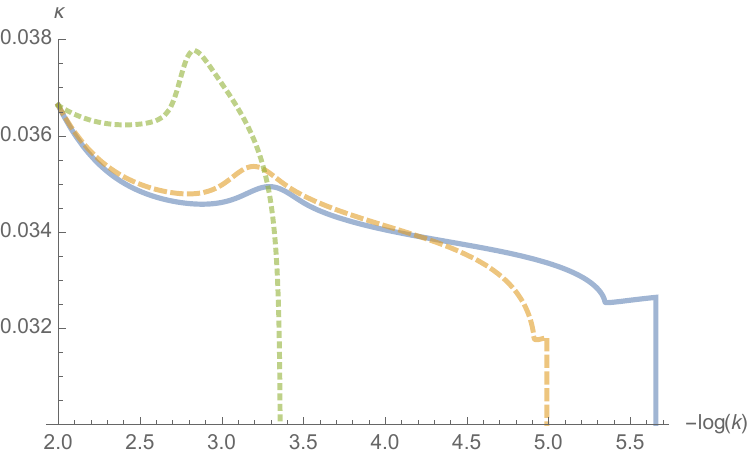}\qquad \includegraphics[scale=0.55]{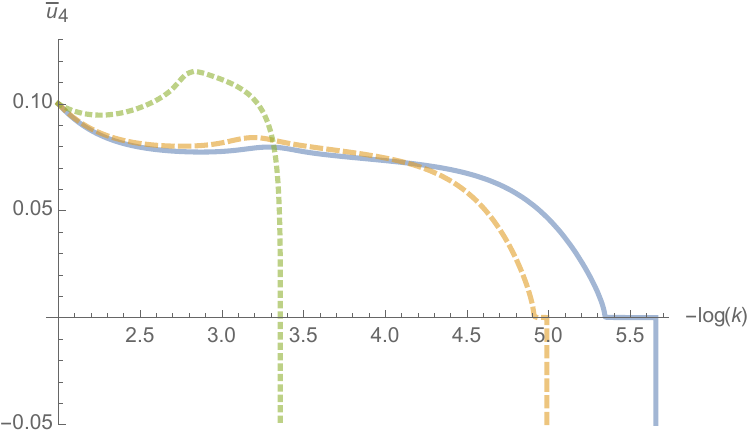}
\caption{Behavior of the RG flow for initial condition $S_0$ and $w_{6,1}(k_0)<-0.0005$. The solid blue curve is for $w_{6,1}(k_0)=-0.0006$, the dashed yellow curve for $w_{6,1}(k_0)=-0.001$, and the green dotted curve for $w_{6,1}(k_0)=-0.005$.}\label{flowNzero}
\end{figure}

Let us investigate the flow equations derived above. The full system of equations is difficult to analyze, even numerically, regarding the existence of a global fixed point. However, we can readily verify that our hypotheses concerning the role of local operators that couple the responses are valid, at least within a region of the phase space sufficiently close to the Gaussian fixed point\footnote{While evaluating this notion of proximity is complex, we will address it in future work, where we intend to establish the phase space of the system precisely, necessitating an analysis of the UV behavior of the flow.}. Let $k_0 := e^{-2} \approx 0.13$ be an infrared scale, and consider the following initial conditions:
\begin{equation}
S_0:=(\bar{\kappa}(k_0)=2\,, \bar{u}_4(k_0)=0.1\,, \bar{u}_6(k_0)=0.1)\,.
\end{equation}
In the absence of disorder, with $\bar{q}^\prime(k_0) = \bar{v}_{4,1} = 0$, the results of the numerical integration are shown in Figure \ref{flowzero}. The flow reaches a quartic fixed point after a few RG steps, and $\kappa$ converges toward a finite positive value, with no singularity occurring along the trajectory. In Figure \ref{flowNzero}, we summarize the same behavior for sufficiently strong disorder ($\bar{w}_{6,1}(k_0) \lesssim -0.0005$), where we recover the finite-scale singularities observed in Section \ref{EVE} and in our previous work \cite{lahoche2024largetimeeffectivekinetics}.
\medskip

\begin{figure}
\begin{center}
\includegraphics[scale=0.55]{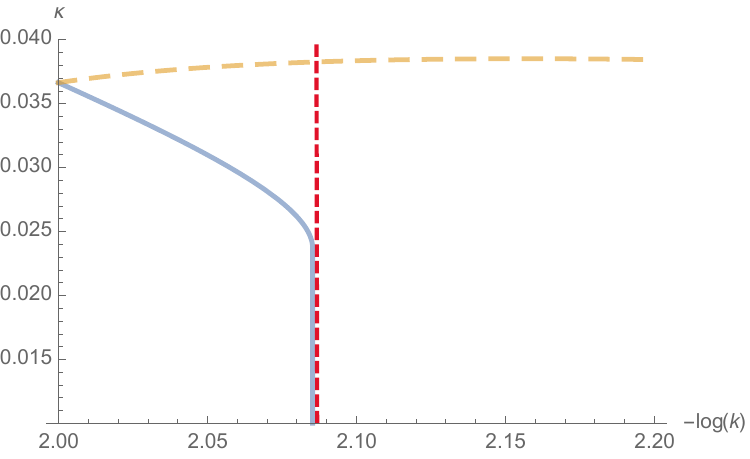}\qquad \includegraphics[scale=0.55]{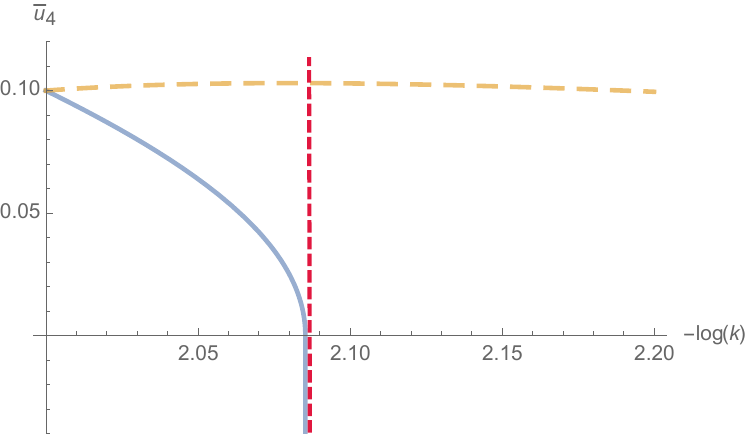}
\end{center}
\caption{Behavior of the RG flow with disorder $\bar{w}_{6,1}(k_0)=-0.01$. The yellow dashed curve is for $\bar{q}^\prime(k_0)=-5.46$ and the solid blue curve for $\bar{q}^\prime(k_0)=0$. On both cases, $\bar{v}_{4,1}(k_0)=1$.}\label{flowNzero1}
\end{figure}

\begin{figure}
\begin{center}
\includegraphics[scale=0.55]{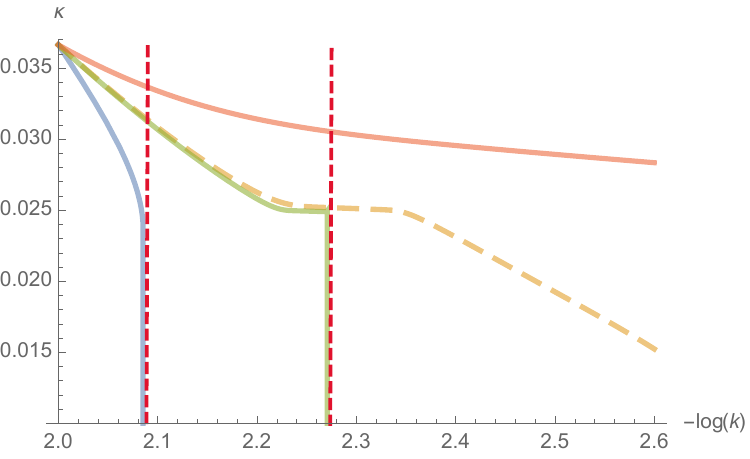}\qquad \includegraphics[scale=0.55]{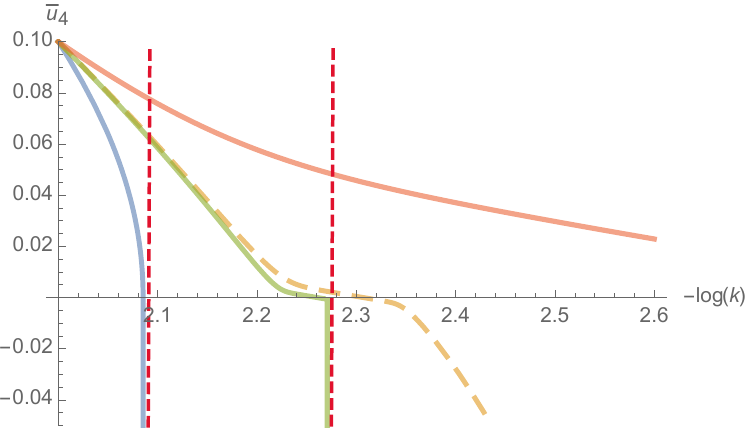}
\end{center}
\caption{Behavior of the RG flow for $\bar{w}_{6,1}(k_0)=-0.01$, $\bar{v}_{4,1}(k_0)=1$ and different values $\bar{q}^\prime(k_0)=0$ (blue curve), $\bar{q}^\prime(k_0)=-2.83$ (green curve), $\bar{q}^\prime(k_0)=-2.89$ (dashed yellow curve) and $\bar{q}^\prime(k_0)=-3.82$ (red curve).}\label{flowNzero2}
\end{figure}

\begin{figure}
\begin{center}
\includegraphics[scale=0.55]{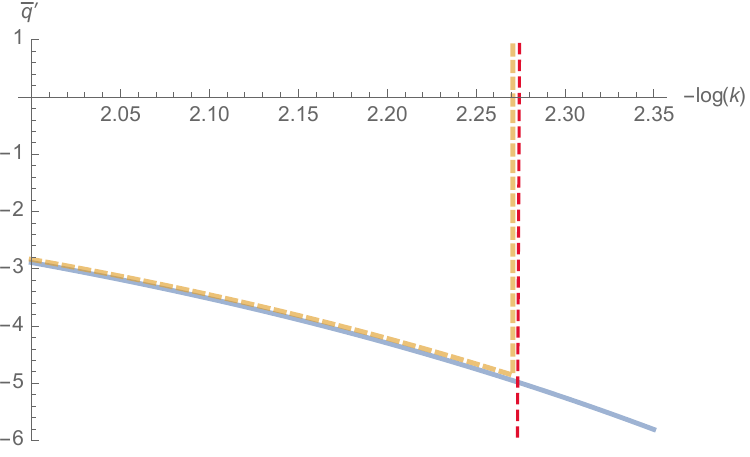}\,\qquad \includegraphics[scale=0.55]{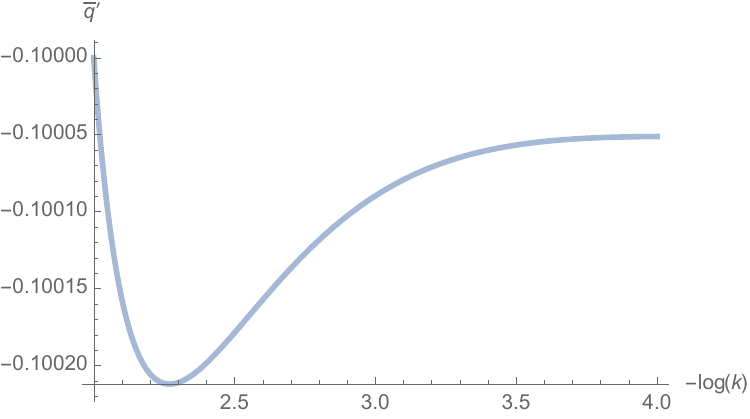}
\end{center}
\caption{On the left: Behavior of the coupling $\bar{q}^\prime(k_0)$ for $\bar{q}^\prime(k_0)=-2.83$ (dashed yellow curve), $\bar{q}^\prime(k_0)=-2.89$ (solid blue curve); the coupling follows a purely scaling law. On the right, behavior of the RG flow for ${q}^\prime(k_0)=-5.46$.}\label{flowNzero3}
\end{figure}

Now, we fix $\bar{w}_{6,1}(k_0) = -0.01$ and $\bar{v}_{4,1}(k_0) = 1$, while varying $\bar{q}^\prime(k_0)$. Note that $\bar{v}_{4,1}$ does not possess the sign required to be interpreted as a physical (time-local) disorder\footnote{A time-local disorder takes the form $\propto \int dt \, J_{i_1\cdots i_n}(t) x_{i_1}(t)\cdots x_{i_n}(t)$.}, in contrast to $q^\prime$. This is imposed by the condition $1 + n\bar{F}(k_0) > 0$. This observation is significant: \textit{the asymptotic flow is not equivalent to a locally multi-disordered system.} The main results are summarized in Figures \ref{flowNzero1} and \ref{flowNzero2}, assuming $n=2$. Consistent with our hypotheses, we observe that beyond a certain critical value $\bar{q}^\prime(k_0) < \bar{q}^\prime_c(k_0) \approx -3.03$, the singularities disappear, and the flow can be extended toward the IR (the green and yellow curves, which lie in the vicinity of the critical value, illustrate this clearly). The evolution of the coupling $\bar{q}^\prime$ is shown in Figure \ref{flowNzero3} for two trajectories. This figure confirms our initial intuition: the singularity is avoided by the introduction of a relevant operator, which persists in the IR and quantifies the obstruction to the factorization of the effective partition function (recalling that the coupling $w_{6,1}$ is irrelevant). Furthermore, $\bar{q}^\prime(k)$ follows an almost perfect scaling law. Upon closer inspection (right panel), we see that the dimensional coupling $q^\prime$ is not strictly constant but approaches a constant value after a transient regime; this value is close to its initial value and remains consistent with our naive estimate of $-0.30$. Notably, however, not all divergences are eliminated. Finally, although we present our numerical analysis for $n=2$, similar results have been obtained for other finite values of $n$.
\medskip

There are still regions of phase space where trajectories diverge, which we put down to our incomplete truncation. For instance, it happens that divergences can be recovered for $-\bar{q}^{\prime}(k_0)$ large enough, even if $1+n \bar{F} >0$. But once again we must not forget that our truncation is very incomplete since it only takes into account strictly local couplings, that the operators only take into account two-replica couplings and that we have limited ourselves to the quartic sector\footnote{Numerically, many of these divergences can be removed by a little shift of the initial value for $\bar{v}_{4,1}(k_0)$, seeming to outline a complicated phase space. It should still be noted, however, that these observations are made based on a very limited truncation.}. We can in a way think that these operators extend the domain of analyticity of the flow, but not maximally, and that a complete reconstruction of the phase space would require more interactions. However, we believe that these results are sufficient to "prove the concept" while waiting for more elaborate methods.
\medskip

\begin{figure}
\begin{center}
\includegraphics[scale=0.7]{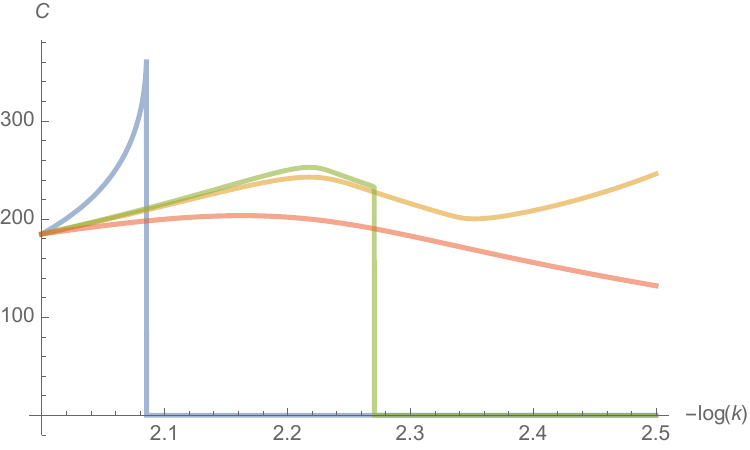}
\end{center}
\caption{Behavior of $C(k)$ for the trajectories we investigated.}\label{flowNzero4}
\end{figure}

\begin{figure}
\begin{center}
\includegraphics[scale=0.7]{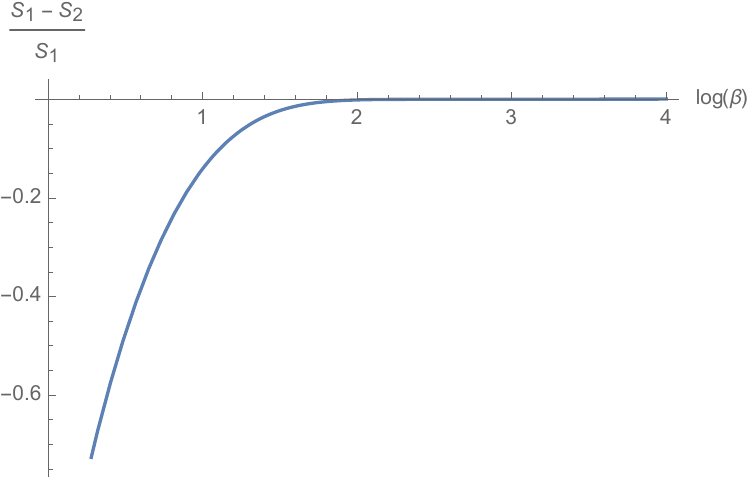}
\end{center}
\caption{Rate of convergence of the sum toward integral.}\label{convrate}
\end{figure}

One limitation arises from the sign of $\bar{F}$, which, in our approximation, depends on only two couplings; we believe this restricts the physical domain of $\bar{q}^\prime$, which could be extended by incorporating additional inter-replica couplings. Another limitation stems from the assumption \eqref{conditioncritical}. In Figure \ref{flowNzero4}, we demonstrate that the trajectories studied here are compatible with an inverse temperature $\beta \approx 100$. Specifically, the function $C(k)$ plotted in the figure is defined as $C(k) := k^{-1} / (\bar{\kappa}^2 \bar{w}_{6,1})$, and these values are consistent with $\beta \sim 100$. At this scale, the continuous approximation remains perfectly valid. Indeed, defining:
\begin{equation}
S_1[\beta]:=\frac{\beta}{2\pi} \int_{-\infty}^{+\infty}  \frac{\dd \omega }{1+\omega^2}=\frac{\beta}{2}\,,
\end{equation}
and,
\begin{equation}
S_2[\beta,n_{\text{max}}]:=\sum_{n=-n_{\text{max}}}^{+n_{\text{max}}} \frac{1}{\Big(\frac{2 \pi n}{\beta}\Big)^2+1} \approx 49.49\,,
\end{equation}
we get $S_1[100]=50$ and $S_2[100,1000] \approx 49.49$, i.e. a difference of just one percent. The rate of convergence is illustrated in Figure \ref{convrate}, for $n_{\text{max}}=10^4$. 
\medskip

Finally, a last limitation may arise from our restriction to a replica-symmetric solution. The nuance of this argument is subtle\footnote{In particular, recall that in our construction, replica symmetry is explicitly broken, as replicas are coupled to different sources. Moreover, the origin of replica symmetry breaking (RSB) is typically linked to the $n \to \infty$ limit.}, especially since we do not take the $n \to 0$ limit. However, one might argue that such a solution is only stable at high temperatures or in regimes where quantum effects are significant. Note that the choice of temperature is constrained within our continuous model, where we assume $\beta$ to be sufficiently large. Nevertheless, we expect the relations between the couplings, rather than their absolute values, to be the primary drivers of the system's behavior. It would certainly be interesting to examine the high-temperature limit more carefully, where summations are replaced by integrals. In this context, it should be noted that the corresponding classical problem can be addressed using a 1-step replica symmetry breaking (1RSB) solution \cite{crisanti1992spherical, crisanti2006spherical}. See also \cite{cugliandolo2001imaginary} for a discussion of the $p$-spin spherical model.
\medskip

With these many red lights in mind, we can see these preliminary results as evidence for the following statement:

\begin{claim}
There exists in the ferromagnetic phase ($\kappa >0$) a region of phase space where there is a first-order phase transition to a solution with replica symmetry $q^\prime \neq 0$ in the vanishing source limit.
\end{claim}

\begin{figure}
\begin{center}
\includegraphics[scale=0.55]{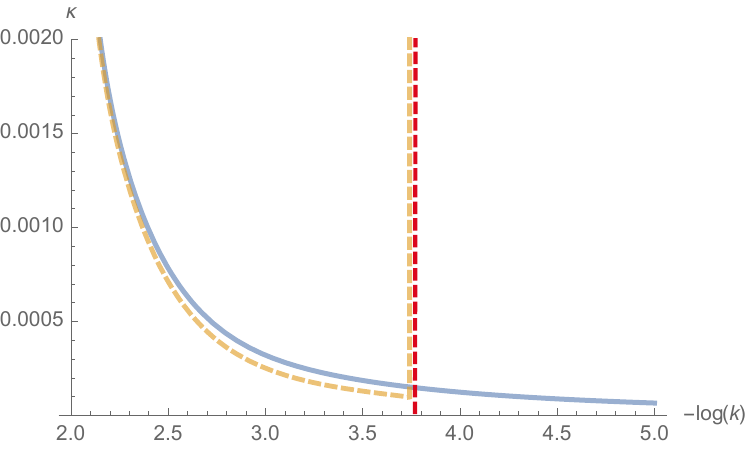}\qquad \includegraphics[scale=0.55]{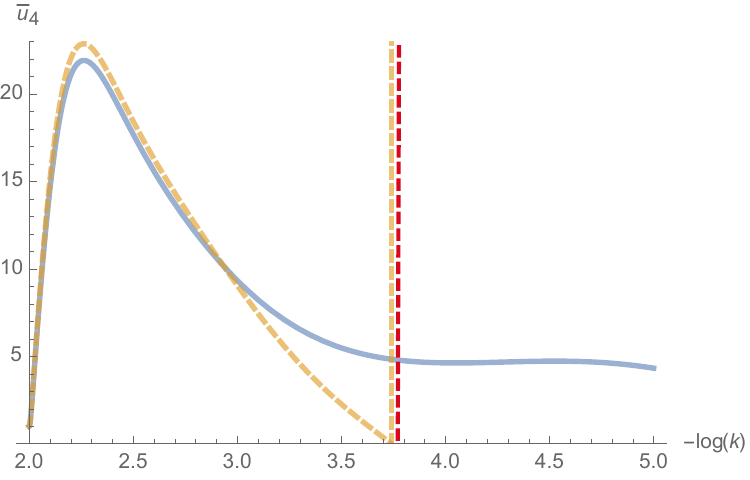}
\end{center}
\caption{Behavior of the RG flow upper and below the critical value $\bar{v}_{4,1,c}(k_0)$.}\label{ninfty}
\end{figure}

Before concluding this section, let us briefly discuss the naive $n \to \infty$ limit considered above. While this study is mathematically motivated—notably by the resulting simplification of the equations—its physical justification is delicate and perhaps warrants further investigation. It should also be noted that other scaling limits could be explored, potentially leading to distinct $n \to \infty$ behaviors. Our current analysis assumes that the couplings (and consequently the fixed-point solutions) do not scale as a power of $n$. This case may not be the most instructive, as the initial values of the couplings appear to be of order $n$, leading to equations where $q^\prime$ decouples from the remainder of the flow; nevertheless, it may have played a significant role upstream. We can therefore interpret this step as an effective regime that incorporates quartic couplings between replicas. From this, we can essentially make two observations:
\medskip

\begin{enumerate}

\item There are no global fixed points (because the coupling $w_{6,1}$ does not renormalizes). Because $\bar{w}_{6,1}\to 0$ asymptotically, it may however exist asymptotic fixed point, and indeed we get:
\begin{equation}
\text{FP1}_{\infty}=\Big(\bar{u}_4= \bar{v}_{4,1}= \bar{u}_6 = 0,\bar{\kappa} = \frac{1}{12 \pi }\Big)\,,
\end{equation}
with critical exponents:
\begin{equation}
\Theta= \Big(\theta_1=0,\theta_2=0,\theta_3=-2\Big)\,.
\end{equation}
The fixed point exhibits two marginal directions and one irrelevant direction. Unfortunately, the reliability of this result is currently limited; we intend to revisit this fixed-point solution in future work.

\item Investigating the RG trajectories, we recover essentially the same conclusions as before in certain regions of the phase space, as summarized in Figure \ref{ninfty}. We consider initial conditions $(\bar{\kappa}(k_0)=1, \bar{u}_4(k_0)=1, \bar{u}_6(k_0)=1, \bar{w}_{6,1}(k_0)=-5)$, with $\bar{v}_{4,1}(k_0)=7$ for the yellow curve and $\bar{v}_{4,1}(k_0)=0$ for the dashed blue curve. For this trajectory, the critical value is $\bar{v}_{4,1,c}(k_0) \approx 7$, above which the divergence disappears\footnote{The construction of this limit suggests a 4PI formalism, in which the four-point function serves as the order parameter.}. As previously observed, however, divergences reappear for larger magnitudes ($\bar{v}_{4,1}(k_0) < 37$ for the chosen initial conditions), which we interpret as a spurious effect of our incomplete truncation. Finally, the limitations noted previously regarding this method remain applicable.
\end{enumerate}
\medskip

\section{Enhanced vertex expansion -- a first look}\label{sec6}

In this section, we analyze the RG flow enhanced by non-local and multi-replica interactions in the symmetric phase, employing a vertex expansion. We revisit the case $p=3$ (where disorder is represented by a rank-3 tensor) and focus on the lowest-order sextic truncation. This investigation complements our previous analysis by focusing on the symmetric phase with respect to the mean field value.
\begin{equation}
M_{i\alpha}=0\,,
\end{equation} 
but it includes interactions that are not generated by standard perturbation theory. As previously noted, this approximation is better suited to non-local interactions, whose flow effectively vanishes under the approximation considered earlier. Furthermore, the vertex expansion allows for a more straightforward treatment of the exact regime, whereas the previous approximation was limited to the deep IR and critical regimes to maintain computational tractability. This section serves as an invitation to future work, which will evaluate the impact of replica interactions more systematically and examine the UV behavior of the flow.

\subsection{Truncation including multi-replica}

Explicitly, the truncation we consider in this section is the following:

\begin{align}
\Gamma_{k,\text{int}}&=u_4\, \vcenter{\hbox{\includegraphics[scale=0.8]{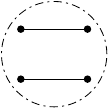}}}\,+\, u_6\, \vcenter{\hbox{\includegraphics[scale=0.8]{V6}}}\,+\,v_{4,1}\,\vcenter{\hbox{\includegraphics[scale=0.8]{Vk41Loc}}}\,+\,w_{6,1}\, \vcenter{\hbox{\includegraphics[scale=0.8]{V33}}}\,.
\end{align}

and for the kinetic part of the effective action:
\begin{align}
\nonumber \Gamma_{k,\text{kin}}&= \frac{1}{2} \int \dd t \sum_{\mu=1}^N \sum_{\alpha=1}^n M_{\mu \alpha}(t)\left(-\frac{\dd^2}{\dd t^2} + p_\mu^2+u_2\right)M_{\mu \alpha}(t) \\
&+\frac{1}{2}\, \int \dd t \, \sum_{\mu=1}^N\sum_{\alpha,\beta} q^\prime(k) M_{\mu \alpha}(t) M_{\mu \beta}(t)\,.
\end{align}
The propagator $G_k$ reads, in Fourier components:
\begin{align}
\nonumber G_k(\omega^2,p^2)&:=\underbrace{\frac{\delta_{\alpha\beta}}{\omega^2+p^2+u_2+R_k(p^2)}}_{\vcenter{\hbox{\includegraphics[scale=1]{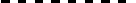}}}}\\
& -\underbrace{\frac{q^\prime}{(\omega^2+p^2+u_2+R_k(p^2))(\omega^2+p^2+u_2+R_k(p^2)+n q^\prime)}}_{\vcenter{\hbox{\includegraphics[scale=1]{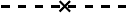}}}}\,.
\end{align}
Graphically, the flow equations write:
\begin{align}
\dot{u}_2\,=\,\vcenter{\hbox{\includegraphics[scale=0.7]{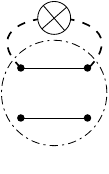}}}\,+\, \vcenter{\hbox{\includegraphics[scale=0.7]{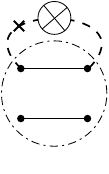}}}\,+\, \vcenter{\hbox{\includegraphics[scale=0.7]{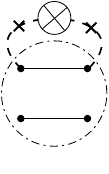}}}\,+\, \vcenter{\hbox{\includegraphics[scale=0.7]{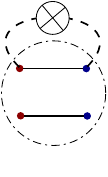}}}\,,
\end{align}

\begin{equation}
\dot{q}^\prime\,=\, \vcenter{\hbox{\includegraphics[scale=0.7]{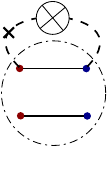}}}\,+\,\vcenter{\hbox{\includegraphics[scale=0.7]{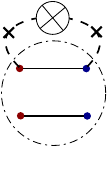}}}
\end{equation}

\begin{align}
\nonumber\dot{u}_4&\,=\,\vcenter{\hbox{\includegraphics[scale=0.7]{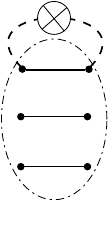}}}\,+\,\vcenter{\hbox{\includegraphics[scale=0.7]{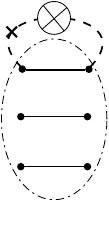}}}\,+\, \vcenter{\hbox{\includegraphics[scale=0.7]{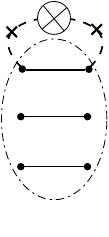}}}\,+\,\vcenter{\hbox{\includegraphics[scale=0.7]{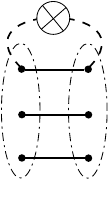}}}\,+\,\vcenter{\hbox{\includegraphics[scale=0.7]{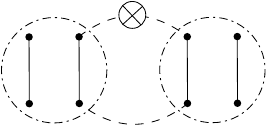}}}\,+\,\vcenter{\hbox{\includegraphics[scale=0.7]{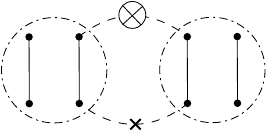}}}\\
&\,+\,\vcenter{\hbox{\includegraphics[scale=0.7]{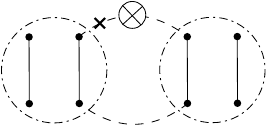}}}\,+\,\vcenter{\hbox{\includegraphics[scale=0.7]{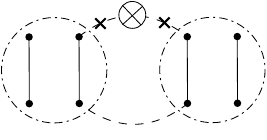}}}\,+\,\vcenter{\hbox{\includegraphics[scale=0.7]{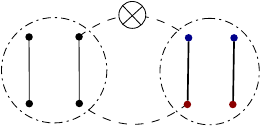}}}\,,
\end{align}

\begin{align}
\dot{v}_{4,1}&\,=\,\vcenter{\hbox{\includegraphics[scale=0.7]{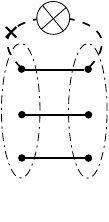}}}\,+\,\vcenter{\hbox{\includegraphics[scale=0.7]{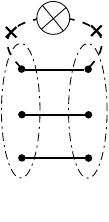}}}\,+\,\vcenter{\hbox{\includegraphics[scale=0.7]{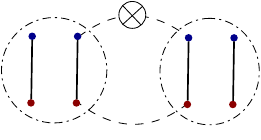}}}\,,
\end{align}


\begin{align}
\dot{w}_{6,1}&\,=\,0\,,
\end{align}

\begin{align}
\nonumber \dot{u}_{6}&\,=\,\vcenter{\hbox{\includegraphics[scale=0.7]{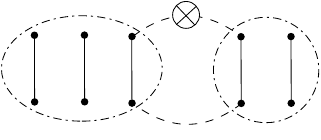}}}\,+\,\vcenter{\hbox{\includegraphics[scale=0.7]{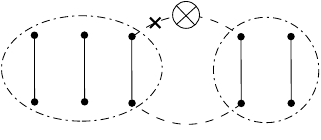}}}\,+\,\vcenter{\hbox{\includegraphics[scale=0.7]{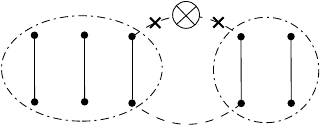}}}\\\nonumber
&\,+\,\vcenter{\hbox{\includegraphics[scale=0.7]{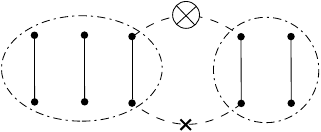}}}\,+\,\vcenter{\hbox{\includegraphics[scale=0.7]{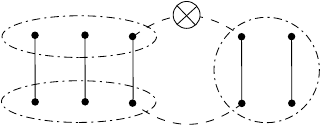}}}\,+\,\vcenter{\hbox{\includegraphics[scale=0.7]{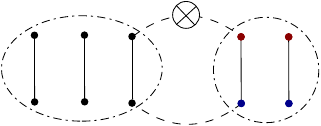}}}\\\nonumber
&\,+\,\vcenter{\hbox{\includegraphics[scale=0.6]{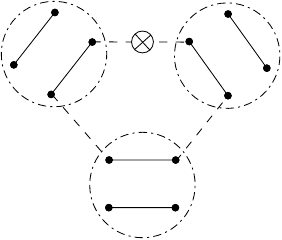}}}\,+\,\vcenter{\hbox{\includegraphics[scale=0.6]{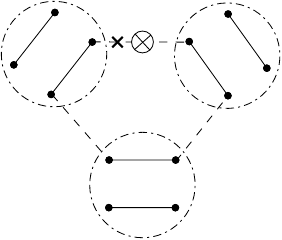}}}\,+\,\vcenter{\hbox{\includegraphics[scale=0.6]{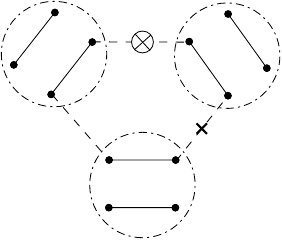}}}\\
&\,+\,\vcenter{\hbox{\includegraphics[scale=0.6]{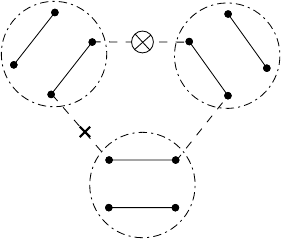}}}\,+\,\vcenter{\hbox{\includegraphics[scale=0.6]{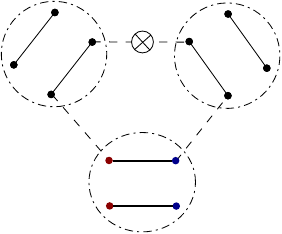}}}\,.
\end{align}

\subsection{Deep IR numerical analysis}

As in the rest of this paper, we focus on the IR regime, in which in particular the regulator is well approximated by:
\begin{equation}
R_k(p^2) \approx (k^2-p^2)\theta(k^2-p^2)\,.
\end{equation}
Our goal here is to demonstrate, once again, that accounting for the relevant observables that correlate the replicas "cures" at least part of the divergences observed in the IR—a detailed study of the UV flow will be deferred to a future work. In the momentum window $0 \leq p^2 \leq k^2$, the effective propagator reduces to:
\begin{align}
\nonumber G_k(\omega^2,p^2)&:=\underbrace{\frac{\delta_{\alpha\beta}}{\omega^2+k^2+u_2}}_{:=G_{k}^{(0)}}-\underbrace{\frac{q^\prime}{(\omega^2+k^2+u_2)(\omega^2+k^2+u_2+n q^\prime)}}_{:=G_k^{(1)}}\,
\end{align}
and the integral becomes singular at $\bar{u}_2 = -1$, where $\bar{u}_2 := u_2 k^{-2}$. Owing to the Litim regulator, the global integration along the effective loop (in the large-$N$ limit) factorizes in front of each term:
\begin{equation}
\Omega(k):= \int \dd p^2 \rho(p^2) \dot{R}_k(p^2) \approx \frac{4 k^5}{3 \pi \sigma^{3/2}}\,,
\end{equation}
assuming $k^2 \ll 4\sigma$. Many of the flow equations involve a trace of the form $\Tr G_k^n$. Moreover, it is straightforward to verify that the $n \times n$ matrix $G_k$ has the eigenvalue $G_k^{(0)} + nG_k^{(1)}$ with multiplicity $1$, and the eigenvalue $G_k^{(0)}$ with multiplicity $n-1$. Thus:
\begin{equation}
\Tr \, (G_k)^n= n (G_{k}^{(0)})^n+\sum_{p=0}^{n-1} (-1)^{n-p}\,C_n^p\, (G_{k}^{(0)})^p\, n^{n-p} \, (G_{k}^{(1)})^{n-p}\,,
\end{equation}
and we need the integral:
\begin{align}
I_{n,p}:=  (-1)^{n-p}\int \frac{\dd \omega}{2 \pi}\,  (G_{k}^{(0)})^p (G_{k}^{(1)})^{n-p} \,,
\end{align}
which can be computed exactly in terms of regularized hypergeometric functions. Denoting $a := k^2 + u_2$ and $b := k^2 + u_2 + nq^\prime$, we have, assuming both $a$ and $b$ are positive:
\begin{align}
\nonumber I_{n,p}(a,b)&:=-\pi  a^{-2 n} b^{n-p} \bigg(\frac{\sqrt{\pi } a^n (a+b n)^{-n+p+\frac{1}{2}} \, _2\tilde{F}_1\left(\frac{1}{2},n;-n+p+\frac{3}{2};\frac{b n}{a}+1\right)}{\Gamma (n-p)}\\
&-\frac{a^{p+\frac{1}{2}} \Gamma \left(2 n-p-\frac{1}{2}\right) \, _2\tilde{F}_1\left(n-p,2 n-p-\frac{1}{2};n-p+\frac{1}{2};\frac{b n}{a}+1\right)}{\Gamma (n)}\bigg)\,.
\end{align}
We recall that regularized hypergeometric functions $_p\tilde{F}_q$, 
\begin{equation}
_p\tilde{F}_q(a_1,\cdots,a_p;b_1,\cdots, b_q; z):= \sum_{n=0}^\infty \, \frac{\prod_{j=1}^p (a_j)_n z^n}{n!\prod_{j=1}^q \Gamma(n+b_j)}\,,
\end{equation}
where $(a_j)_k$ are the standard \textit{Pochhammer symbols}:
\begin{equation}
(a_j)_0:=1\,,\qquad (a_j)_n=a_j(a_j+1)\cdots (a_j+n-1)\,,\qquad n>0\,.
\end{equation}
Because we assume that $q^\prime$ and $u_2$ have the same dimensions, i.e., $\bar{u}_2 := k^{-2} u_2$ and $\bar{q}^\prime := k^{-2} q^\prime$, we can define the dimensionless function $\bar{I}_{n,p}(\bar{a}, \bar{b})$ in terms of the dimensionless parameters $\bar{a} := 1 + \bar{u}_2$ and $\bar{b} := 1 + \bar{u}_2 + n\bar{q}^\prime$, such that:
\begin{equation}
\bar{I}_{n,p}(\bar{a},\bar{b}):=  k^{2n-1}I_{n,p}({a},{b})\,.
\end{equation}
Computing each diagrams, using the definition \eqref{localsectorGamma2n} for couplings, we get (see also \cite{lahoche2024largetimeeffectivekinetics}):
As in the rest of this paper, we focus on the IR regime, in which in particular the regulator is well approximated by:
\begin{equation}
R_k(p^2) \approx (k^2-p^2)\theta(k^2-p^2)\,.
\end{equation}
Our goal here is to demonstrate, once again, that accounting for the relevant observables that correlate the replicas "cures" at least part of the divergences observed in the IR; a detailed study of the UV flow will be deferred to future work. In the momentum window $0 \leq p^2 \leq k^2$, the effective propagator reduces to:
\begin{align}
\nonumber G_k(\omega^2,p^2)&:=\underbrace{\frac{\delta_{\alpha\beta}}{\omega^2+k^2+u_2}}_{:=G_{k}^{(0)}}-\underbrace{\frac{q^\prime}{(\omega^2+k^2+u_2)(\omega^2+k^2+u_2+n q^\prime)}}_{:=G_k^{(1)}}\,
\end{align}
and the integral becomes singular at $\bar{u}_2 = -1$, where $\bar{u}_2 := u_2 k^{-2}$. Because we use the Litim regulator, the global integration along the effective loop (in the large-$N$ limit) factorizes in front of each term:
\begin{equation}
\Omega(k):= \int \dd p^2 \rho(p^2) \dot{R}_k(p^2) \approx \frac{4 k^5}{3 \pi \sigma^{3/2}}\,,
\end{equation}
assuming $k^2 \ll 4\sigma$. Many of the flow equations involve a trace of the form $\Tr G_k^n$. Moreover, it is straightforward to verify that the $n \times n$ matrix $G_k$ has the eigenvalues $G_k^{(0)} + nG_k^{(1)}$ with multiplicity $1$ and $G_k^{(0)}$ with multiplicity $n-1$. Thus:"
\begin{equation}
\Tr \, (G_k)^n= n (G_{k}^{(0)})^n+\sum_{p=0}^{n-1} (-1)^{n-p}\,C_n^p\, (G_{k}^{(0)})^p\, n^{n-p} \, (G_{k}^{(1)})^{n-p}\,,
\end{equation}
and we need to the integral:
\begin{align}
I_{n,p}:=  (-1)^{n-p}\int \frac{\dd \omega}{2 \pi}\,  (G_{k}^{(0)})^p (G_{k}^{(1)})^{n-p} \,,
\end{align}
which can be computed exactly in terms of regularized hypergeometric functions. Denoting as $a:=k^2+u_2$ and $b:=q^\prime$, we have, assuming $a$ and $b$ are both positives:
\begin{align}
\nonumber I_{n,p}(a,b)&:=-\pi  a^{-2 n} b^{n-p} \bigg(\frac{\sqrt{\pi } a^n (a+b n)^{-n+p+\frac{1}{2}} \, _2\tilde{F}_1\left(\frac{1}{2},n;-n+p+\frac{3}{2};\frac{b n}{a}+1\right)}{\Gamma (n-p)}\\
&-\frac{a^{p+\frac{1}{2}} \Gamma \left(2 n-p-\frac{1}{2}\right) \, _2\tilde{F}_1\left(n-p,2 n-p-\frac{1}{2};n-p+\frac{1}{2};\frac{b n}{a}+1\right)}{\Gamma (n)}\bigg)\,.
\end{align}
We recall that regularized hypergeometric functions $_p\tilde{F}_q$, 
\begin{equation}
_p\tilde{F}_q(a_1,\cdots,a_p;b_1,\cdots, b_q; z):= \sum_{n=0}^\infty \, \frac{\prod_{j=1}^p (a_j)_n z^n}{n!\prod_{j=1}^q \Gamma(n+b_j)}\,,
\end{equation}
where $(a_j)_k$ are the standard Pochhammer symbols:
\begin{equation}
(a_j)_0:=1\,,\qquad (a_j)_n=a_j(a_j+1)\cdots (a_j+n-1)\,,\qquad n>0\,.
\end{equation}
Because we assume that $q^prime$ and $u_2$ have the same dimensions i.e. $\bar{u}_2:=k^{-2} u_2$ and $\bar{q}^\prime:=k^{-2} q^\prime$, we can define the dimensionless function $\bar{I}_{n,p}(\bar{a},\bar{b})$, expressed in terms of the dimensionless $\bar{a}:=1+\bar{u}_2$ and $\bar{b}:=\bar{q}^\prime$, such that
\begin{equation}
\bar{I}_{n,p}(\bar{a},\bar{b}):=  k^{2n-1}I_{n,p}({a},{b})\,.
\end{equation}
Computing each diagrams, using the definition \eqref{localsectorGamma2n} for couplings, we get (see also \cite{lahoche2024largetimeeffectivekinetics}):
\begin{equation}
\dot{\bar{u}}_2=-2\bar{u}_2-\frac{\bar{u}_4}{18\pi} (\bar{I}_{2,2}+2\bar{I}_{2,1}+n\bar{I}_{2,0})-\frac{\bar{v}_{4,1}}{36\pi}\frac{1}{(1+\bar{u}_2)^{\frac{3}{2}}}\,,
\end{equation}
\begin{equation}
\dot{\bar{q}}^\prime=-2 \bar{q}^\prime - \frac{\bar{v}_{4,1}}{18\pi} (2\bar{I}_{2,1}+n\bar{I}_{2,0})
\end{equation}
\begin{align}
\nonumber \dot{\bar{u}}_4=&- \frac{\bar{\tilde{u}}_6 }{30\pi^2}\frac{1}{(1+\bar{u}_2)^2}-\frac{\bar{u}_6 }{30\pi}(\bar{I}_{2,2}+2\bar{I}_{2,1}+n\bar{I}_{2,0})+ \frac{\bar{v}_{4,1} \bar{u}_4}{6\pi} \frac{1}{(1+\bar{u}_2)^{\frac{5}{2}}} \\
&+\frac{2\bar{u}_4^2}{9\pi} (\bar{I}_{3,3}+3\bar{I}_{3,2}+n\bar{I}_{3,1})\,.
\end{align}
\begin{align}
\dot{\bar{v}}_{4,1}&=\frac{\bar{\tilde{u}}_6 }{30\pi^2(1+\bar{u}_2)^2}\left[\frac{2\bar{q}^\prime}{(1+\bar{u}_2+n\bar{q}^\prime)}-\frac{n(\bar{q}^\prime)^2}{(1+\bar{u}_2+n \bar{q}^\prime)^2}\right]+\frac{\bar{v}_{4,1}^2}{12\pi} \frac{1}{(1+\bar{u}_2)^{\frac{5}{2}}}
\end{align}
\begin{align}
\nonumber\dot{\bar{{u}}}_6&=2\, {\bar{{u}}}_6+\frac{192 \bar{u}_4\bar{u}_6}{5\pi} (\bar{I}_{3,3}+3\bar{I}_{3,2}+n\bar{I}_{3,1})+\frac{4}{5\pi^2} \frac{\bar{u}_4\bar{\tilde{u}}_6}{(1+\bar{u}_2)^{3}}+\frac{72}{5\pi} \frac{\bar{v}_{4,1}\bar{u}_6}{(1+\bar{u}_2)^{\frac{5}{2}}}\\
&-\frac{8 \bar{u}_4^3}{9\pi} (\bar{I}_{4,4}+4\bar{I}_{4,3})-\frac{15}{18 \pi} \frac{\bar{u}^2_4\bar{v}_{4,1}}{(1+\bar{u}_2)^{\frac{7}{2}}}\,,
\end{align}
and:
\begin{equation}
\dot{\bar{\tilde{u}}}_6=\, {\bar{\tilde{u}}}_6\,.
\end{equation}

\begin{figure}
\begin{center}
\includegraphics[scale=0.55]{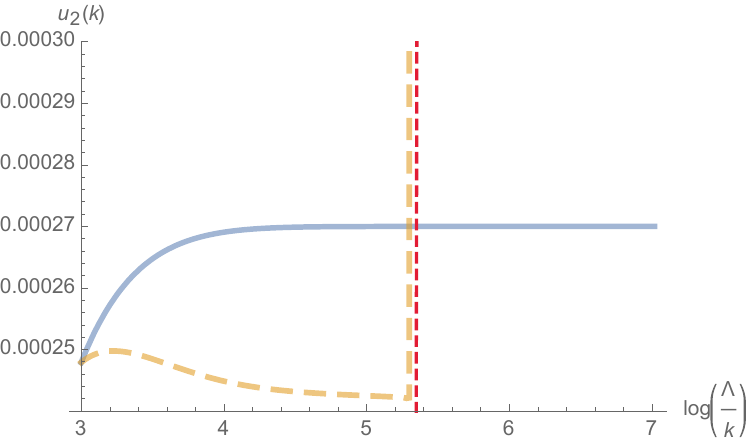}\qquad \includegraphics[scale=0.55]{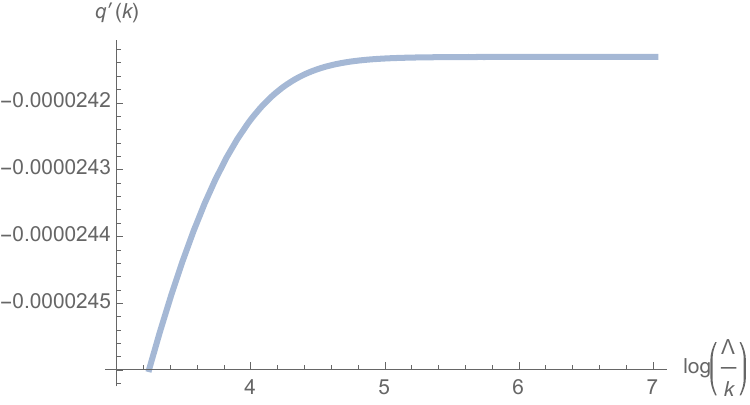}
\end{center}
\caption{Left: Renormalization group (RG) trajectory for the dimensional mass $u_2(k)$ with initial conditions $S_1(k_0)$ at $k_0 = 0.05$. The dashed yellow curve corresponds to $\bar{q}'(k_0) = \bar{v}_{4,1}(k_0) = 0$ and $\bar{\tilde{u}}_6(k_0) = 3.68$, exhibiting a singularity at $k_c \approx 0.005$. The blue curve corresponds to $\bar{q}'(k_0) = -0.01$ and $\bar{v}_{4,1}(k_0) = 0.01$. Right: RG trajectory of $q'(k)$ along the blue path shown in the left panel. Calculations were performed for $n = 2$.}\label{figkey}
\end{figure}

Note that, as in Section \ref{EVE}, we denote the magnitude of the original sextic disorder interaction by $\tilde{u}_6$. We begin by verifying that we recover the conjectured results. We choose the trajectory with the following initial conditions:"
\begin{equation}
S_1(k_0):=(\bar{u}_2(k_0):=0.1, \bar{u}_4(k_0)=1,\bar{u}_6(k_0)=1)\,,
\end{equation}
This trajectory resides in the symmetric phase; for $k_0 = 0.05$, it exhibits a singularity at $k_c \approx 0.005$ as soon as the disorder exceeds the critical value $\bar{\tilde{u}}_6(k_0) \approx 3.68$. The results are summarized in Figure \ref{figkey}, and we recover essentially the same behavior as before: for a fixed $\bar{v}_{4,1}(k_0)$, the singularity is removed when $-\bar{q}^\prime(k_0)$ is sufficiently large (for the initial conditions chosen for the figure, we find $-\bar{q}^\prime(k_0) > 10^{-5}$), and $q^\prime(k_0)$ approaches a very small, nearly constant value. Remarkably, the values of $\bar{q}^\prime$ and $\bar{v}_{4,1}$ are much smaller than those observed in the symmetric phase, which we attribute to the competition between the "magnetization" and the inter-replica correlations. More precisely, we hypothesize that the inertia of the correlations associated with the vacuum of the local potential opposes the establishment of these new correlations.

\section{Ward identities and anomalous dimension}\label{sec2}

In the first part of this section, we discuss the constraints imposed on the RG flow by the Ward identities. In particular, due to the gauge-fixing condition \eqref{Gaugefix}, the $O(N)$ symmetry is explicitly broken by the kinetic term, rendering the Ward identities associated with the original $O(N)$ symmetry non-trivial. This part builds upon the discussions in \cite{lahoche20242}, which were themselves inspired by \cite{Lahochebeyond}. In the second part, we discuss an improvement to the standard vertex expansion considered in \cite{lahoche2024largetimeeffectivekinetics}, known as the effective vertex expansion. This approach exploits the large-$N$ relations between observables to close the hierarchy of flow equations—see \cite{Lahochebeyond,lahoche2023functional}.

\subsection{Ward identities for internal symmetries}

The Ward identities related to some classical transformations can be classified into two
categories:

\begin{itemize}
\item Type $1$ transformations, which left the path integral measure invariant but not the classical action.
\item Type $2$ transformations, which left both the path integral
measure and the classical action invariant.
\end{itemize}

Usually, these two kinds of transformation provide different kinds of Ward identities. For transformations of the second kind, the variation of the path integral reads:
\begin{align}
\nonumber \langle \delta \mathcal{J} \rangle &= \int_{-\infty}^\infty\,\dd t\,\sum_{k=1,\alpha=1}^{N,n} L_{k,\alpha}(t)\delta \langle x_{k,\alpha}(t) \rangle \\
&= \int_{-\infty}^\infty\,\dd t\,\sum_{k=1,\alpha=1}^{N,n} \frac{\delta \Gamma}{\delta \langle x_{k,\alpha}(t) \rangle } \delta\langle x_{k,\alpha}(t) \rangle= \delta \Gamma=0\,,\label{variationGamma}
\end{align}
but we are mainly interested in transformations arising from transformations of the first kind in this paper (see remark \ref{remarkWard} at the end of the subsection).
\medskip

The original model—prior to averaging and gauge fixing—is invariant under $O(N)$ transformations due to the formal translation invariance of the path integral measure and the $O(N)$ invariance of the probability densities for $J$ and $K$. As previously stated, the gauge-fixing condition explicitly breaks the $O(N)$ symmetry of the classical action. Note that the averaging over $J$ introduces an additional subtlety. The interaction part of the classical action, which we denote by $\overline{S_{\text{cl,int}}}[\{\bm{x}_\alpha\}]$ (i.e., the terms involving powers of the field higher than $2$), remains invariant under global rotations acting independently on each replica:
\begin{equation}
x_{\mu,\alpha}\to x_{\mu,\alpha}^\prime=\sum_{\nu=1}^N O_{\mu\nu} x_{\nu,\alpha}\,, \qquad \forall \,\alpha\,.\label{GlobalGauge}
\end{equation}
In addition one can consider rotations acting differently on the replica, what we call \textit{local rotations}, namely:
\begin{equation}
x_{\mu,\alpha}\to x_{\mu,\alpha}^\prime=\sum_{\nu=1}^N O_{\mu\nu}^{(\alpha)} x_{\nu,\alpha}\,.\label{LocalGauge}
\end{equation}
The interaction part of the classical action, $\overline{S_{\text{cl,int}}}[\{\bm{x}_\alpha\}]$, is invariant under global rotations, but not under local ones. However, the partition function remains invariant in both cases due to the formal translation invariance of the path integral measure \cite{grosse2009progress,lahoche2018unitary}. Let us formalize these observations.
\medskip

To begin, let us consider the global transformation \eqref{GlobalGauge}. We will compute the infinitesimal transformation of the partition function associated with an infinitesimal rotation:
\begin{equation}
O_{\mu \nu}= \delta_{\mu\nu}+ \varepsilon_{\mu\nu}+\mathcal{O}(\varepsilon^2)\,,
\end{equation}
where $\varepsilon_{\mu\nu}$ are the entries of a skew-symmetric matrix, $\varepsilon_{\mu\nu}=-\varepsilon_{\nu\mu}$. Denoting as $\overline{S_{\text{cl,int}}}[\{\bm{x}_\alpha\}]$ and $\overline{S_{\text{cl,kin}}}[\{\bm{x}_\alpha\}]$ respectively the interacting and the kinetic parts of the classical action $\overline{S_{\text{cl,int}}}[\{\bm{x}_\alpha\}]$, the variation (at order $\varepsilon$) of the partition function $\overline{\mathcal{Z}^n_{\beta,k}[K,J,\{\bm L_\alpha\}]}$ reads (let us recall that $\hbar=1$):
\begin{align}
\delta \overline{\mathcal{Z}^n_{\beta}[K,J,\{\bm L_\alpha\}]}=\int \, \prod_{\alpha=1}^n[\mathcal{D} x_\alpha(t)]\, \left[-\delta \overline{S_{\text{cl,kin}}}[\{\bm{x}_\alpha\}]+\delta \mathcal{J} \right] e^{-\left(\overline{S_{\text{cl}}}[\{\bm{x}\}]+\Delta S_k[\{\bm{x}_\alpha\}]\right)+\mathcal{J}}\equiv 0\,,
\end{align}
The last equality arises from the formal invariance of the path integral measure under global translations. Furthermore, $\delta \overline{S_{\text{cl,int}}}[\{\bm{x}_\alpha\}] = 0$ because the interactions are explicitly invariant. We must then compute each variation explicitly. For the source term $\mathcal{J}$, we obtain:
\begin{align}
\nonumber \delta \mathcal{J}&=\int_{-\beta/2}^{+\beta/2}\,\dd t\,\sum_{\mu,\nu,\alpha=1}^{N,n} L_{\mu,\alpha}(t)x_{\nu,\alpha}(t) \varepsilon_{\mu\nu}\\
&=\frac{1}{2}\int_{-\beta/2}^{+\beta/2}\,\dd t\,\sum_{\mu,\nu,\alpha=1}^{N,n} \left(L_{\mu,\alpha}(t)x_{\nu,\alpha}(t) - L_{\nu,\alpha}(t)x_{\mu,\alpha}(t) \right)\varepsilon_{\mu\nu}\,,
\end{align}
where, for the second line, we exploited the symmetries of the matrix $\varepsilon$. Now, let us compute the variation of the kinetic action. Since the $\dot{\bm x}^2$ contribution and the mass term are $O(N)$-invariant, the relevant contribution arises from the regulator $\Delta S_k$ and the effective kinetic term originating from matrix $K$. We obtain:
\begin{align}
\nonumber\delta \overline{S_{\text{cl,kin}}}[\{\bm{x}_\alpha\}]&=\int_{-\beta/2}^{+\beta/2}\,\dd t\, \sum_{\mu,\alpha}\, \left(p_\mu^2+R_k^{(N)}(p_\mu^2)\right) x_\nu(t) x_\mu(t) \varepsilon_{\mu \nu} \\
&=\frac{1}{2}\int_{-\beta/2}^{+\beta/2}\,\dd t\, \sum_{\mu,\nu,\alpha}\, \left(p_\mu^2+R_k^{(N)}(p_\mu^2)-p_\nu^2-R_k^{(N)}(p_\nu^2)\right) x_\nu(t) x_\mu(t) \varepsilon_{\mu \nu}\,,
\end{align}
and the total variation of the partition function reads finally:
\begin{align}
\nonumber \int_{-\beta/2}^{+\beta/2}&\,\dd t\,\sum_{\mu,\nu,\alpha} \varepsilon_{\mu\nu}\Bigg[ \left(p_\mu^2+R_k^{(N)}(p_\mu^2)-p_\nu^2-R_k^{(N)}(p_\nu^2)\right) \frac{\delta^2}{\delta L_{\mu,\alpha}(t) \delta L_{\nu,\alpha}(t)}-\\
&-L_{\mu,\alpha}(t) \frac{\partial}{\partial L_{\nu,\alpha}(t)}+L_{\nu,\alpha}(t) \frac{\partial}{\partial L_{\mu,\alpha}(t)}\Bigg] \exp \left(\mathcal{W}^{(n)}_{\beta,k}[K,J,\mathcal{L}]\right)=0\,.
\end{align}
This result should be true for any infinitesimal element of the Lie algebra $\varepsilon \in \mathfrak{so}(N)$, and because the variation in front has been suitably skew-symmetrized, we have the following statement:
\begin{theorem}\label{Wardth}
The observable of the gauged and averaged theory satisfies the identity:
\begin{align}
\nonumber \int_{-\beta/2}^{+\beta/2}&\,\dd t\,\sum_{\alpha} \Bigg[ \left(p_\mu^2+R_k^{(N)}(p_\mu^2)-p_\nu^2-R_k^{(N)}(p_\nu^2)\right) \frac{\delta^2}{\delta L_{\mu,\alpha}(t) \delta L_{\nu,\alpha}(t)}\\
&-L_{\mu,\alpha}(t) \frac{\partial}{\partial L_{\nu,\alpha}(t)}+L_{\nu,\alpha}(t) \frac{\partial}{\partial L_{\mu,\alpha}(t)}\Bigg] \exp \left(\mathcal{W}^{(n)}_{\beta,k}[K,J,\mathcal{L}]\right)=0\,.
\end{align}
\end{theorem}

Now, let us consider local rotations. The group of local rotations is $(O(N))^n$, i.e., $n$ copies of the rotation group; a local rotation is a set $\mathcal{R}_O^{(n)} := (O_1, O_2, \dots, O_n)$ of $n$ independent rotations. We then consider an infinitesimal rotation acting only on the component $\alpha$:
\begin{equation}
\mathcal{R}_O^{(n)}=(\mathrm{I},\cdots \mathrm{I}+\varepsilon_\alpha , \cdots \mathrm{I})+\mathcal{O}(\varepsilon^2)\,,
\end{equation}
where $\mathrm{I}$ denotes the $N \times N$ identity matrix. The deterministic potential remains invariant, but the non-local one does not. The variation of the interacting action, to first order in $\varepsilon_\alpha$, is given by:
\begin{equation}
\overline{S_{\text{cl,int}}}[\{\bm{x}_\alpha\}]=q \lambda \times  \int \dd t \dd t^\prime\sum_{\beta,\mu,\nu}(X_{\alpha\beta})^{q-1} ( x_{\mu^\prime,\alpha}(t)x_{\mu,\beta}(t^\prime)- x_{\mu,\alpha}(t)(t)x_{\mu^\prime,\beta}(t^\prime)) \varepsilon_{\alpha,\mu\nu}\,,
\end{equation}
where:
\begin{equation}
X_{\alpha\beta}(t,t^\prime):=\frac{1}{N}\,\sum_{\mu=1}^N \, x_{\mu,\alpha}(t)x_{\mu,\beta}(t^\prime)\,.
\end{equation}
This term then contributes to the global variation; furthermore, because we have skew-symmetrized the contracted tensor with $\varepsilon_{\alpha,\mu\nu}$, we arrive at the following result:

\begin{theorem}
Due to the breaking of local gauge invariance resulting from the averaging over disorder, the gauged and averaged theory further satisfies the identity:
\begin{align}
\nonumber &\int_{-\beta/2}^{+\beta/2}\,\dd t\, \Bigg[ \left(p_\mu^2+R_k^{(N)}(p_\mu^2)-p_\nu^2-R_k^{(N)}(p_\nu^2)\right) \frac{\delta^2}{\delta L_{\mu,\alpha}(t) \delta L_{\nu,\alpha}(t)}\\\nonumber 
&-L_{\mu,\alpha}(t) \frac{\partial}{\partial L_{\nu,\alpha}(t)}+L_{\nu,\alpha}(t) \frac{\partial}{\partial L_{\mu,\alpha}(t)}\Bigg] \exp \left(\mathcal{W}^{(n)}_{\beta,k}[K,J,\mathcal{L}]\right)\\
&-q \lambda \times  \int \dd t \dd t^\prime\sum_{\gamma} \big\langle (X_{\alpha\gamma})^{q-1} ( x_{\mu^\prime,\alpha}(t)x_{\mu,\gamma}(t^\prime)- x_{\mu,\alpha}(t)(t)x_{\mu^\prime,\gamma}(t^\prime))\Big \rangle=0\,.
\end{align}
\end{theorem}

\begin{remark}
In addition to internal global and local $O(N)$ symmetries, there are external symmetries arising from the extra structure provided by the space-time manifold. In our case, since the space-time dimension is 1, these transformations reduce to translations, dilations, or general time reparametrizations. Note that, classically, the choice of the time parameter is closely linked to the existence of a preferred equilibrium thermodynamic state, and within non-relativistic quantum mechanics, it must be viewed as a property of the von Neumann algebra \cite{connes1994neumann, rovelli1993statistical}. The theory can also be formulated in a purely covariant manner (i.e., by including time in the set of coordinates). Parametrization invariance then imposes a constraint on the Hilbert space, and solving this constraint is equivalent to imposing the Schrödinger equation; this essentially leads to translation or dilation transformations:
\begin{equation}
t\to t^\prime= t+t_0\,, \qquad t\to t^\prime= K_0\times t\,.
\end{equation}
To these transformations we have to add also Galilean transformations, mixing both time and the coordinates field $\bm{x}$:
\begin{equation}
x_\mu \to x_\mu^\prime=x_\mu- (\bm V \cdot \bm e_\mu) t\,,
\end{equation}
where $\bm e_\mu$ is the unit vector along direction $\mu$ and $\bm V$ the velocity of the relative frame. 
\end{remark}\label{remarkWard}

\subsection{Time reversal symmetric relations between observables}

\begin{figure}
\begin{center}
\includegraphics[scale=0.8]{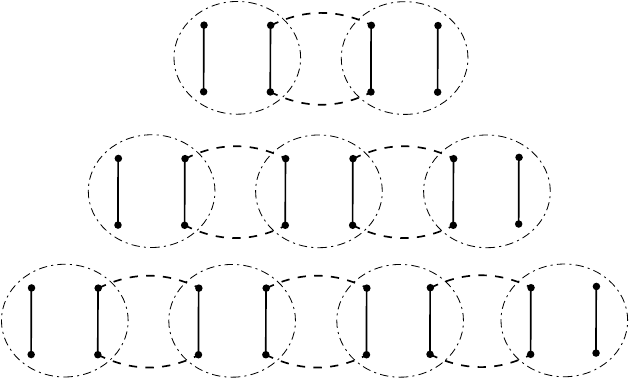}
\end{center}
\caption{Leading order contribution to the quartic effective vertex.}\label{figLO4pts}
\end{figure}

The global Ward identity given by theorem \ref{Wardth} can be rewritten as, using Fourier variables:
\begin{align}
\nonumber \sum_{\omega}\,\sum_{\alpha} \Bigg[ \delta E_k(p_\mu^2,p_\nu^2)& (G_{k}(\omega,p_\mu,p_\nu)+M_{\mu,\alpha}(\omega)M_{\nu,\alpha}(-\omega))-\\
&-L_{\mu,\alpha}(\omega) M_{\nu,\alpha}(-\omega)+L_{\nu,\alpha}(\omega) M_{\mu,\alpha}(-\omega)\Bigg]=0\,,
\end{align}
where:
\begin{equation}
\delta E_k(p_\mu^2,p_\nu^2):=p_\mu^2+R_k^{(N)}(p_\mu^2)-p_\nu^2-R_k^{(N)}(p_\nu^2)\,,
\end{equation}
and:
\begin{equation}
M_{\mu,\alpha}(\omega):=\frac{\delta \mathcal{W}^{(n)}_{\beta,k}}{\delta L_{\mu,\alpha}(-\omega)}\,,\quad G_{k}(\omega,p_\mu,p_\nu)\delta_{\omega,-\omega^\prime}:= \frac{\delta^2 \mathcal{W}^{(n)}_{\beta,k}}{\delta L_{\mu,\alpha}(-\omega^\prime)\delta L_{\nu,\alpha}(\omega)}\,.
\end{equation}
Note that we have omitted the index $\alpha$ used in the definition \eqref{defG} to simplify the notation. Using this definition and taking derivatives with respect to $M_{\mu_1, \alpha}(\omega_1)$ and $M_{\mu_2, \alpha}(\omega_2)$, we obtain\footnote{We assume that $G_{k}(\omega, p_\mu, p_\nu)$ is an even function with respect to its variables and depends on $\omega^2$, $p_\mu^2$, and $p_\nu^2$. Similarly, we assume that $\Gamma_k^{(2)}(p^2, \omega)$ depends only on $\omega^2$ (time-reversal symmetry).}:
\begin{align}
\nonumber &\sum_{\omega}\,\sum_{p_\sigma,p_{\sigma^\prime}} \Bigg[ \delta E_k(p_\mu^2,p_\nu^2) \bigg(-G_k(\omega,p_\mu,p_\sigma) \Gamma_{k,p_\sigma,p_{\sigma^\prime} p_{\mu_1},p_{\mu_2}}^{(4)}(\omega,-\omega,\omega_1,\omega_2)G_k(\omega,p_{\sigma^\prime},p_\sigma)\\\nonumber
&+(\delta_{\mu \mu_1}\delta_{\nu\mu_2}\delta_{\omega,\omega_1}\delta_{-\omega,\omega_2}+\delta_{\mu \mu_2}\delta_{\nu\mu_1}\delta_{-\omega,\omega_1}\delta_{\omega,\omega_2})\bigg)-\frac{\delta L_{\mu,\alpha}(\omega)}{\delta M_{\mu_1,\alpha}(\omega_1)} \delta_{\nu\mu_2}\delta_{\omega_2,-\omega}\\
&-\frac{\delta L_{\mu,\alpha}(\omega)}{\delta M_{\mu_2,\alpha}(\omega_2)} \delta_{\nu\mu_1}\delta_{\omega_1,-\omega}+\frac{\delta L_{\nu,\alpha}(-\omega)}{\delta M_{\mu_1,\alpha}(\omega_1)} \delta_{\mu\mu_2}\delta_{\omega_2,\omega}+\frac{\delta L_{\nu,\alpha}(-\omega)}{\delta M_{\mu_2,\alpha}(\omega_2)} \delta_{\mu\mu_1}\delta_{\omega_1,\omega}\Bigg]=0\,.\label{secondderivWard}
\end{align}
The functional derivatives of the source field (evaluated at the fixed classical field) can be computed from the properties of the Legendre transform \cite{zia2009making}. For instance, imposing that the external field vanishes on the right-hand side, we obtain:
\begin{equation}
\frac{\delta L_{\mu,\alpha}(\omega)}{\delta M_{\mu_2,\alpha}(\omega_2)}=\delta_{\mu,\mu_2}\delta_{\omega,\omega_2}\left(\Gamma_k^{(2)}(p_\mu^2,\omega)+R_k^{(N)}(p_\mu^2)\right)\,,
\end{equation}
and imposing that source field vanishes on the left-hand side of equation \eqref{secondderivWard}, we get:

\begin{align}
\nonumber &\sum_{\omega}\,\sum_{p_\sigma,p_{\sigma^\prime}} \bigg[ \delta E_k(p_\mu^2,p_\nu^2) \bigg(-G_k(\omega,p_\mu,p_\sigma) \Gamma_{k,p_\sigma,p_{\sigma^\prime} p_{\mu_1},p_{\mu_2}}^{(4)}(\omega_1,\omega_2,\omega,-\omega,)G_k(\omega,p_{\sigma^\prime},p_\sigma)\\\nonumber
&+(\delta_{\mu \mu_1}\delta_{\nu\mu_2}\delta_{\omega,\omega_1}\delta_{-\omega,\omega_2}+\delta_{\mu \mu_2}\delta_{\nu\mu_1}\delta_{-\omega,\omega_1}\delta_{\omega,\omega_2})\bigg)\\\nonumber
&+\delta_{\mu \mu_1}\delta_{\nu\mu_2}\delta_{\omega,\omega_1}\delta_{-\omega,\omega_2} \left(\Gamma_k^{(2)}(p_\nu^2,\omega)+R_k^{(N)}(p_\nu^2)-\Gamma_k^{(2)}(p_\mu^2,\omega)-R_k^{(N)}(p_\mu^2)\right)\\
&+\delta_{\mu \mu_2}\delta_{\nu\mu_1}\delta_{-\omega,\omega_1}\delta_{\omega,\omega_2} \left(\Gamma_k^{(2)}(p_\nu^2,\omega)+R_k^{(N)}(p_\nu^2)-\Gamma_k^{(2)}(p_\mu^2,\omega)-R_k^{(N)}(p_\mu^2)\right)\bigg]\,.\label{secondderivWardstep2}
\end{align}
In the symmetric phase, the propagator is diagonal with respect to generalized momenta, 
\begin{equation}
G_{k}(\omega,p_\mu,p_\nu)=: \tilde{G}(p_\mu^2,\omega^2) \delta_{\mu\nu}\,;
\end{equation}
then, sending $p_\mu^2-p_\nu^2=\delta p^2 \to 0$ we get, assuming $\omega_1=-\omega_2$:
\begin{align}
\nonumber &(\delta_{\mu \mu_1}\delta_{\nu\mu_2}+\delta_{\mu \mu_2}\delta_{\nu\mu_1}) \bigg[\bigg(1+\frac{R_k^{(N)}(p_\mu^2)-R_k^{(N)}(p_\nu^2)}{\delta p^2}\bigg)\, \tilde{G}_k(p_\mu^2,\omega^2_1)\\
&\times A_{k,p_\mu,p_{\nu} p_{\mu_1},p_{\mu_2}}(\omega_1,\omega_2,\omega) \tilde{G}_k(p_\nu^2,\omega^2_2) - \frac{\Sigma_k(p_\mu^2,\omega^2_1)-\Sigma_k(p_\nu^2,\omega^2_1)}{\delta p^2}=0\bigg]\,,\label{secondderivWardstep2}
\end{align}
where we used the standard definition for self-energy:
\begin{equation}
\Gamma_k^{(2)}(p_\mu^2,\omega^2)=\omega^2+p_\mu^2+\mu_1-\Sigma(p_\mu^2,\omega^2)\,.
\end{equation}
Furthermore, we assumed the frequency conservation for $\Gamma_k^{(4)}$:
\begin{align}
\nonumber \Gamma_{k,p_{\mu_1},p_{\mu_2,} p_{\mu_3},p_{\mu_4}}^{(4)}&(\omega_1,\omega_2,\omega_3,\omega_4):=\delta(\omega_1+\omega_2+\omega_3+\omega_4)\,A_{k,p_{\mu_1},p_{\mu_2,} p_{\mu_3},p_{\mu_4}}(\omega_1,\omega_2,\omega_3)\,.
\end{align}
In the local potential approximation, 
\begin{equation}
A_{k,p_{\mu_1},p_{\mu_2,} p_{\mu_3},p_{\mu_4}}(\omega_1,\omega_2,\omega_3)=:\frac{8u_4(k)}{N} \left(\delta_{\mu_1\mu_2}\delta_{\mu_3\mu_4}+\delta_{\mu_1\mu_3}\delta_{\mu_2\mu_4}+\delta_{\mu_1\mu_4}\delta_{\mu_2\mu_3}\right)\,.\label{definitionu4}
\end{equation}
This approximation may seem questionable, as it neglects the dependence of the vertices on external frequencies. However, because the theory is just renormalizable, one can expect these contributions to be irrelevant; thus, the approximation is justified in the vicinity of the Gaussian fixed point and asymptotically in the IR \cite{bagnuls2001exact}. Note moreover that in the large-$N$ limit, no dependence on the external generalized momenta is expected, due to the structure of the leading-order diagrams (see Figure \ref{figLO4pts} and \cite{lahoche4}). Setting $\mu_1 = \mu$ and $\mu_2 = \nu$ (with $\mu \neq \nu$), the local approximation leads to the following, at leading order in $\delta p^2$
\begin{align}
\boxed{\frac{8 u_4(k)}{N}\left(1+\frac{R_k(p_\mu^2)-R_k(p_\nu^2)}{\delta p^2}\right)  \tilde{G}_k^2(p_\mu^2,\omega^2_1) - \frac{\Sigma_k(p_{\mu}^2,\omega^2_1)-\Sigma_k(p_{\nu}^2,\omega^2_1)}{\delta p^2}=\mathcal{O}(\delta p^2/N)\,.}\label{secondderivWardstep2}
\end{align}
This equation shows that the variation of the self-energy with respect to the generalized momentum is of next-to-leading order ($\mathcal{O}(1/N)$), as expected. Due to the $1/N$ factor on the left-hand side, the propagator can be computed at leading order,
\begin{equation}
\tilde{G}_k(p_\mu^2,\omega^2_1)=\frac{1}{\omega_1^2+p_{\mu_1}^2+u_2(k)}\,,
\end{equation}
where $u_2(k)$ is the large $N$ effective mass, which is fixed by a closed equation recalled in \cite{lahoche2024largetimeeffectivekinetics}. Taking the limit $\delta p^2\to 0$, and defining the wave function renormalization $Z(k)$ as:
\begin{equation}
Z(k,\omega^2):=\frac{\dd}{\dd p^2_\mu} \tilde{G}^{-1}(p_\mu^2,\omega^2)\Big\vert_{p_\mu=0}=1-\frac{\dd}{\dd p^2_\mu} \Sigma(p_\mu^2,\omega^2)\Big\vert_{p_\mu=0}\,.
\end{equation}
Finally, setting to zero the external momenta:
\begin{equation}
\boxed{\frac{8 u_4(k)}{N} \frac{1+R^\prime_k(0)}{(\omega^2+\mu_2(k))^2}+Z(k,\omega^2)-1=0\,.}
\end{equation}

\section{Conclusion and open issues}\label{sec5}

In this paper, we have investigated the behavior of the renormalization group (RG) flow for a "$2+p$" quantum spin glass model, constructed by integrating out degrees of freedom labeled by the eigenvalues of the matrix-like disorder spectrum. In our approach, the effective coupling constants are defined with a specific normalization that ensures the effective number of sites on the underlying network—representing the system’s volume—remains constant. Crucially, we take the infinite-volume limit prior to performing the coarse-graining procedure.

This paper presents three main results:

\begin{itemize}

\item The first result concerns the Effective Vertex Expansion (EVE) method, which allows us to close the hierarchy around sextic interactions in the symmetric phase at the leading order of the derivative expansion. This method, initially introduced for tensorial field theories \cite{Lahochebeyond}, is particularly well-suited to our model due to its specific non-locality. Our conclusions align with our previous findings \cite{lahoche2024largetimeeffectivekinetics}, confirming the existence of finite-scale singularities driven by disorder, which we interpret as a signal that the perturbative theory is incomplete. The primary steps in constructing this approximation are detailed in the companion paper \cite{lahoche2024frequency}.

\item The second and primary contribution of this paper is a formalism designed to go beyond standard vertex expansions. This approach resembles a formal expansion around the vacuum of the Local Potential Approximation (LPA), where the local potential flow is "dressed" by the flow of non-local interactions. We derived the flow equations for various couplings, focusing on the next-to-leading order beyond the marginally renormalizable sector and examining asymptotic fixed-point solutions. Our numerical investigations indicate that divergences in the symmetric phase may cancel out; furthermore, in regimes where these divergences persist, the flow of non-local couplings becomes relevant—contrary to standard perturbative expectations. These results provide robust evidence for a phase transition between a regime where replicas are uncorrelated and one where their correlations are non-zero.

\item The third and last pertains to a non-trivial Ward identity arising from the gauge fixing required to construct the RG flow, which breaks the model's underlying $O(N)$ invariance. Owing to these Ward identities, the flow of the field strength renormalization $Z(k)$ in the symmetric phase is entirely determined by the $\mathcal{O}(N^0)$ RG flow of the quartic coupling.
\end{itemize}

The ultimate challenge remains the reconstruction of the complete theory space from an RG perspective to decipher the glassy transition. Several intermediate objectives are essential for this endeavor, particularly concerning the physical properties of the transition. Our upcoming work will focus on the 2PI formalism, which is particularly appropriate when the $2$-point function serves as the order parameter.

Looking ahead, we plan to address the following open issues:

\begin{enumerate}
\item Evaluating the reliability of the vertex expansion by extending it beyond the leading order of the derivative expansion.
\item Incorporating coarse-graining across both eigenvalue and temporal dimensions.
\item Developing approximation schemes valid beyond the deep IR.
\item Assessing the accuracy of the derivative expansion by explicitly computing anomalous dimensions and more systematically incorporating non-local interactions into the truncations.
\item Investigating the spontaneous breaking of time-translation invariance.
\end{enumerate}
Finally, we intend to explore the UV regime, where the non-trivial scaling resulting from coarse-graining over the Wigner spectrum likely plays a significant role. Specifically, outside the critical regime investigated in \cite{lahoche2024largetimeeffectivekinetics}, the canonical dimension becomes dependent on the relevant couplings $u_2$ and $q^\prime$, presenting a novel challenge in the construction of the effective phase space.

\noindent
\section*{Acknowledgment:} Vincent Lahoche and the authors warmly thank Mrs. Dalila Derdar for these very stimulating and inspiring exchanges from the first stages of the realization of this work. Without these highly stimulating exchanges, this work would not have been the same.
\pagebreak

\appendix

\section{Classical solution for $p=0$}\label{App2}

\begin{figure}
\begin{center}
\includegraphics[scale=0.8]{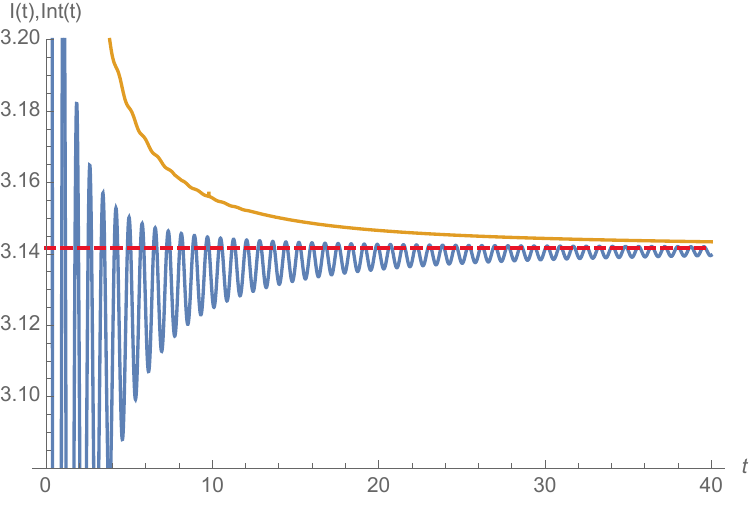}
\end{center}
\caption{Behavior of the function $\text{I}(t)$ (blue curve) and the average $\text{Int}(t)$ (yellow). The red curve is for the value $\pi$.}\label{figplotI}
\end{figure}

In this section, we examine the solution to the classical equation of motion for $p = 0$; the corresponding quantum problem was solved in the large-$N$ limit in our previous work \cite{lahoche2024largetimeeffectivekinetics}. The method proposed here is based on \cite{bray1994theory} and has also been considered for the dynamics of classical spin glasses in \cite{Dominicis, lahoche2023low}. The equation of motion for the classical particle, given a quartic potential, reads:
\begin{equation}
m_0 \frac{d^2 x_i}{dt^2}= - \sum_{j=1}^N K_{ij} x_j-h_1 x_i-\frac{h_2}{6} \frac{\bm x^2}{N} x_i\,.
\end{equation}
It is suitable to work in the basis $\{u^{(\mu)}\}$, where the matrix $K$ is diagonal, and the move equation becomes:
\begin{equation}
m_0 \frac{d^2 x_\mu}{dt^2}= - \left(\xi_\mu+h_1+\frac{h_2}{6} \frac{\bm x^2}{N}\right) x_\mu\,,\label{classicalmovebis}
\end{equation}
where $x_\mu := \sum_{i=1}^N x_i u_i^{(\mu)}$ and $\{\xi_\mu\}$ denote the eigenvalues of $K$ (see equation \eqref{eigenequation}). The equation of motion \eqref{classicalmovebis} is non-linear and difficult to solve exactly. However, in the large-$N$ regime, it is reasonable to assume that $a(t) := \bm{x}^2(t)/N$ self-averages (a phenomenon known as the quenched regime) and decouples from the explicit $x$-dependence in the equation of motion:
\begin{equation}
\ddot{x}_\mu= - \omega_\mu^2(t) x_\mu\,,\label{effclassicalmove}
\end{equation}
where:
\begin{equation}
\omega_\mu^2:=\frac{p_\mu^2+m^2+\frac{h_2}{6} a(t)}{m_0}\,,
\end{equation}
where we have used definitions \eqref{generalizedmomentadef} and \eqref{generalizedmassdef} for $p_\mu^2$ and $m^2$, respectively. Each component $x_\mu$ evolves according to the equation of motion for a time-dependent harmonic oscillator—a problem encountered in various fields of physics, such as cosmology, plasma physics, optics, and geophysics (see \cite{fiore2022time} and references therein). We will focus on the simpler regime where $a(t)$ is nearly constant and employ perturbation theory:
\begin{equation}
a(t)=a_0+\epsilon(t)\,,
\end{equation}
assuming $\vert \epsilon(t) \vert \ll 1$. Let us decomposes the solution $x_\mu(t)$ in series:
\begin{equation}
x_\mu(t)=x_\mu^{(0)}(t)+x_\mu^{(1)}(t)+x_\mu^{(2)}(t)+\cdots\,,
\end{equation}
where $x_\mu^{(n)}(t)$ is assumed to be of order $\epsilon^n(t)$. The solution $x_\mu^{(0)}(t)$ is:
\begin{equation}
x_\mu^{(0)}(t)=A \cos \left( \omega_\mu^{(0)}t+\phi_0\right)\,,
\end{equation}
where we used $\omega_\mu^{(0)}$ defined as:
\begin{equation}
\omega_{\mu}^{(0)}:=\frac{p_\mu^2+m^2+\frac{h_2}{6} a_0}{m_0}\,.
\end{equation}
In the large $N$ limit, we must have:
\begin{equation}
\frac{1}{N}\sum_\mu\, (x_\mu^{(0)}(t))^2 \to \int \rho(p^2) (x^{(0)}(p^2,t))^2=:2\pi A^2I(t)\,,
\end{equation}
where the function $x^{(0)}(p^2,t)$ is the continuum limit of the series $x_\mu^{(0)}(t)$. The computation of the integral involves the regularized confluent hypergeometric function:
\begin{equation}
I(t)=\pi \left[  \, _0\tilde{F}_1\left(;2;-\frac{4 t^2}{m_0^2}\right) \cos \left(2 \left(\frac{t \left(\frac{a_0 h_2}{6}+m^2+2\right)}{m_0}+\phi _0\right)\right)+1\right]\,,
\end{equation}
and it shows on Figure \ref{figplotI} for the equilibrium condition (low classical energy configuration):
\begin{equation}
a_0=-\frac{6 m^2}{h_2}\,.
\end{equation}
We furthermore define the time averaging:
\begin{equation}
\text{Int}(t)=\frac{1}{t}\int_0^t\,I(t^\prime) \dd t^\prime\,,
\end{equation}
also plotted on Figure \ref{figplotI}, and we have:
\begin{equation}
\lim_{t\to \infty} \text{Int}(t)=\pi\,,
\end{equation}
then, if we assume that the trajectory oscillates weakly around $a_0$, we get:
\begin{equation}
A=\sqrt{2a_0}=\sqrt{-\frac{12 m^2}{h_2}}\,,
\end{equation}
and we have for the zero-order solution:
\begin{equation}
x_\mu^{(0)}(t)=\sqrt{-\frac{12 m^2}{h_2}}\, \cos \left( \omega_\mu^{(0)}t+\phi_0\right)\,.
\end{equation}

\pagebreak
\printbibliography[title={Bibliography}]

\end{document}